\documentclass[12pt]{article}
\usepackage{amsmath,epsf,amssymb,latexsym,pifont,enumerate,cite}
\newtheorem{theorem}{Theorem}
\newtheorem{prop}{Proposition}

\newif\iffigs\figstrue

\DeclareFontFamily{U}{rsf}{}
\DeclareFontShape{U}{rsf}{m}{n}{
  <5> <6> rsfs5 <7> <8> <9> rsfs7 <10-> rsfs10}{}
\DeclareMathAlphabet\Scr{U}{rsf}{m}{n}

\def\pplogo{\vbox{\kern-\headheight\kern -29pt
\halign{##&##\hfil\cr&{
\ppnumber}\cr\rule{0pt}{2.5ex}&\ppdate\cr}
}}
\makeatletter
\def\ps@firstpage{\ps@empty \def\@oddhead{\hss\pplogo}%
  \let\@evenhead\@oddhead 
}
\def\maketitle{\par
 \begingroup
 \def\thefootnote{\fnsymbol{footnote}}
 \def\@makefnmark{\hbox{$^{\@thefnmark}$\hss}}
 \if@twocolumn
 \twocolumn[\@maketitle]
 \else \newpage
 \global\@topnum\z@ \@maketitle \fi\thispagestyle{firstpage}\@thanks
 \endgroup
 \setcounter{footnote}{0}
 \let\maketitle\relax
 \let\@maketitle\relax
 \gdef\@thanks{}\gdef\@author{}\gdef\@title{}\let\thanks\relax}
\makeatother

\def\C{{\mathbb C}}
\def\P{{\mathbb P}}

\def\R{{\mathbb R}}
\def\Z{{\mathbb Z}}
\def\H{{\mathbb H}}
\def\Hom{\operatorname{Hom}}
\def\Hol{\operatorname{Hol}}

\def\Vol{\operatorname{Vol}}

\def\tr{\operatorname{tr}}
\def\Pic{\operatorname{Pic}}

\def\Img{\operatorname{Im}}

\def\SO{\operatorname{SO}}
\def\Sl{\operatorname{SL}}
\def\Gl{\operatorname{GL}}
\def\GO{\operatorname{O{}}}
\def\SU{\operatorname{SU}}
\def\GU{\operatorname{U{}}}
\def\Sp{\operatorname{Sp}}

\def\rank{\operatorname{rank}}
\def\Spin{\operatorname{Spin}}

\def\HS#1{{\bf F}_{#1}}

\def\sm{$\sigma$-model}
\def\nlsm{non-linear \sm}

\def\CY{Calabi--Yau}

\def\cM{{\Scr M}}

\def\cH{{\Scr H}}

\def\cF{{\Scr F}}
\def\cX{{\Scr X}}
\def\cG{{\Scr G}}
\def\ff#1#2{{\textstyle\frac{#1}{#2}}}
\def\RoR{$R\leftrightarrow1/R$}
\def\spnh{\Spin(32)/\Z_2}

\def\labto#1{\mathrel{\mathop\to^{#1}}}

\newenvironment{difficult}{\begin{list}{\ding{36}}
	{\setlength{\leftmargin}{0.6\leftmargini}}
	\item\footnotesize}
  {\end{list}}

\begin{document}
\setcounter{page}0
\def\ppnumber{\vbox{\baselineskip14pt\hbox{DUKE-CGTP-00-01}
\hbox{hep-th/0001001}}}
\def\ppdate{January 2000} \date{}

\title{\LARGE Compactification, Geometry and Duality: $N=2$.\\[10mm]}
\author{
Paul S. Aspinwall\\[10mm]
\normalsize Center for Geometry and Theoretical Physics, \\
\normalsize Box 90318, \\
\normalsize Duke University, \\
\normalsize Durham, NC 27708-0318\\[10mm]
}

{\hfuzz=10cm\maketitle}

\def\Large{\large}
\def\LARGE{\large\bf}


\begin{abstract}
We review the geometry of the moduli space of $N=2$ theories in four
dimensions from the point of view of superstring compactification. The
cases of a type IIA or type IIB string compactified on a \CY\
threefold and the heterotic string compactified on K3$\times T^2$ are
each considered in detail. We pay specific attention to the
differences between $N=2$ theories and $N>2$ theories. The moduli
spaces of vector multiplets and the moduli spaces of hypermultiplets
are reviewed. In the case of hypermultiplets this review is limited by
the poor state of our current understanding. Some peculiarities such
as ``mixed instantons'' and the non-existence of a universal
hypermultiplet are discussed.
\end{abstract}

\vfil\break

\enlargethispage{2\baselineskip}\thispagestyle{empty}
\tableofcontents


\section{Introduction}    \label{s:int}

One of the most basic properties one may study about a class of string
compactifications is its moduli space of vacua. If the class is
suitably chosen one may find this a challenging subject which probes
deeply into our understanding of string theory. In four dimensions it
is the $N=2$ cases which provide the ``Goldilocks'' theories to
study. As we will see, $N=4$ supersymmetry is too constraining and
determines the moduli space exactly, leaving no room for interesting
corrections from quantum effects. $N=1$ supersymmetry is highly
unconstrained leaving the possibility that our supposed moduli acquire
mass ruining the moduli space completely. $N=2$ however is {\em just right\/}
--- quantum effects are not potent enough to kill the moduli but they
can affect the structure of the moduli space. (It is therefore a pity
that the real world does not have $N=2$ supersymmetry --- such a
theory is necessarily non-chiral.)

The subject of $N=2$ compactifications is enormous and we will present
here only a rather biased set of highlights. These lectures will be
sometimes closely-related to a set of lectures I gave at TASI 3 years ago
\cite{me:lK3}. Having said that, the focus of these lectures differs
from the former and the set of topics covered is not identical.
I will however often refer to \cite{me:lK3} for details of certain subjects.

Related to the problem of finding the moduli space of a class of 
theories is the
following problem in string duality. Consider these
four possibilities for obtaining an $N=2$ theory in four
dimensions:
\begin{enumerate}
\item A type IIA string compactified on a \CY\ threefold $X$.
\item A type IIB string compactified on a \CY\ threefold $Y$.
\item An $E_8\times E_8$ heterotic string compactified on K3$\times T^2$.
\item A $\spnh$ heterotic string compactified on K3$\times T^2$.
\end{enumerate}
Can we find cases where the resulting 4 dimensional physics is
identical for two or more of these possibilities and, if so, how do
we match the moduli of these theories to each other? 
This is a story that began with \cite{KV:N=2,FHSV:N=2} some time ago
but many details are poorly-understood to this day.
One might suppose
that knowing the moduli space of each theory listed above is a
prerequisite for solving this problem but actually it is often useful,
as we will see,
to consider this duality problem at the same time as the moduli space
problem. Note that there are other possibilities for
producing $N=2$ theories in four dimensions such as the type I open
string on K3$\times T^2$. We will stick with the four listed above in
these lectures as they are quite sufficient for our purposes.

As will be discussed shortly this problem breaks up into two
pieces. One factor of the moduli space consists of the vector
multiplet moduli space and the other factor consists of the
hypermultiplet moduli space. By most criteria the moduli space of
vector multiplets is well-understood today. This complex K\"ahler
space can be modeled exactly in terms of the deformation space of a
\CY\ threefold. We will therefore be able to review this subject
fairly extensively.

In contrast the hypermultiplet moduli space remains a subject of
research very much ``in progress''. We will only be able to discuss in
detail the classical boundaries of these moduli spaces. The interior
of these spaces may offer considerable insight into string theory but
we will only be able to cover some tantalizing hints of such
possibilities.

These lectures divide naturally into three sections. In section
\ref{s:gen} we discuss generalities about moduli spaces of various
numbers of supersymmetries in various numbers of dimensions. Although
these lectures are intended to focus on the case of $N=2$ in four
dimensions, there are highly relevant observations that can be made by
considering other possibilities. Of particular note in this section is
the rigid structure which emerges with more supersymmetry than the
case in question.

The heart of these lectures then consists of a discussion of the
vector multiplet moduli space in section~\ref{s:vec} and then the
hypermultiplet moduli space in section~\ref{s:hyp}. It is perhaps
worth mentioning again that these lectures do not do justice to this
vast subject and should be viewed as a biased account. Topic such as
open strings, D-branes and M-theory have been neglected only because of the
author's groundless prejudices.

The paragraphs starting with a ``\ding{36}'' are technical and can be
skipped if the reader does not wish to be embroiled in subtleties.


\section{General Structure}  \label{s:gen}


\subsection{Holonomy}  \label{ss:hol}

We begin this section with a well-known derivation of key properties
of moduli spaces based on $R$-symmetries and holonomy arguments. We
should warn the more mathematically-inclined reader that we shall not
endeavour to make completely watertight rigorous statements in the
following. There may be a few pathological special cases which
circumvent some of our (possibly implicit) assumptions.

Suppose we are given a vector bundle, $V$, with a connection. We may define
the ``holonomy'', $\Hol(V)$, of this bundle as the group generated by
parallel transport around loops in the base with respect to this
connection. (A choice of basepoint is unimportant.) We may also define
the {\em restricted\/} holonomy group $\Hol^0(V)$ to be generated by {\em
contractable\/} loops.

This notion can be very useful when applied to supersymmetric field
theories as noted in \cite{BagW:N=2}. First let us consider the moduli
space of a given class of 
theories. We will consider the moduli space as the base space of a
bundle. Note that the moduli space of theories, $\cM$, is equipped with a
natural metric --- that of the sigma-model. The tangent directions in
the moduli space are given by the massless scalar fields with completely flat
potentials. These massless fields may thus be given vacuum expectation
values leading to a deformation of the theory. Let us denote these
moduli fields $\phi^i$, $i=1,\ldots,\dim(\cM)$. The low-energy effective
action in the uncompactified space-time is then given by
\begin{equation}
  \int d^dx\,\sqrt{g} G_{ij}
  \partial^\mu\phi^i\partial_\mu\phi^j+\ldots  \label{eq:mod}
\end{equation}
where $G_{ij}$ is our desired metric on $\cM$.
We therefore have a natural torsion-free connection 
on the tangent bundle of $\cM$
given by the Levi-Civita connection with respect to this
metric. 

Now consider the supersymmetry generators given by spinors
$Q_A$, $A=1,\ldots,N$, where as usual $N$ denotes the number of
supersymmetries (we suppress the spinor index). These objects are
representations of $\Spin(1,d-1)$ and are
\begin{itemize}
\item {\em Real\/} if $d=1,2,3\mod8$
\item {\em Complex\/} if $d=0,4\mod8$
\item {\em Quaternionic\/} (or {\em symplectic Majorana\/}) if $d=5,6,7\mod8$.
\end{itemize}

The bundle of supersymmetry generators over $\cM$ will also
have a natural connection related to that on the tangent bundle. 
The key relation in supersymmetry is the equation
\begin{equation}
  \gamma_{\alpha\beta}^\mu\{\overline Q_{\bar A}^\alpha,
	Q_B^\beta\} = \delta_{\bar AB}P^\mu,  \label{eq:susy}
\end{equation}
where $\gamma$ are the usual gamma matrices, $P$ is translation and
the bars in this equation are to be interpreted 
according to whether the spinors are real, complex or quaternionic.%
\footnote{We ignore central charges which are irrelevant for this argument.}
Because parallel transport must preserve $\delta_{\bar AB}$ in
(\ref{eq:susy}) we see 
immediately that under holonomy $Q_A$ must transform as a fundamental 
representation of 
\begin{itemize}
\item $\SO(N)$ if $d=1,2,3\mod8$
\item $\GU(N)$ if $d=0,4\mod8$
\item $\Sp(N)$ if $d=5,6,7\mod8$,
\end{itemize}
if the loop around which we transport is contractable.
These groups are the ``$R$-symmetries'' of the supersymmetric field
theories and give $\Hol^0$ of this bundle. We also note that in $4M+2$
dimensions, for integer $M$, the supersymmetries are chiral. This
means that we consider left and right supersymmetries separately as we
will illustrate in some examples below.

The massless scalar fields live in supermultiplets. Within each
supermultiplet the set of scalar fields will form a particular
(possibly trivial) representation of the $R$-symmetry. We refer to
\cite{Strath:N} for a detailed account of this. Occasionally the 
supermultiplet contains only one scalar component and this then 
transforms trivially under $R$. So long as this is not the case the
holonomy of our tangent bundle is related to the $R$-symmetry. 

We may be more precise than this. As we go around a loop in $\cM$ the
scalars within every given supermultiplet will be mixed simultaneously
by the
$R$-symmetry. The supermultiplets themselves may also be mixed as a
whole into each other by holonomy. This implies that, so long as the 
scalars transform nontrivially under $R$, {\em the holonomy of
the tangent bundle is factorized with the $R$-symmetry forming one factor.}
It is important to note however that we
may not mix a scalar from one supermultiplet freely with any scalar
from another supermultiplet in a way that violates this factorization.
This is incompatible with the detailed supersymmetry transformation
laws (as the reader might verify if they are unconvinced).

Note in particular that the scalars within two different {\em types\/} of
supermultiplets can never mix under holonomy. This is a useful
observation given the following due to De Rham (see, for example,
\cite{Besse:E}) 
\begin{theorem}
If a Riemannian manifold is complete, simply connected and if the
holonomy of its tangent bundle with respect to the Levi-Civita
connection is reducible, then this manifold is a product metrically.
\end{theorem}
Thus if $\cM$ is simply-connected we see that $\cM$ 
factorizes exactly into parts labelled by the type of supermultiplet
containing the massless scalars. If $\cM$ is not simply-connected we
may pass to the universal cover and use this theorem again. The
general statement is therefore that the moduli space factorizes up to
the quotient of a discrete group acting on the product.

\begin{difficult}
Actually we should treat the word ``complete'' in the above theorem
with a little more care. There are nasty points at finite distance in
the moduli space where the manifold structure breaks down. These
points also lead to a breakdown in the factorization of the
moduli space. These extremal transitions will be studied more in
section~\ref{ss:ex}. We should only say that the moduli space
factorizes locally away from such points.
\end{difficult}

We may now analyze the structure of each factor of $\cM$ from the
Berger-Simons theorem (for an account of this we refer again to
\cite{Besse:E}) which states that the manifold must appear as a row in the
following list:\footnote{The notation $\Sp(1).\Sp(n)$ means
$\Sp(1)\times\Sp(n)$ divided by the diagonal central $\Z_2$.}
\begin{equation}
\begin{array}{c|c}
\Hol^0&\dim(\cM)\\
\hline
\SO(n)&n\\
\GU(n)&2n\\
\SU(n)&2n\\
\Sp(1).\Sp(n)&4n\\
\Sp(n)&4n\\
\Spin(7)&8\\
G_2&7
\end{array}   \label{eq:nonsym}
\end{equation}
or be a ``symmetric space'' (which we will define shortly).
Note that the following names are given to some of these holonomies:
\begin{itemize}
\item {\em K\"ahler\/} if $\Hol\subset\GU(n)$,
\item {\em Ricci-flat K\"ahler\/} if $\Hol\subset\SU(n)$,
\item {\em Quaternionic K\"ahler\/} if $\Hol\subset\Sp(1).\Sp(n)$,
\item {\em Hyperk\"ahler\/} if $\Hol\subset\Sp(n)$.
\end{itemize}

\begin{table}[t]
\begin{center}
\renewcommand\arraystretch{1.1}
\begin{tabular}{|l|c|c|}
\hline
&Manifold&Real Dimension\\
\hline
{\bf A I}&$\Sl(n,\R)/\SO(n)$&$\ff12(n-1)(n+2)$\\
{\bf A II}&$\SU^*(2n)/\Sp(n)$&$(n-1)(2n+1)$\\
{\bf A III}&$\SU(p,q)/\mathrm{S}(\GU(p)\times\GU(q))$&$2pq$\\
{\bf BD I}&$\SO_0(p,q)/(\SO(p)\times\SO(q))$&$pq$\\
{\bf D III}&$\SO^*(2n)/\GU(n)$&$n(n-1)$\\
{\bf C I}&$\Sp(n,\R)/\GU(n)$&$n(n+1)$\\
{\bf C II}&$\Sp(p,q)/(\Sp(p)\times \Sp(q))$&$4pq$\\
{\bf E I}&$E_{6(6)}/\Sp(4)$&42\\
{\bf E II}&$E_{6(2)}/(\SU(6)\times\SU(2))$&40\\
{\bf E III}&$E_{6(-14)}/(\SO(10)\times\GU(1))$&32\\
{\bf E IV}&$E_{6(-26)}/F_4$&26\\
{\bf E V}&$E_{7(7)}/\SU(8)$&70\\
{\bf E VI}&$E_{7(-5)}/(\SO(12)\times\SU(2))$&64\\
{\bf E VII}&$E_{7(-25)}/(E_6\times\GU(1))$&54\\
{\bf E VIII}&$E_{8(8)}/\SO(16)$&128\\
{\bf E IX}&$E_{8(-24)}/(E_7\times\SU(2))$&112\\
{\bf F I}&$F_{4(4)}/(\Sp(3)\times\SU(2))$&28\\
{\bf F II}&$F_{4(-20)}/\SO(9)$&16\\
{\bf G I}&$G_{2(2)}/(\SU(2)\times\SU(2))$&8\\
\hline
\end{tabular}
\end{center}
\caption{Symmetric Spaces} \label{tab:s}
\end{table}

A symmetric space is a Riemannian manifold which admits a ``parity''
$\Z_2$-symmetry about every point. This parity symmetry acts as $-1$
in every direction on the tangent space. All symmetric spaces are of
the form $G/H$ for groups $G$ and $H$, where the holonomy is given by
$H$. They have been classified by 
E.~Cartan and we list all the noncompact forms in table \ref{tab:s}.%
\footnote{We have been sloppy about the precise global form of the group
$H$. As listed one often needs to quotient by a finite group to get the correct
answer. For example in entry ``\textbf{E V}'', $\SU(8)$ is not a
subgroup of $E_{7(7)}$ --- the correct form of $H$ should be
$\SU(8)/\Z_2$. The notation $\SO_0(p,q)$ refers to the part of the Lie
group connected to the identity.}
The noncompact forms are the ones relevant to moduli spaces.

A key point to note here is that the symmetric spaces are {\em rigid\/} ---
they have no deformations of the metric which would preserve the
holonomy. The same is not true for the non-symmetric spaces listed in
(\ref{eq:nonsym}). Thus if
the holonomy is of a type which forces a symmetric space as the only
possibility we will refer to this as a {\em rigid\/} case.

Let us consider a few examples. 
\begin{itemize}
\item $N=(1,1)$ in 6
dimensions (i.e., one left-moving supersymmetry and one right-moving
supersymmetry). This implies that the $R$-symmetry is
$\Sp(1)\times\Sp(1)=\SO(4)$ (up to irrelevant discrete
groups). Analysis of the supermultiplets shows that matter
supermultiplets have 4 scalars transforming as a {\bf 4} of
$\SO(4)$. If we only had one such supermultiplet we could say nothing
about the moduli space as a generic Riemannian manifold of 4
dimensions has holonomy $\SO(4)$. If we have a generic number, $n$, of
supermultiplets and assuming the moduli space doesn't factorize
unnaturally then a quick look at the list above shows that the only
possibility is the symmetric space.
$\SO_0(4,n)/(\SO(4)\times\SO(n))$. Thus this case is rigid. The
gravity supermultiplet has a single scalar giving an additional factor
of $\R$ to the moduli space.
\item $N=(2,0)$ in 6
dimensions. This implies that the $R$-symmetry is
$\Sp(2)=\SO(5)$ (up to irrelevant discrete
groups). Analysis of the supermultiplets shows that matter
supermultiplets have 5 scalars transforming as a {\bf 5} of
$\SO(5)$. If we have a generic number, $n$, of
supermultiplets and assuming the moduli space doesn't factorize
unnaturally then the list above shows that the only
possibility is the symmetric space
$\SO_0(5,n)/(\SO(5)\times\SO(n))$. Thus this case is rigid. There are
no further moduli.
\item $N=4$ in 4
dimensions. This implies that the $R$-symmetry is
$\GU(4)=\SO(6)\times\GU(1)$ (up to irrelevant discrete
groups). Analysis of the supermultiplets shows that matter
supermultiplets have 6 scalars transforming as a {\bf 6} of
$\SO(6)$. If we have a generic number, $n$, of
supermultiplets and assuming the moduli space doesn't factorize
unnaturally then the only
possibility for this factor is the symmetric space
$\SO_0(6,n)/(\SO(6)\times\SO(n))$. Thus this case is rigid. The
gravity multiplet contains a complex scalar transforming under the
$\GU(1)$ factor of the holonomy. By holonomy arguments this contributes a
complex K\"ahler factor to the moduli space. Closer analysis of this
supergravity shows that this factor is actually $\Sl(2,\R)/\GU(1)$.
\end{itemize}

This last example demonstrates an important point. Analysis of the
$R$-symmetry may be sufficient to imply that we have a rigid moduli
space but sometimes the moduli space is rigid even when the holonomy
may imply otherwise. A more detailed analysis of the supergravity
action is required in some cases to show that we indeed have a
symmetric space. The rule of thumb is as follows: {\em If we have
maximal (32 supercharges, e.g. $N=8$ in four dimensions) or
half-maximal (16 supercharges, e.g. $N=4$ in
four dimensions) supersymmetry then, and usually only then, is the moduli
space rigid.} Note that there are a few strange examples such as
\cite{FHSV:N=2} where the moduli space is rigid even when there are
fewer than 16 supercharges.


\subsection{U-Duality}     \label{ss:U}

In this section we will focus on global properties of the rigid moduli spaces.
The analysis of the moduli spaces so far is not quite complete. The
problem is that the moduli space need not be a {\em manifold}. There
may be singular points corresponding to the theories with special
properties. In the rigid case however the fact that the moduli space
is symmetric wherever it is not singular is a very powerful
constraint.

Let us suppose first that we have an {\em orbifold\/} point. That is a region
in the moduli space which looks {\em locally\/} like a manifold
divided by a discrete group fixing some point $x$. Away from the fixed
point set the moduli space is symmetric and thus ``homogeneous''. That
is, there exist a transitive set of translation symmetries. Assuming
geodesic completeness of the moduli space, these translations may be
used to extend the local orbifold property to a global one. That is,
the moduli space is {\em globally\/} of the form of a manifold
divided by a discrete group \cite{AP:T}.

This homogeneous structure of the moduli space may also be used to
rule out other possibilities of singularities which occur at finite
distance. Consider beginning at a smooth point in moduli space and
approaching a singularity. The homogeneous structure implies that
nothing about the local structure of the moduli space may change as
you approach the singularity --- everything happens suddenly as you
hit the singularity. This rules out every other type of ``reasonable''
singularity that one may try to put in the moduli space. To be completely
rigorous would require us to make precise technical
definitions about the allowed geometry of the moduli space. Instead we
shall just assert here that any type of singularity at finite distance
that one might think of (such as a conifold) would ruin the
homogeneous nature of the moduli space and so is not allowed in the
rigid case.

We therefore arrive at the conclusion that the {\bf only allowed global
form of a rigid moduli space is of a symmetric
space divided by a discrete group.}

This implies that any analysis of the moduli space of string theories
in the case of maximal or half-maximal supersymmetry comes down to
question of this discrete group. This group is precisely the group
known as S-duality, T-duality or U-duality depending on the context.

Many examples of such dualities were discussed in \cite{me:lK3} and we
refer the reader there for details as well as references. For example
the general rule is that a space locally of the form
$\SO_0(p,q)/(\SO(p)\times\SO(q))$ becomes
\begin{equation}
  \GO(\Upsilon_{p,q})\backslash\GO(p,q)/(\GO(p)\times\GO(q)),
\end{equation}
where $\Upsilon_{p,q}$ is some lattice (often even and unimodular) of
signature $(p,q)$ and $\GO(\Upsilon_{p,q})$ is its discrete group of isometries.

Indeed the only interesting question one may ask about the moduli
space in the rigid case is what exactly this discrete group is! Any
quantum corrections to the local structure are not allowed due to
rigidity. It is not therefore surprising that S, T and U-duality are
so ubiquitous when studying theories with a good deal of
supersymmetry.
As we will we see however, the picture becomes quite different when
the supersymmetry is less that half-maximal.

One final word of warning here. We have not been clear about what we
mean by a ``class'' of string theories. If we determine the moduli
space of some kind of string compactified on some kind of space, up to
topology, then our moduli space may have numerous disconnected
components. In this case the above results apply to each component
separately. This reducibility often happens when we have half-maximal
supersymmetry.


\subsection{Eight supercharges}     \label{ss:N2}

Now we turn our attention to theories with quarter-maximal
supersymmetry, or a total of 8 supercharges. Here we will also specify
how one might obtain such a theory from string theory.

If we compactify a ten-dimensional supersymmetric theory on
$\R^{1,d-1}\times M$, where $M$ is some compact manifold, then
holonomy arguments may be again used to determine the number of
unbroken supersymmetries in $\R^{1,d-1}$. This time it is the holonomy
of the compact space $M$ rather than the moduli space which we
analyze. The basic idea is roughly that a symmetry in ten dimensions will be
broken by the holonomy of (a suitable bundle on) $M$ to the {\em
centralizer\/} of this holonomy group. That is, a symmetry in
uncompactified space is broken if it can be transformed by parallel
transport around a loop in the internal compactified dimensions.%
\footnote{While this seems a very reasonable statement it is probably
not rigorous. Breaking the gauge group of the heterotic string in this
way does not always lead to the correct global form.}

We begin with $N=(1,0)$ in six dimensions. We may obtain this by
compactifying a heterotic string theory on a four-dimensional manifold
with holonomy $\SU(2)$. The only such manifold is a K3 surface. We refer to
\cite{me:lK3} for an explanation of these points.

The $R$-symmetry in this case is $\Sp(1)$. An analysis of
supermultiplets show that scalars may occur in either of two types:
\begin{enumerate}
\item The {\em Hypermultiplet\/} contains 4 scalars which we may view
as a quaternion. The holonomy $\Sp(1)$ may then be viewed as
multiplication on the left by another quaternion of unit norm.
\item The {\em Tensor\/} multiplet contains a single scalar. Thus
holonomy tells us nothing interesting.
\end{enumerate}
Note that the vector supermultiplet contains no scalars. The moduli
space of such theories will locally factorize into a moduli space
of hypermultiplets, which will be {\em quaternionic K\"ahler\/} and a
moduli space of tensor multiplets.

Now let us consider $N=2$ theories in four dimensions. We may obtain
this by compactifying a heterotic string on a six-dimensional manifold
with holonomy $\SU(2)$. The only such manifold is a product of a K3
surface and a 2-torus (or a finite quotient thereof). Alternatively we
may compactify a type IIA or IIB superstring (which has twice as much
supersymmetry than the heterotic string) on a manifold of holonomy
$\SU(3)$. As usual, we will refer to such a Ricci-flat K\"ahler
manifold as a ``\CY\ threefold''. Again we refer to \cite{me:lK3} for
extensive details and references on these points.

The $R$-symmetry in this case is $\GU(2)=\Sp(1)\times\GU(1)$.
An analysis of
supermultiplets show that scalars may occur in either of two types:
\begin{enumerate}
\item The {\em Hypermultiplet\/} contains 4 scalars which we may view
as a quaternion. The holonomy $\Sp(1)$ may then be viewed as
multiplication on the left by another quaternion of unit norm.
\item The {\em Vector\/} multiplet contains 2 scalars transforming
as a complex scalar under the $\GU(1)$ factor of the holonomy.
\end{enumerate}
Because of this the moduli
space of such theories locally factorizes into a moduli space
of hypermultiplets, which is {\em quaternionic K\"ahler\/} and
we denote $\cM_H$, and a moduli space of vector multiplets which
is K\"ahler and we denote $\cM_V$.

Although we will see below that $\cM_V$ is not any old complex K\"ahler
manifold and $\cM_H$ is not any old quaternionic K\"ahler manifold, it
is true that they are not completely determined by supersymmetry and
consequently have
deformations. Thus in contrast to the rigid cases with a lot of
supersymmetry, $N=2$ theories in 4 dimensions (and $N=(1,0)$ theories
in six dimensions, etc.) can have interesting quantum corrections which warp
the moduli space away from that which would be expected classically.

Further analysis of $\cM_H$ by Bagger and Witten \cite{BagW:N=2}
yielded a property which is worth noting. They showed that the
scalar curvature of $\cM_H$ is negative and proportional to
the gravitational coupling constant. Thus $\cM_H$ is not
hyperk\"ahler unless the gravitational coupling is taken to zero.

What is particularly nice about $N=2$ theories is that their moduli cannot
gain mass through quantum effects. This is to be contrasted with the
$N=1$ case in four dimensions where the moduli can become
massive. This is discussed in M.~Dine's lectures in this volume.


\subsection{Type II compactification} \label{ss:II}

Let us now turn to the coarse structure of the moduli space of type
IIA and type IIB compactifications on \CY\ threefolds.

Begin with a type IIA string compactified on a \CY\ threefold $X$. To
leading order we demand that the metric be Ricci-flat. Actually this
statement is not exact and receives quantum corrections. At higher
loop in the \nlsm\ we discuss below, the metric is warped away from
the Ricci-flat solution \cite{GVZ:4loop} and when 
one takes nonperturbative effects into account it is unlikely that one
can faithfully represent $X$ in terms of a Riemannian metric at
all. We will see this breakdown of Riemannian geometry later in
section~\ref{sss:Tsnap}.

What is true however is that if the \CY\ is large then the Ricci-flat
metric is a good approximation. Thus we may at least get the dimension
of the moduli space correct using this metric. Thanks to Yau's proof
of the Calabi conjecture \cite{Yau:} we do not have to undertake the
unpleasant (and as yet unsolved) problem of explicitly constructing
the Ricci-flat metric. We may instead assert its unique existence
given a complex structure on $X$ together with the cohomology class of
its K\"ahler form, $[J]\in H^2(X,\R)$.

Deformation of the complex structure of $X$ yields $h^{2,1}(X)$ {\em
complex\/} moduli whereas the K\"ahler form degree of freedom yields
$h^{1,1}(X)$ {\em real\/} degrees of freedom. (We refer to section 2
of \cite{Greene:TASI} for a discussion of the classical geometry we
use here.) The deformation of the Ricci-flat metric on $X$ thus
produces $h^{1,1}(X) + 2h^{2,1}(X)$ real moduli.

There is also the ten-dimensional dilaton, $\Phi$, which controls the
string coupling constant. This contributes one real modulus.

All the remaining moduli arise from objects which na\"\i vely appear as
$p$-forms in the ten-dimensional type II theory. The basic idea is
that both type II strings (and indeed all closed string theories)
have a ``$B$-field'' 2-form arising in the NS-NS spectrum while the
type IIA string also contains a 1-form and a 3-form from the R-R sector and
the type IIB string contains a 0-form, a 2-form and a self-dual 4-form
from the R-R sector. We refer to \cite{Pol:books} for this basic
property of string theory.

This description of these ten-dimensional fields in terms of de Rham
cohomology is rather vague and unfortunately does not really tell us
the full truth about these objects and the resulting degrees of
freedom they yield as moduli. The aspects which are poorly described
concern both what happens when $X$ is singular and
the discrete degrees of freedom (arising from torsion in the
cohomology for example).

We cannot pretend to understand the basic nature of string theory
until we have a better description of the geometry of these fields. At
present there are two leading contenders namely ``gerbes'' and
``K-theory''. Unfortunately at the time of writing, neither of these
theories together with its application to string theory is completely
understood although the subject is maturing rapidly.

The idea of a gerbe is best understood by first considering the R-R
1-forms of the type IIA string. These 1-forms are believed to describe
a $\GU(1)$ gauge theory in the ten-dimensional spacetime given by the
type IIA string theory. Thus this R-R 1-form is actually a connection
on a $\GU(1)$-bundle and is the vector potential of some
ten-dimensional ``photon''.

Consider now what happens if we compactify this $\GU(1)$ gauge theory on a
manifold $X$. This requires a choice of $\GU(1)$-bundle $V\to X$
satisfying the Yang-Mills equations of motion. Such a bundle may have
a first Chern class $c_1(V)$ corresponding to ``magnetic monopoles''
in $X$. If we demand that there are no such monopoles then our bundle
must be flat.

A flat bundle is over $X$ is described purely by the monodromy of the
bundle around the various non-contractable loops in $X$. That is, by a
{\em homomorphism\/} from $\pi_1(X)$ to $\GU(1)$ --- an element
of $\Hom(\pi_1(X),\GU(1))$. This is equal to $\Hom(H_1(X),\GU(1))$ as
$\GU(1)$ is an abelian group. Using the universal coefficients theorem
this in turn is equal to $H^1(X,\GU(1))$. 

We arrive at the conclusion
that {\em the moduli space of flat $\GU(1)$ bundles over $X$ is given
by $H^1(X,\GU(1))$}. This then would be the contribution to the moduli
space of the R-R 1-form of the type IIA string.

The idea of gerbes is to extend the notion of a $\GU(1)$-bundle with a
connection to an object whose connection is a form of degree greater
than one. Thus the $B$-field of string theory is treated as some
connection on a gerbe where the string itself carries unit electric
charge with respect to this generalized gauge theory. The theory of
gerbes is described clearly in terms of \v Cech cohomology by Hitchin
in \cite{Hit:gerb}. (See also \cite{Sh:gerb1,Sh:gerb2}, for example, 
for further discussion.)

The basic property which we require is
\begin{prop}
The moduli space of flat gerbes over $X$ whose connection corresponds to a
$p$-form is given by $H^p(X,\GU(1))$.
\end{prop}

Thus assuming no solitons in the background corresponding to a gerbe
curvature (such as an ``$H$-monopole'') this yields the desired
moduli space. 

The exact sequence
\begin{equation}
  0\to\Z\to\R\to\GU(1)\to0
\end{equation}
yields the exact sequence
\begin{equation}
 H^p(X,\Z)\to H^p(X,\R)\to H^p(X,\GU(1))
 \to H^{p+1}(X,\Z)\to H^{p+1}(X,\R).
\end{equation}
Thus if the cohomology of $X$ is torsion-free, we have
\begin{equation}
  H^p(X,\GU(1)) \cong\frac{H^p(X,\R)}{H^p(X,\Z)},
\end{equation}
which is a torus whose dimension is given by the Betti number $b_p$. Torsion
in $H^{p+1}(X,\Z)$ will extend this moduli space although it will not
change the dimension.

It is worth mentioning that discrete degrees of
freedom associated to torsion are very poorly understood at present
even though they will appear whenever a type IIA string is
compactified on a non-simply-connected \CY\ threefold. In the work of
\cite{FHSV:N=2} (see also \cite{me:flower} for a fuller description of
the degrees of freedom) these discrete choices were not treated as
choices at all and were fixed by a process known as ``black hole level
matching''. Clearly more work needs to be done to understand this better.

Alternatively one associates the R-R $p$-forms to the associated
electrically-charged D-branes of dimension $p-1$. The R-R field then
measures the phase one associates to a D-brane instanton. This gives a
nice physical picture of the meaning of the R-R moduli. It is believed
however following the work of Minasian and Moore \cite{MM:K} and
Witten \cite{W:K} that the charges of D-branes are classified by {\em
K-theory\/} and not cohomology (see also the lectures by J.~Schwarz 
\cite{Schw:Klect}).

One might therefore suppose that the R-R moduli spaces may be given by
something more in the spirit of K-theory like, for example,
$K^1(X,\GU(1))$ for the type IIA string and $K^0(X,\GU(1))$ for the type
IIB string.\footnote{K-theory may be regarded as a generalized
cohomology theory based on the Eilenberg-Steenrod axioms for
cohomology. To define $K^p(X,\GU(1))$ we may assert that
$K^{\mathrm{even}}$ for a point is $\GU(1)$ whereas 
$K^{\mathrm{odd}}$ for a point is 0 and all cohomology axioms are satisfied.} 

The Chern character gives an isomorphism over the rational numbers
between $K^0$ and $H^{\textrm{even}}$ and between $K^1$ and
$H^{\textrm{odd}}$. This means that as far as simple dimension counting is
concerned the moduli space of R-R fields is the same whether we use
the gerbe picture or whether we use the K-theory picture. These
pictures may not be equivalent globally over the entire moduli space
however. Again more work is needed here.

Either way, for the type II string compactified on a \CY\ threefold we have the
dimension of the moduli space given by certain Betti numbers. 
The $B$-field gives us $b_2(X)=h^{1,1}(X)$ real degrees of
freedom. These can be paired up with the K\"ahler form degrees of
freedom to produce $h^{1,1}(X)$ {\em complex\/} degrees of
freedom. This complexification of the K\"ahler form is seen clearly
from mirror symmetry as we will see in section~\ref{sss:mmap}.
The uncompactified components of the $B$-field give an antisymmetric
tensor field in four dimensions. Such a field may be dualized in the
usual way to produce a real scalar. This is usually called the
``axion'' and is paired up with the dilaton to form a complex degree
of freedom.

Analysis of the R-R fields is as follows.
For the type IIA string on $X$ we have $b_1(X)+b_3(X)$ real moduli. A manifold
with precisely $\SU(3)$ holonomy has $b_1(X)=0$. (One way of seeing
this is that a nonzero number would imply a continuous isometry
leading to a torus factor.) We also have $h^{3,0}=1$ from the
holomorphic 3-form which is nonzero and unique up to isometry. {\em Thus in
total we have $2+2h^{2,1}$ degrees of freedom from the R-R sector.}

All told we have produced $2h^{1,1}+4(h^{2,1}+1)$ real moduli. Since
we expect our moduli space to factorize (up to a discrete quotienting)
as $\cM_H\times\cM_V$, we need to label these moduli as to whether
they form scalar fields in hypermultiplets or vector
multiplets. A careful analysis of this was performed in
\cite{VW:pairs} but we may obtain the same result by a simple crude
argument as follows. Clearly the type of field determines which kind of
supermultiplet it lives in. For example, all the R-R fields must be in
hypermultiplets or they must all be in vector multiplets. 
We also do not expect the labelling to depend on the specific values
of $h^{1,1}(X)$ and $h^{2,1}(X)$.
These facts
together with the fact that the dimension of $\cM_H$ is a multiple of
four immediately tells is that
\begin{prop}
For the type IIA string compactified on a \CY\ threefold $X$ we have
\begin{itemize}
\item $\cM_V$ is spanned by the deformation of the complexified
K\"ahler form and has complex dimension $h^{1,1}(X)$.
\item $\cM_H$ is spanned by the deformations of complex structure of
$X$, the dilaton-axion, and the R-R fields. It has quaternionic
dimension $h^{2,1}(X)+1$.
\end{itemize}
\end{prop}

A similar analysis for the type IIB string differs only in the fact
that the R-R fields consist of even forms rather than odd forms. This
results in
\begin{prop}
For the type IIB string compactified on a \CY\ threefold $Y$ we have
\begin{itemize}
\item $\cM_V$ is spanned by the deformation of the complex
structure of $Y$ and has complex dimension $h^{2,1}(Y)$.
\item $\cM_H$ is spanned by the deformations of the complexified
K\"ahler form, the dilaton-axion, and the R-R fields. It has quaternionic
dimension $h^{1,1}(Y)+1$.
\end{itemize}
\end{prop}

We emphasize that these results are subject to quantum
corrections. That is we may find the dimensions of these moduli spaces
and the forms of these moduli spaces around some limit point using the
above results, but the precise geometry of the moduli space may
vary. We will discuss this in detail shortly.


\subsection{Heterotic compactification} \label{ss:Het}

Now we deal with the heterotic string compactified on a product of a K3
surface, which we denote $S_H$, and a 2-torus (or ``elliptic curve''),
which we denote $E_H$. (We may also take a quotient of this product by
a finite group preserving the $\SU(2)$ holonomy. This makes a little
difference to the analysis below but we ignore this possibility for
the sake of exposition.)

Again, to leading order, one of the things we are required to specify is
a Ricci-flat metric on $S_H\times E_H$. In the case of $E_H$ this is
easy as a Ricci-flat metric is a {\em flat\/} metric. We simply give
one complex parameter specifying the complex structure of $E_H$ and a
real number specifying the area of $E_H$.

The moduli space of Ricci-flat metrics on $S_H$ is well-understood but
a full explanation is rather lengthy. It is described in detail in
\cite{me:lK3}. One of the most important points is that, unlike the
threefold case, it does {\em not\/} factorize into a product of
deformations of the complex structure and deformations of the K\"ahler
form. This can be traced to the fact that a K3 surface has a
hyperk\"ahler structure which allows for a choice (parametrized by an
$S^2$) of complex structures for a fixed Ricci-flat metric. Indeed,
this choice allows for a deformation of complex structure to be
reinterpreted as a deformation of the K\"ahler form. The result which
we quote here is as follows. Let $\Gamma_{3,19}$ be an even self-dual
lattice of signature $(3,19)$ representing $H^2(S_H,\Z)$ together with
its cup product. The moduli space of Ricci-flat metrics on a K3
surface is then
\begin{equation}
  \R_+\times\GO(\Gamma_{3,19})\backslash\GO(3,19)/(\GO(3)\times\GO(19)).
\end{equation}
where the $\R_+$ factor represents the total volume of $S_H$.

As in the type II strings, the heterotic string has a dilaton which is
complexified by adding the axion originating in the uncompactified
parts of the $B$-field.

Next we come to one of the awkward and interesting parts of the
heterotic string --- the ``vector bundle''. Na\"\i vely stated
we take a smooth principal $\cG_0$-bundle, $V$, on $S_H\times
E_H$. The group $\cG_0$ should be $(E_8\times E_8)\rtimes\Z_2$ or
$\spnh$ according to which heterotic string we use.\footnote{We
explain this mouthful in the former case at the end of section
\ref{sss:EH}.} This bundle is used 
to ``compactify'' the gauge degrees of freedom of the ten-dimensional
heterotic string.

The vector bundle $V$ is equipped with a connection and this must
satisfy certain conditions for the equations of motion of the string
theory to be satisfied. Such a connection should be
considered analogous to the Levi-Civita connection on the tangent
bundle derived from the metric. Indeed, a simply ansatz frequently
used is to embed the holonomy of the tangent bundle into $\cG_0$ and
obtain an effective choice for $V$. This process is often referred to
as ``embedding the spin connection in the gauge group'' and was used
in the earliest models of the heterotic string \cite{CHSW:}.

The equations of motion imply a certain topological constraint
on $V$. This
topological condition can also be interpreted as that required for the
cancellation of gravitational and Yang-Mills anomalies. We explain
this shortly. In abstract terms, the homotopy
class of our $\cG_0$-bundle determines a characteristic class in
$H^4(S_H\times E_H,\pi_3(\cG_0))$ which may be thought of as the
generalization of the second Chern class or the first Pontryagin
class of $V$. This must be equal to the second Chern class of the tangent
bundle of the base space $S_H\times E_H$.

To leading order, the condition that $V$ must satisfy is simply that
it obeys the Yang-Mills equations of motion. Fortunately the moduli
space of solutions to these equations over a compact K\"ahler manifold
is a well-studied problem in the mathematics literature. The trick is
to complexify the problem giving a holomorphic bundle whose structure
group lies in the {\em complexification\/} of $\cG_0$. In most of what follows
we will assume this complexification process implicitly and still refer
to $V$ as a $\cG_0$-bundle. 

The situation is easiest to explain in the case that $V$ is a
$\GU(n)$-vector bundle. In this case the complexification is a generic
holomorphic vector bundle whose structure group is $\Gl(n,\C)$. The
Hermitian-Yang-Mills equations of interest then impose that the
connection satisfies $F_{\bar\imath\bar\jmath}=F_{ij}=0$ and
$g_{i\bar\jmath}F^{i\bar\jmath}=0$, where $g_{i\bar\jmath}$ is the
K\"ahler metric 
on the base manifold. We may integrate the equation
$g_{i\bar\jmath}F^{i\bar\jmath}=0$ to obtain the necessary condition
on the ``degree'' of $V$:
\begin{equation}
  \deg(V) = \int c_1(V)\wedge (*J)=0,  \label{eq:mu}
\end{equation}
where $J$ is the K\"ahler form. 

We also need to explain what is meant by a ``stable'' vector
bundle. To a given bundle $E$ we associate its ``slope''
\begin{equation}
\mu(E) = \frac{\deg(E)}{\rank(E)} .
\end{equation}
A bundle $E$ is said to be {\em stable\/} if every coherent
subsheaf\footnote{We could almost say ``subbundle'' here.} $\cF$ of
lower rank satisfies $\mu(\cF)<\mu(E)$. A ``semistable'' bundle is
allowed to satisfy $\mu(\cF)\leq\mu(E)$.

We then have the following theorem due to Donaldson, Uhlenbeck and Yau
\cite{Don:YM,UY:YM}\footnote{The original form of this theorem does
not restrict attention to curvatures satisfying
$g_{i\bar\jmath}F^{i\bar\jmath}=0$. Instead the case of constant 
$g_{i\bar\jmath}F^{i\bar\jmath}$ is considered. This is often called
``Hermitian-Einstein'' and is analogous to the case of an Einstein
metric as opposed to a Ricci-flat metric.}
\begin{theorem}
A bundle is stable and satisfies (\ref{eq:mu}) if and only if it
admits an irreducible Hermitian-Yang-Mills connection. This connection
is unique.
\end{theorem}

This reduces the difficult problem of finding the moduli space of
bundles in terms of solution sets of differential equations to a more
algebraic problem of finding the moduli space of holomorphic vector
bundles. {\em This is exactly analogous to replacing the problem of finding
the moduli space of Ricci-flat metrics for a \CY\ manifold to that of
finding the moduli space of complex structures.}

Note that the theorem above imposes that the connection be
irreducible. In many cases this is a little strong a we need to
consider semistable bundles. This is discussed in \cite{FMW:F}.

Continuing the analogy of solving the Yang-Mills equations for $V$ to
the finding of a Ricci-flat metric we might suppose that looking for
higher-order corrections to the equations of motion may require
corrections to be made to the connection. These corrections will
affect our moduli space problem. In addition we should
expect that worldsheet instantons might ruin the very interpretation
of these degrees of freedom of the heterotic string as coming from a
vector bundle.

The act of replacing the differential geometry problem of finding
vector bundles satisfying the Yang-Mills equations by the algebraic
geometry question of looking at the moduli space of stable holomorphic
bundles might actually be seen as moving a step closer to the truth in
string theory. As we move around the moduli space we will often
encounter degenerations of the bundle data which can be interpreted
easily in the algebraic picture by using the language of
``sheaves''. See \cite{DGM:sh,KS:toricF,AD:tang} for example for more
on this.

Anyway, to return to our problem, we require a (semi)stable
holomorphic bundle over the product of a K3 surface and an elliptic
curve. The first simplification is to assume that this bundle
factorizes nicely. That is, we have two bundles
\begin{equation}
\begin{split}
  V_S &\to S_H\\
  V_E &\to E_H.
\end{split}
\end{equation}
Let the structure group of $V_S$ be $\cG_S$ and let the structure
group of $V_E$ be $\cG_E$. This is then a special case of a
$\cG_0$-bundle over $S_H\times E_H$ if $\cG_0\supset\cG_S\times\cG_E$.
Our problem nicely factorizes into finding the moduli space of $V_S$
and the moduli space of $V_E$.

Finally we come to the other interesting part of the heterotic string
--- the $B$-field. In the case of the heterotic string a
deep understanding of this object is even more troublesome than the
$B$-field of the type II string. 
This was analyzed recently by Witten \cite{W:K3inst,W:hADE}. 
Let us assume the heterotic string is compactified on a generic \CY\
space $Z$.
We can make a simple statement --- the number of real degrees of
freedom of the $B$-field is given by $\dim H^2(Z)$ as it was for the type
II strings. Beyond this simple
dimension counting we have to work harder.
The general idea is that anomaly cancellation in the heterotic string
requires an equation in differential forms as follows.
\begin{equation}
  H = dB + \frac{\alpha'}{4\pi}(\omega_Y-\omega_T),
\end{equation}
where $H$ is the physically significant, and thus gauge-invariant, field
strength associated to the heterotic string. The terms $\omega_Y$ and 
$\omega_L$ are Chern--Simons 3-forms associated to the connections of
the Yang--Mills gauge bundle and the tangent bundle respectively.
We refer to \cite{GSW:book} for a general review of these facts.

Note that the exterior derivative of this formula gives
\begin{equation}
  dH = \frac{\alpha'}{4\pi}\left(\tr R\wedge R-\tr F\wedge F\right),
		\label{eq:dH}
\end{equation}
where $R$ and $F$ are the curvatures of the tangent bundle and $V$
respectively. Taking cohomology classes this gives the topological
constraint on $V$ discussed above.\footnote{This argument using De
Rham cohomology misses the torsion part.}
 
The fact that $\omega_Y$ and $\omega_L$ are not gauge invariant
objects implies that $B$ will have some nontrivial transformation
properties. 

An effect of $B$, as in the type II string, is to weight instantons
as will be explained briefly in section \ref{sss:mmap}. Namely, if a
2-sphere $S$ in the target space represents a worldsheet instanton
then the action is weighted by a factor given by
\begin{equation}
c=\exp\left(2\pi i\int_S B\right).  \label{eq:hp}
\end{equation}
In the simpler case of the type IIA string this phase is determined by
the homology class of $S$. That is, $B\in\Hom(H_2(Z),\GU(1))\cong
H^2(Z,\GU(1))$. 
Witten noted the following awkward property of this phase when dealing
with the heterotic string. Suppose we
have a family of rational curves in the target space. For simplicity
we assume the space contains $\P^1\times C$ for some complex curve $C$. Fix a
particular $\P^1$ in this family. Let $c_0$ be the phase associated to
this curve given by (\ref{eq:hp}). Now move this curve in a
contractable loop $\gamma$ within $C$. Let $W\subset C$ be a disc in
$C$ with boundary $\gamma$. When we return back to our original $\P^1$
one finds the phase induced by the $B$-field equal to
\begin{equation}
  c_1=\exp\left(-2\pi i\int_{W\times\P^1}dH\right)\,c_0,
\end{equation}
where $dH$ is given by (\ref{eq:dH}).
Thus unless $dH=0$ the contribution of the $B$-field to the phase
factor in the instanton is not single-valued. Physically the theory
is OK because there is another contribution to the phase of the
instanton action given by a Pfaffian associated to the worldsheet
fermions in the heterotic string. This exactly cancels the above
holonomy \cite{W:hADE}.

Instead of taking a single \CY\ target space with a family of
2-spheres we may take a family of \CY\ target spaces containing
a given 2-sphere. The above analysis holds with little modification and shows
that going around a contractable loop in the moduli space of \CY\ spaces can
introduce an ambiguity in the associated $B$-field phase. In other
words {\em the $B$-field does not live in the flat bundle
$H^2(Z,\GU(1))$ over the moduli space!} All we can say in
general is that the $B$-field lives in a bundle over the moduli space
whose generic fibre is a torus of dimension $\dim H^2(Z)$.

One way of avoiding this nastiness is by ``embedding the spin
connection in the gauge group''. In this 
very special case, the above holonomies disappear. One may also get
the holonomies to vanish by taking the sizes to infinity
by taking $\alpha'\to0$. In both of these cases $B$ really does
live in $H^2(Z,\GU(1))$.

\subsubsection{$E_H$ and its bundle}  \label{sss:EH}

Let us deal first with the bundle $V_E$ over the fixed torus
$E_H$. This case is rather easy to analyze as the only
bundles over $E_H$ which solve the equations of motion are ones which
are {\em flat}. Because the tangent bundle and gauge bundle are flat
we have $dH=0$ and avoid any curvature of the bundle in which the
$B$-field lives. Thus $B\in H^2(E_H,\GU(1))$.

We are required to find the moduli space of flat
$\cG_E$-bundles over $E_H$. 
Let us assume that $\cG_E$ is simply-connected.
This problem was extensively analyzed in
\cite{FMW:F}.

A flat $\cG_E$-bundle over $E_H$ is specified by its ``Wilson
lines''. That is, we specify a homomorphism $\pi_1(E_H)\to\cG_E$ up to
conjugation by $\cG_E$. Since $\pi_1(E_H)$ is the abelian group
$\Z\oplus\Z$, we need to specify two commuting elements 
of $\cG_E$. A useful result of Borel \cite{Bor:T2} states that any two
commuting elements of $\cG_0$ may be conjugated simultaneously into
the maximal Cartan subgroup $T\subset\cG_0$. This implies that our
desired moduli space is $T\times T$ divided by any remnants of the
conjugation equivalence. The latter is given precisely by the Weyl
group $W(\cG_0)$. The desired moduli space of bundles over a fixed
$E_H$ is therefore
\begin{equation}
  \frac{T\times T}{W(\cG_0)}. \label{eq:bcE}
\end{equation}

Now consider supplementing this data by the moduli of $E_H$ to get the
full moduli space related to $E_H$. The K\"ahler form and $B$-field
classically live in $\R_+\times\R/\Z$ which may be exponentiated to
give $\C^*$. The moduli space of complex structures is given by the
upper half plane, $\mathsf{H}$, divided by $\Sl(2,\Z)$ as is
well-known. Note that 
this $\Sl(2,\Z)$ acts on the generators of $\pi_1(E_H)$ and thus on
(\ref{eq:bcE}) by mixing the two $T$'s.
We thus have
\begin{prop}
The classical moduli space of $\cG_E$-bundles on $E_H$ together with
the moduli space of Ricci-flat metrics and $B$-fields on $E_H$ is
given by
\begin{equation}
  \Sl(2,\Z)\backslash\left(\mathsf{H}\times 
	\frac{T\times T}{W(\cG_0)}\right) \times\C^*.  \label{eq:cl-T2}
\end{equation}
\end{prop}

This rather ugly-looking result becomes more pleasant when
stringy considerations are taken into account. For example, let us
divert our attention briefly to the case of a heterotic string
compactified {\em only\/} on $E_H$.\footnote{To be precise, we
consider the component of the moduli space containing the trivial
bundle.} This implies
$\cG_E=\cG_0$. This case was studied by Narain
\cite{N:torus} (see also \cite{HNW:torus}). The {\em exact\/} result
is that the moduli space is given by
\begin{equation}
\GO(\Gamma_{2,18})\backslash\GO(2,18)/(\GO(2)\times\GO(18)),
	\label{eq:N2-18}
\end{equation}
(times a real line for the dilaton). The lattice $\Gamma_{2,18}$ is
the even self-dual lattice of signature $(2,18)$ which is given by the
root lattice of $E_8\times E_8$, or $\SO(32)$, supplemented by two
orthogonal copies of $U$. 
We use the standard notation $U$ for the even
self-dual lattice of signature $(1,1)$. 

Note that the heterotic string compactified on a single 2-torus has
half-maximal supersymmetry and indeed the moduli space
(\ref{eq:N2-18}) is of a form promised in section \ref{ss:U}. The
{\em only\/} way that (\ref{eq:N2-18}) differs from the classical
statement (\ref{eq:cl-T2}) is that there are extra discrete
identifications. See, for example, section 3.5 of \cite{me:lK3} for
details of how these moduli spaces are mapped to each other.
These extra identifications in the exact case are called ``T-Dualities''. 

These T-Dualities include the familiar \RoR\ dualities of the torus
as well as dualities which mix moduli corresponding to the bundle with
moduli corresponding to the base.

When compactifying the heterotic string on $S_H\times E_H$ we will
have fewer supersymmetries and so we have every reason to expect that
quantum effects will have a more serious effect on the classical
moduli space of vector bundles. We will see that this is so.

\begin{difficult}
Let us return again for a moment to the eight dimensional case of the
heterotic string only compactified on $E_H$.
It is known that the moduli space of flat $\cG_0$-bundles on a 2-torus is not
connected. In the case of the $E_8\times E_8$ heterotic string it is
believed to be a valid string model if the two $E_8$ factors are
exchanged under holonomy around a non-contractable loop in the
torus. These models were explored in \cite{CHL:bigN,CP:ao}. Such a
bundle is not really an $E_8\times E_8$-bundle but is more accurately
described as an $(E_8\times E_8)\rtimes\Z_2$-bundle where this latter
$\Z_2$ acts to exchange the two $E_8$ factors. Pedants who like to say
``$\spnh$ heterotic string'' rather than $\SO(32)$ heterotic string''
should by all rights be expected to say ``$(E_8\times
E_8)\rtimes\Z_2$ heterotic string'' rather than ``$E_8\times E_8$
heterotic string''!

Similarly a $\spnh$-bundle may have a nontrivial second
Stiefel-Whitney class over the torus. Such a bundle is not homotopic
to the trivial bundle and so lies in a different component of the
moduli space.

These classes of bundles have been studied in
\cite{LSMT:w2,W:w2t,BPS:Fquan}. In particular, a connection between
these two classes was discussed in \cite{W:w2t}. See also
\cite{BFM:w2} for a nice mathematical treatment of these issues.

We should expect the same kind of effects for various possibilities of
$\cG_E$ when we now compactify down to four dimensions. Monodromy can
be expected to play a r\^ole around the cycles in $E_H$ whenever
$\cG_E$ admits an outer automorphism (possibly even if this outer
automorphism was not induced by an endomorphism of $\cG_0$). We may
also obtain second Stiefel-Whitney classes whenever $\cG_E$ is not
simply-connected. 

It is probably fair to say that we do not have a full understanding of these
disconnected components of the moduli space
in the context of string duality at the present time. We will ignore
this problem in these notes and implicitly assume that the flat bundles on
$E_H$ are always homotopic to the trivial bundle.

See also \cite{McI:spin} where another issue to do with the global
form of the gauge group is raised.

\end{difficult}

\subsubsection{$S_H$ and its bundle}

We now need to consider the bundle $V_S\to S_H$ subject to the anomaly
cancellation condition. In the case that $V_S$ is an $\SU(n)$ bundle
this would amount to $c_2(V_S)=24$ for example. In general this is a
much harder problem to solve than the preceding case. 
Having said that, the bundle part of the problem is not too bad so
long as we ignore quantum corrections. Work
by Mukai \cite{Muk:bun} (see also \cite{Muk:K3} 
for a nice account of this work) tells us that we may put the hyperk\"ahler
structure of the K3 surface $S_H$ to good use.

The basic result we will use is that the moduli space of stable vector
bundles over $S_H$ will also have a hyperk\"ahler structure. In fact,
Mukai has shown that in many cases one may obtain a moduli space which is
itself another K3 surface! The relationship between $S_H$ and this
latter K3 surface may be viewed as a kind of mirror symmetry
in some cases \cite{Mor:SYZ}.

We will have more to say about the bundle $V_S$ and its
moduli space in the case that $S_H$ is an elliptic fibration in
section~\ref{sss:E8l} but for now we will just content ourselves with the
knowledge that the moduli space has a hyperk\"ahler structure.

The moduli space of Ricci-flat metrics and $B$-fields on $S_H$ is
given by \cite{Sei:K3,AM:K3p}
\begin{equation}
\GO(\Gamma_{4,20})\backslash\GO(4,20)/(\GO(4)\times\GO(20)).
	\label{eq:K3mod}
\end{equation}
See \cite{me:lK3} for more details. The fact that it is a symmetric space
may be deduced from its appearance in the moduli space of a type IIA
string compactified on a K3 surface --- which has half-maximal
supersymmetry. 

Our complete moduli space of deformations of $S_H$ together with its
bundle $V_S$ may therefore itself be viewed as a fibration. The base
space of this fibration is given by (\ref{eq:K3mod}) (or perhaps only
some subspace of it) while the fibre
is given by the hyperk\"ahler moduli space of the bundle $V_S$. 

Note that (\ref{eq:K3mod}) may be viewed as a quaternionic K\"ahler
manifold (well, orbifold to be precise) from the fact that
$\Sp(1).\Sp(20)\supset\SO(4)\times\SO(20)$ (up to finite groups). 
Assuming the moduli space of $S_V$ varies over this space in a way
compatible with this quaternionic structure we see that the total
moduli space will also have a quaternionic K\"ahler structure.

Our crude counting argument tells us immediately that this total
moduli space of $V_S\to S_H$ 
should be identified with $\cM_H$ leaving the remaining moduli in
$\cM_V$. Again one may be more careful along the lines of
\cite{VW:pairs} if one wishes. Anyway, to recap we have
\begin{prop}
For the heterotic string compactified on $(V_S\to S_H)\times
(V_E\to E_H)$  we have
\begin{itemize}
\item $\cM_V$ is spanned by the deformations of $V_E\to E_H$ (i.e.,
deformations of $V_E$ {\em and\/} deformations of the complex
structure {\em and\/} complexified K\"ahler form on $E_H$) and by the
dilaton-axion. It has complex dimension $\rank(V_E)+3$.
\item $\cM_H$ is spanned by the deformations of $V_S\to S_H$.
\end{itemize}
\end{prop}
The dimension of the space $\cM_H$ depends on several considerations
and we do not compute it here. Note in particular that certain bundles 
put constraints on the K3 they live on and the complete form of
(\ref{eq:K3mod}) may not be seen.

\subsection{Who gets corrected?}  \label{ss:who}

So far we have listed the degrees of freedom present in a given string
theory and then determined the classical picture of the resulting
moduli space. This is not expected to be exact however --- there will
be corrections from various sources.

To specify exactly how these corrections arise will again strongly
test our knowledge about what string theory is exactly. Even though we
don't really know what string theory is, we do know enough to make statements
about where we might expect quantum corrections to arise.

An irrefutable statement about string theory is that it contains at
least two limits in which we expect quantum field theory to provide a
good picture (at least most of the time). The first quantum field
theory consist of the two-dimensional worldsheet conformal field
theory, i.e., the ``pre-duality'' picture of string theory. Indeed
this picture gives us the ``stringiness'' in string theory! Secondly
we have the effective quantum field theory which lives in the
target spacetime dimensions.

Consider first the worldsheet quantum field theory. This has an
action \cite{Pol:books}
\begin{equation}
  \frac1{4\pi\alpha'}\int_\Sigma d^2\sigma\sqrt{\gamma}\Bigl(
  \gamma^{ab}g_{\mu\nu}(x) + i\epsilon^{ab}B_{\mu\nu}(x)\Bigr)
  \partial_ax^\mu\partial_bx^\nu +\frac1{4\pi}\int_\Sigma
  d^2\sigma\sqrt{\gamma} \Scr{R}\Phi_0+\ldots,
      \label{eq:nlsm} 
\end{equation}
where $x$ maps the worldsheet $\Sigma$ into ten-dimensional
spacetime. We have a 
worldsheet metric $\gamma_{ab}$, and target space metric and $B$-field
given by $g_{\mu\nu}$ and $B_{\mu\nu}$. 
In addition  $\Scr{R}$ represents the worldsheet scalar curvature
and $\Phi_0$ is the
dilaton which we assume to be independent of $x$.
The difficulty in analyzing this model is that the metric and
$B$-field vary as a function of the position in target space, $x$.
The important point to note is that
$\alpha'$ (which sets the ``string scale'' in units of area) acts as a
coupling constant. If $\sqrt{\alpha'}$ is much less than a
characteristic distance scale, $R$, of variations in the metric and
$B$-field then $x$ represents almost ``free'' fields. We can then use
a perturbation theory expanding in powers of $\alpha'/R^2$. We may
also have nonperturbative effects due to {\em worldsheet instantons\/} which
contribute towards correlation functions as
$\exp(-R^2/\alpha')$. These instantons are the maps $x$ which solve
the equations of motion of (\ref{eq:nlsm}) and are given in our context
as {\em holomorphic\/} maps \cite{DSWW:}.

To compute any correlation function using this worldsheet field theory
version of string theory it is necessary to integrate over all
worldsheets. This includes a sum over all genera with genus zero
corresponding to tree-level, genus one giving one loop, etc. Such
summands will be weighted by a relative factor of $\exp(g\Phi_0)$, where
$g$ is the genus of $\Sigma$, thanks to the last term in (\ref{eq:nlsm}).

This picture of string theory induces an effective spacetime action
proportional to\footnote{We are being thoroughly negligent with
factors of 2 etc., and we have omitted an overall factor. See section
3.7 of \cite{Pol:books} for a better discussion.} 
\begin{equation}
  \int d^{10}x\,\sqrt{g}e^{-2\Phi_0}\Bigl(R_g + |\nabla\Phi_0|^2 +
  |dB|^2\Bigr) +\ldots  \label{eq:qft}
\end{equation}
We may use this as the basis of a spacetime quantum field theory.
The important thing to note here is that $\lambda=\exp(\Phi_0)$ appears
as a coupling constant in this quantum field theory. This is hardly
surprising given that the number of loops in this field theory
corresponds to the genus of the worldsheet in the previous field
theory. $\lambda$ is often called the ``string coupling''.

At the heart of the subtlety of string theory is that each of these
field theories above contains the seeds for the other field theory's
downfall! As we have already mentioned, there are good reasons for
believing that worldsheet instanton effects in the worldsheet
conformal field theory make a complete understanding of spacetime in
terms of Riemannian geometry unlikely. Thus the spacetime quantum
field theory cannot really be considered in the form of the action
(\ref{eq:qft}). Equally, nonperturbative effects, such as instantons,
coming from the spacetime field theory cannot be understood in terms
of the genus expansion of the worldsheet theory. The best we can do is
to assume that true string theory knows about both of these field
theories and includes the nonperturbative effects from both
simultaneously. This idea will become very important in section 
\ref{sss:mixed}.

The worldsheet picture of string theory can only really be considered
to be an accurate picture of string theory when $\lambda\to0$ and
equally the spacetime effective action point of view can only be
relied upon safely when $\alpha'\to0$. These are the two limits of
string theory where we really understand what is going on.

We need to look at the moduli spaces of the previous section and ask
how they may be warped by corrections coming from quantum effects of
either of our two field theories. Fortunately it is not the case that
all of the moduli spaces are affected by both corrections. We can see
this from the holonomy argument in section \ref{ss:hol} that the
moduli space factorizes as $\cM_H\times\cM_V$ {\em exactly}.

Let us consider $\lambda$-corrections first from the spacetime field
theory. These must vanish as $\lambda\to0$. Because of this they
cannot affect the factor of the moduli space which does not contain
the dilaton. Similarly the $\alpha'$-corrections must disappear in the
large radius limit of the compactification and so cannot affect a
factor of the moduli space which does not know about sizes.

One may try to argue that the moduli space of complex structures of a
\CY\ threefold does not know about size. Algebraic geometers can
compute the moduli space of an algebraic variety without knowing about
feet and inches! On the other hand it is the K\"ahler form which
determines the volume of the threefold and so we might expect its
moduli space to be subject to $\alpha'$-corrections. One should be a
little careful with this argument as varying the complex structure can
vary volumes of object such as minimal 3-cycles in the
threefold. That being said, this argument can be shown to be
rigorously correct. For example, one may use topological field theory
methods to show that the moduli space of complex structures is
unaffected by quantum corrections from the worldsheet field theory
\cite{W:AB}. 

The results for which parts of the moduli spaces are affected by
quantum corrections are given in table~\ref{fig:qc}. 
We should note that some entries in this table may only be valid if 
only one of the coupling constants $\alpha'$ or $\lambda$ is nonzero. 
For example if $\lambda=0$ then the moduli space $\cM_{V}$ for the 
heterotic string is not prone to $\alpha'$-corrections but this may 
not be true when $\lambda$ is nonzero.

\begin{table}
\begin{center}
\begin{tabular}{|c||c|c|}
\hline
&$\cM_H$-corrections&$\cM_V$-corrections\\
\hline
IIA on $X$&$\lambda$&$\alpha'$\\
IIB on $Y$&$\lambda$ and $\alpha'$&Exact\\
Het on $S_H\times E_H$&$\alpha'$&$\lambda$\\
\hline
\end{tabular}
\end{center}
\caption{Quantum corrections.}
        \label{fig:qc}
\end{table}

Upon compactification on a space $X$ to flat $d$-dimensional spacetime
we obtain the spacetime effective action
\begin{equation}
  \int d^dx\,\sqrt{g}e^{-2\Phi}\Bigl(R_g + |\nabla\Phi|^2 + |dB|^2\Bigr) +
  \int d^dx\,\sqrt{g}g^{\mu\nu} G_{ij}
  \partial_\mu\phi^i\partial_\nu\phi^j+\ldots,  \label{eq:qftd}
\end{equation}
from (\ref{eq:qft})
where $\phi^i$ are coordinates on the moduli space as in
(\ref{eq:mod}).
The quantity $\Phi$ represents the effective $d$-dimensional dilaton
and is given basically by $\Phi_0-\ff12\log\Vol(X)$.
In the compactification scenario {\em this field theory is declared
to be accurate\/}. Because this part
of spacetime is flat Minkowski space (or very nearly) we assert that worldsheet
instantons are not allowed to spoil this field theory. This assumption
is implicit in all of these lectures. Of course, this means that we are
not allowed to ask questions about the $d$-dimensional physics which
might probe effects such as quantum gravity. Then the compactification
model would be invalid.


\section{The Moduli Space of Vector Multiplets} \label{s:vec}


\subsection{The special K\"ahler geometry of $\cM_V$}   \label{ss:sK}

In order to discuss quantum corrections we need to establish limits on
how much we are allowed to warp the moduli spaces consistent with the
supersymmetry. We have said that $\cM_V$ is K\"ahler and we can now
put further limits on the structure of this moduli space.

We wish to exploit the fact that the moduli space factor $\cM_V$
for the type IIB string compactified on a \CY\ space $Y$ is not warped
at all by quantum corrections. The fact that $\cM_V$ is given exactly in the
form of a moduli space of complex structures on a \CY\ threefold will
allow us to ask more detailed questions about the differential
geometry of $\cM_V$.

The deformations of complex structure of $Y$ are best thought of as
{\em variations of Hodge structure\/} as we now explain. Any \CY\
threefold has Hodge numbers $h^{p,q}$ in the form of a Hodge diamond
\begin{equation}
\newcommand{\m}[1]{\multicolumn{2}{c}{#1}}
\arraycolsep=7pt
\begin{array}{*8{c}}
    &&&\m1&&& \\ &&\m0&\m0&& \\ &\m0&\m{h^{1,1}}&\m0& \\
    \m1&\m{h^{2,1}}&\m{h^{2,1}}&\m1 \\ &\m0&\m{h^{1,1}}&\m0& \\
    &&\m0&\m0&& \\ &&&\m1&&&
\end{array}
\end{equation}
Of interest to us is the middle row of this diamond which relates to
$H^3(Y)$. In particular we have a relationship between the Dolbeault
cohomology groups and the integral cohomology:
\begin{equation}
  H^3(Y,\C) = H^{3,0}(Y)\oplus H^{2,1}(Y)\oplus H^{1,2}(Y)\oplus
  H^{0,3}(Y) = H^3(Y,\Z)\otimes_\Z\C.
\end{equation}
As we vary the complex structure the way in which the lattice
$H^3(Y,\Z)$ embeds itself into the space $H^3(Y,\C)$ ``rotates'' with
respect to the decomposition of $H^3(Y,\C)$ into the Dolbeault
cohomology groups.

Consider a holomorphic 3-form $\Omega\in H^{3,0}(Y)$. This is never
zero anywhere on $Y$ and is uniquely defined up to a constant multiple
thanks to the \CY\ condition. Now consider a symplectic basis for
$H_3(Y)$ given by $A^a$ and $B_a$ for $a=1,\ldots,h^{2,1}+1$ with
intersections $A^a\cap B_b=\delta^a_b$. Define the {\em periods}
\begin{equation}
t^a = \int_{A^a}\!\Omega \quad\text{and}\quad \cF_a=\int_{B_a}\!\Omega.
	\label{eq:periods}
\end{equation}

These periods ``measure'' the complex structure of $Y$. Since $Y$ has
only $h^{2,1}$ deformations of complex structure it is clear that not
all of these periods may be independent. Firstly we have noted that
$\Omega$ is defined only up to a constant multiple so the periods can
at best only be homogeneous coordinates in a projective
space. Secondly it was shown by Bryant and Griffiths \cite{BG:period}
that, given all the $t^a$'s, all the $\cF_a$'s are determined. That
is, we may express the $\cF_a$'s as functions of the $t^a$'s. Thus we
are locally modeling the moduli space by $\P^{h^{2,1}}$. This gives
us the correct dimension for the moduli space. (Note that the topology
of the moduli space is unlikely to be that of a projective space as we have
ignored the subtleties of degenerations so far. Also, the metric on
the moduli space will not be the Fubini-Study metric. One way of
seeing this is that some degenerations will be an infinite distance
away from generic points in the moduli space.)

It is then not hard to show, see for example section 3 of
\cite{Cand:mir}, that we may define a function $\cF$ locally on the
moduli space such that
\begin{equation}
\begin{split}
  \cF &= \frac12\sum_ct^c\cF_c\\
  \cF(\lambda t^0, \lambda t^1, \ldots) &= \lambda^2\cF(t^0, t^1,
  \ldots)\\
  \cF_a &= \frac{\partial \cF}{\partial t^a}.
\end{split} \label{eq:pre}
\end{equation}

We may rephrase this more globally in terms of bundle language
following Strominger \cite{Strom:S}. The moduli space $\cM_V$ has an
``ample'' line bundle $L$ 
such that $c_1(L)$ is given by the cohomology class of the K\"ahler
form on $\cM_V$. We also have an $\Sp(h^{2,1}+1)$-bundle $\cH$ over
$\cM_V$ whose fibre is given by $H^3(Y,\C)$ in the fundamental
representation. We then have sections
\begin{equation}
\begin{split}
  \Omega&\in\Gamma(\cH\otimes L)\\
  \cF &\in\Gamma(L^2).
\end{split}
\end{equation}

The important point is that the function $\cF$, which is called the
``prepotential'' contains all the useful information we will need. The
geometry of $\cM_V$ is completely determined by it. This fact shows
that $\cM_V$ cannot be any old K\"ahler manifold. It is conventional
to denote the special property that we have a prepotential by saying
that $\cM_V$ is ``special K\"ahler''.

Our discussion above takes the point of view that special K\"ahler
geometry appears from the moduli space of complex structure on \CY\
threefolds. This is not the original definition however. Special
K\"ahler was first used to denote the geometry of the moduli space of
scalar fields in vector multiplets of arbitrary $N=2$ supersymmetric
field theories coupled to gravity in four dimensions as in
\cite{dWvP:spec}. 
In this context, the projective coordinates $t^a$ are known as
``special'' or ``flat'' coordinates.
The link between these points of view is that the
K\"ahler metric on $\cM_V$ given in the effective action (\ref{eq:qftd}) is
given by
\begin{equation}
\begin{split}
  G_{a\bar b} &= \frac{\partial^2K}{\partial t^a\bar\partial t^b}\\
  K &= -\log\Img\left(\bar t^a\frac{\partial \cF}{\partial t^a}\right).
\end{split}  \label{eq:metric}
\end{equation}

The remarkable fact, as proved by Strominger in \cite{Strom:S} (see
also a discussion of this in \cite{Craps:S}), is that these points of
view are equivalent. That is to say {\em the local conditions arising
from differential geometry for deformations of Hodge structure of a
\CY\ threefold are identical to the conditions on the moduli space of
vector moduli in an $N=2$ theory of supergravity in four dimensions.}

It is well worth pausing to reflect on the implications of this
statement. Since we have approached the question of supergravity in
lower dimensions from the point of view of string theorists this
statement may not seem particularly stunning --- it is just a
confirmation that things are working out nicely. Our moduli space of
compact \CY\ manifolds ties in nicely with the geometry of the moduli
space of vacuum expectation values of the massless scalar particles in
the uncompactified dimensions. This statement of equivalence does not
depend on string theory however. What would we have made of it if we
had not yet discovered string theory? It is as if the $N=2$ theories
of supergravity in four dimensions ``knew'' that they were related in
some way the \CY\ threefolds. 
String theory, or at least ten-dimensional supergravity,
provides this link nicely via compactification.
Even if string theory turns out to be wrong for some reason, this link
between $N=2$ theories and \CY\ threefolds is irrefutable. 

We should provide a word of caution about the strength of the
statements above. Just because a moduli space is consistent with these
conditions that it be a deformation of Hodge structure of a \CY\
manifold, it does not imply that such a \CY\ manifold must exist. 

It is perhaps worthwhile to mention the following structure about
special K\"ahler geometry which gives a hint as to why the $N=2$
theory ``knows'' about the \CY\ 3-fold. Consider the following
Hermitian form on 
$H^3(Y,\C)$:
\begin{equation}
  H_Y(\omega_1,\omega_2) = 2i\int_Y \omega_1\wedge\bar\omega_2.
\end{equation}
It is easy to show (see \cite{Ty:Jac} for example)
that the imaginary part of this form coincides with
the usual cup product structure when restricted to $H^3(Y,\Z)$. One
may also show that
\begin{itemize}
  \item on $H^{3,0}(Y)$ the form $H_Y$ is negative definite, and
  \item on $H^{2,1}(Y)$ the form $H_Y$ is positive definite.
\end{itemize}
One may show \cite{Cand:mir} that this is reflected in the fact that
the the matrix
\begin{equation}
  \Img\left(\frac{\partial^2\cF}{\partial t^a\partial t^b}\right),
\end{equation}
has signature $(1,h^{2,1})$. This signature nicely ``separates'' the
$H^{3,0}$ part of the cohomology from the $H^{2,1}$ part.

We can now argue (very) roughly as follows.  If we had an even number
of dimensions to our compact space then we wouldn't have the right
symplectic structure (e.g. the $\Sp(h^{2,1}+1)$ group above) on the
middle dimension cohomology to see the correct variation of Hodge
structure. If the compact space were complex dimension one, we
wouldn't have ``enough room'' in the Hodge diamond for an indefinite
Hermitian form. If we had five or more complex compact directions we
would expect something more complicated. Thus three dimensions is the
most natural. We also obtain $h^{3,0}=1$ from the signature telling us
that we must have a \CY!\footnote{Of course, we could do something
like take a fivefold with Hodge numbers $h^{5,0}=0$ and $h^{4,1}=1$
which might give us the right structure. We consider this less natural
than the \CY\ threefold.}

We would like to emphasize again the fact that this discussion of
special K\"ahler geometry depends on the exactness of the effective
action (\ref{eq:qftd}). If true quantum gravity effects in four
dimensions are considered we may expect much of this discussion to
break down. Indeed, the statement that we have a moduli space in the
form of a Riemannian manifold (or orbifold etc.) would then be
suspect.


\subsection{$\cM_V$ in the type IIA string}  \label{ss:mir}

Now we wish to look at the way that $\cM_V$ is seen in the type IIA
string on the \CY\ threefold $X$. This involves the old work of {\em
mirror symmetry.} Since there have been numerous reviews of mirror
symmetry we will be fairly brief here and focus only the warping of
the special geometry of $\cM_V$.


\subsubsection{Before corrections and five dimensions}    \label{sss:BC}

As noted in section \ref{ss:who} $\cM_V$ consists of the moduli
space of the K\"ahler form on $X$ but is subject to corrections coming
from worldsheet instantons. Let us first establish what it would look
like if there were no quantum corrections.

One may approach this directly as in \cite{Cand:mir} or one may
proceed in a slightly different way via M-theory. We will do the
latter (as most string theory students these days are perhaps even
better acquainted with M-theory than with string theory!). The first
thing to note is that an $N=2$ theory in four dimensions may be
obtained by compactifying an $N=1$ theory in {\em five\/}
dimensions on a circle. 
Then if we reinterpret the IIA string theory as M-theory
on a circle we see that this five-dimensional theory may be obtained
by compactifying M-theory on the \CY\ threefold $X$.

To put it another way we may consider the limit of a type IIA string
on $X$ as the string coupling becomes very strong. In the same way
that the ten-dimensional type IIA string becomes eleven-dimensional
M-theory in this limit, the four dimensional $N=2$ theory will turn
into the five-dimensional $N=1$ theory.

The useful thing about this limit for our purposes is that the
effective string scale given by $\alpha'$ tends to zero in this
limit. Thus stringy effects such as worldsheet instantons are
completely suppressed. This is explained nicely by Witten
\cite{W:MF}. The general idea is that the metric in the uncompactified
directions needs to be rescaled as we change dimension (just as it is
going from ten dimensions to eleven dimensions \cite{W:dyn}). This
rescaling is infinite taking the string scale to zero size. 

A vector multiplet in five dimensions contains only one real scalar as
opposed to the two scalars coming from the four dimensional vector
multiplet. The rescaling between the type IIA theory and M-theory also
causes a slight reshuffling of moduli as explained in \cite{CCAF:M3}.
The result is that we have a moduli space $\cM_V^5$ of vector multiplets
which is a real space of dimension $h^{1,1}-1$. This is the classical
moduli space of cohomology classes K\"ahler forms on $X$ {\em of fixed
volume.} The deformation corresponding to the volume defects to the
hypermultiplets replacing the lost dilaton 
leaving the moduli space $\cM_H$ unchanged between four and five dimensions.

The compactification of M-theory on a smooth \CY\ threefold yields
$h^{1,1}$ vector fields from the M-theory 3-form in eleven
dimensions. This yields a supersymmetric $\GU(1)^{h^{1,1}}$ gauge
theory (with gravity) in five dimensions. The action for such a field
theory contains the interesting ``Chern-Simons''-like term
\begin{equation}
  \int d^5x\,\kappa_{abc}F^a\wedge F^b\wedge A^c
		\label{eq:CS5}
\end{equation}
where $\kappa_{abc}$ is symmetric in its indices
$a,b,c=1,\ldots,h^{1,1}(X)$. As usual with these topological types of
terms in field theory one may compute $\kappa_{abc}$ from the
intersection theory of $X$. In this case one discovers that
\begin{equation}
  \kappa_{abc} = \int_X e_a\wedge e_b\wedge e_c,
\end{equation}
where the $e_a$'s are the generators of (the free part of)
$H^2(X,\Z)$. Equivalently we may use 4-cycles $D_a$ in $H_4(X,\Z)$
dual to $e_a$ and obtain intersection numbers:
\begin{equation}
  \kappa_{abc} = D_a\cap D_b\cap D_c.    \label{eq:kcap}
\end{equation}

Furthermore, as explained in \cite{GST:sK5}, we may put homogeneous
coordinates $\xi^a$ on $\cM_V^5$ such that the metric is given by
\begin{equation}
  G_{ab} = \frac{\partial^2}{\partial \xi^a\partial \xi^b}\log\left(
      \kappa_{cde}\xi^c\xi^d\xi^e\right).
\end{equation}
This should be regarded as the ``special'' real geometry of $\cM_V^5$
where the ``prepotential'' is given by a pure cubic
$\cF_5=\kappa_{cde}\xi^c\xi^d\xi^e$. Relationships of this to Jordan
algebras are discussed in \cite{GST:sK5,GST:J}. 

We may instead regard $\xi^a$ as the {\em affine\/} coordinates in
$H^2(X,\R)=\R^{h^{1,1}}$. The K\"ahler form is then given by
\begin{equation}
  J = \xi^a e_a,
\end{equation}
and the condition that we fix the volume to, say one, for $\cM_V^5$,
is given by
\begin{equation}
  \int J\wedge J\wedge J=\cF_5=1.
\end{equation}
This latter condition can also be seen directly from supergravity
without reference to 
the geometrical interpretation of $X$ as in \cite{GST:J}. Thus again
we see strong hints that five dimensional $N=1$ supergravity ``knows''
that it has something to do with K\"ahler threefolds even without
direct reference to M-theory.

Note that the moduli space $\cM_V^5$ is not the complete hypersurface
$\cF_5=1$ in $H^2(X,\R)$. It turns out that phase transitions occur
precisely on the walls of the K\"ahler cone to truncate this
hypersurface to lie completely within the K\"ahler cone. This is
discussed in \cite{W:MF,MS:five,DKV:dP,IMS:5deg} for example but we
will not pursue it here.

We may now perform crude dimensional reduction of this
five-dimensional field theory to render it a four-dimensional theory.
Recall that dimensional reduction simply asserts that the fields have
no dependence on motion in the directions we wish to lose and we
decompose the vectors, tensors, etc accordingly into lower-dimensional
objects.

Performing this operation is a lengthy operation but the result for
the moduli space is straight-forward. 
As required, we obtain special K\"ahler geometry for $\cM_V$ in four
dimensions. 
Now we have {\em complex\/} homogeneous
coordinates $t^0,t^a$, where $a$ still runs $1,\ldots,h^{1,1}$ and a
prepotential
\begin{equation}
  \cF_0 = \frac{\kappa_{abc}t^at^bt^c}{t^0}.  \label{eq:F0}
\end{equation}

The very important point to realize however is that dimensional
reduction is {\em not\/} necessarily the same thing as
compactification on a circle (as emphasized in \cite{SW:3d} for
example). The problem is that solitons present in the five-dimensional
theory can become instantons in the four dimensional theory and add
quantum corrections to the picture. We may only regard (\ref{eq:F0})
as the classical contribution to the prepotential. We may expect
quantum corrections to appear with respect to $\alpha'$ as noted
earlier. Note that (\ref{eq:F0}) may be computed as the classical
contribution directly without a foray into five dimensional physics
\cite{Cand:mir}. The five dimensional picture is probably worth being
aware of however as it offers many insights.


\subsubsection{Mirror Pairs}   \label{sss:mir}

The easiest way of computing the quantum corrections to the
prepotential of the type IIA string compactified on $X$ is to use a
duality argument in the form of mirror symmetry. That is, can we find
a \CY\ threefold $Y$ such that the type IIB string compactified on $Y$
yields the same physics in four dimensions as the type IIA string
compactified on $X$? 
If this is the case $X$ and $Y$ are said to be ``mirror'' \CY\ threefolds.
Given the current state of our knowledge of
string theory it is probably not possible to rigorously prove that any
such pairs $X$ and $Y$ satisfy this condition. We can come fairly
close however. The reason is that because the dilaton of each of the
two type II strings appears in the moduli space of hypermultiplets in
a similar way, we may choose both strings to be simultaneously very
weakly coupled over the whole moduli space $\cM_V$. This allows us to
reliably use the worldsheet approach to analyze mirror pairs.

Thus the construction of mirror pairs of \CY\ can be reduced to a
conformal field theory question in two dimensions. We will then assume
that if two theories are mirror at this conformal field theory level
then they will be mirror pairs in the full nonperturbative string
theory picture.

The canonical example of mirror pairs of conformal field theories is
provided by the Greene--Plesser construction \cite{GP:orb}. An
explanation of this would require a major diversion into the
subtleties of conformal field theories which would take us way beyond
the scope of these lectures. We will then content ourselves to quote
their result. We refer to \cite{Greene:TASI,Greene:mir} for more details.

Consider the weighted projective space
$\P^4_{\{w_0,w_1,w_2,w_3,w_4\}}$ with homogeneous coordinates 
\begin{equation}
  [x_0,x_1,\ldots,x_4] \cong [\lambda^{w_0}x_0,\lambda^{w_1}x_1,
      \ldots,\lambda^{w_4}x_4],
\end{equation}
for $\lambda\in\C^*$. We may now consider the hypersurface $X$ given by
\begin{equation}
  x_0^{\frac{d}{w_0}}+x_1^{\frac{d}{w_1}}+\ldots+x_4^{\frac{d}{w_4}}=0,
\end{equation}
where $d=\sum w_i$ and we impose the condition
\begin{equation}
  \frac{d}{w_i}\in\Z,\quad\text{for all $i$}.  \label{eq:CYc}
\end{equation}
The projective space will generically have orbifold
singularities along subspaces. These orbifold loci may be blown-up to
smooth the space and we assume that $X$ is transformed suitably along
with this blowing up process to render it smooth. 

The Greene--Plesser statement is then 
\begin{prop}
  $X$ is mirror to $Y$, where $Y$ is the (blown-up) orbifold $X/G$ and
  $G$ is the group with elements
\begin{equation}
  g:[x_0,x_1,\ldots,x_4]\mapsto[\alpha_0x_0,\alpha_1x_1,\ldots,\alpha_4x_4],
\end{equation}
 where we impose $\alpha_i^{\frac{d}{w_i}}=1$ for all $i$ and
 $\prod\alpha_i=1$.    \label{prop:GP}
\end{prop}

The essence of this statement can be generalized considerably to
hypersurfaces in toric varieties and to complete intersections in toric
varieties as was done by Batyrev \cite{Bat:m} and Borisov
\cite{Boris:m,BB:mir}. This Batyrev--Borisov statement is not yet
understood at the level of conformal field theory but the evidence
that it does produce mirror pairs is very compelling. Thus there are a
very large number of candidate mirror pairs of \CY\ threefolds.

\subsubsection{The mirror map}   \label{sss:mmap}

Knowing the mirror partner $Y$ of $X$ is a good start to knowing how to
compute the quantum corrections to the prepotential of the type IIA
compactification but we need a little more information. Namely, we
need to know exactly how to map the coordinates on our special K\"ahler
moduli spaces between the type IIA and the type IIB picture. This is
known as the ``mirror map''.

The most direct way of finding the mirror map is to take a little peek
into the moduli space of hypermultiplets even though we are only
concerned with vector multiplets in this section. The fact we need to
borrow from hypermultiplets is that the Ramond-Ramond moduli must be
mapped into each other under mirror symmetry. For the hypermultiplet
moduli spaces we wrote down in section \ref{ss:II} this shows that
$H^{\textrm{odd}}(X,\GU(1))$ is mapped to
$H^{\textrm{even}}(Y,\GU(1))$.

The next statement we need concerns the symmetry of mirror pairs
themselves. We may state this as
\begin{prop}
If $X$ and $Y$ are mirror pairs then so are $Y$ and $X$.
\end{prop}
That is, the type IIA string on $Y$ is physically equivalent to the
type IIB string on $X$. This statement is completely trivial 
in terms of the definition of mirror pairs at the level of conformal
field theory. See for example \cite{Greene:TASI} for more
details. Here, since we are trying to be careful about not specifying our
definition of string theory, we will just have to assume that this proposition 
is true.

We therefore may assume that $H^{\textrm{even}}(X,\GU(1))$ is mapped
to $H^{\textrm{odd}}(Y,\GU(1))$ under the mirror map. This implies
some map between $H^{\textrm{even}}(X,\Z)$ and
$H^{\textrm{odd}}(Y,\Z)$. 

This map between the {\em integral\/} structures of the even
cohomology of $X$ and the odd cohomology of $Y$ is very interesting
and forms one of the most powerful ideas in mirror symmetry. Clearly
it cannot be an exact statement at the level of classical
geometry. This is because as we wander about the moduli space of
complex structures on $Y$ we may induce monodromy on
$H^{\textrm{odd}}(Y,\Z)=H^3(Y,\Z)$. That is, if we pick a certain basis for
integral 3-cycles in $Y$ we may go around a closed loop in moduli
space which maps this basis nontrivially back into itself. If this
statement were then mapped into a statement about the type IIA string
on $X$ we would conclude that some even-dimensional cycle, such as a
0-cycle, could magically transmute into a 2-cycle as we move about the
moduli space of complexified K\"ahler forms. Clearly this does not
happen!

\begin{difficult}
To explain this effect in geometric terms Kontsevich \cite{Kon:mir} has a very
interesting proposal based on some ideas by Mukai
\cite{Muk:FM}. Rather than thinking in terms of $H^{\textrm{even}}(X,\Z)$
directly one may consider $\mathbf{D}(X)$, the {\em derived category
of coherent sheaves\/} on $X$. Objects in $\mathbf{D}(X)$ are basically
complexes of 
sheaves of the form $\ldots\to\cF_1\to\cF_2\to\cF_3\to\ldots$. The
automorphisms of $H^3(Y,\Z)$ induced by monodromy can then be
understood in many cases in terms of automorphisms of $\mathbf{D}(X)$
\cite{Horj:DX}. Objects in $\mathbf{D}(X)$ can then be mapped into 
$H^{\textrm{even}}(X,\Z)$ essentially by using the Chern character.
This is a fascinating subject somewhat in its infancy
that promises much insight into mirror symmetry and stringy geometry.
\end{difficult}

Instead this statement must only be true classically at the large
radius limit of $X$ and thus the corresponding ``large complex
structure'' limit of $Y$. Somehow near this limit point, and in
particular monodromy about this point, these two integral structures
must align classically. This was first suggested in \cite{AL:qag} and
then explained more clearly in \cite{Mor:gid}. 

Let us suppose we fix a point in the moduli space of complex
structures, $\cM_V$, on $Y$ which will be our candidate limit point. As this is
a limit point it is natural to expect that $Y$ will be singular
here. Actually one expects to
find singular $Y$'s along complex co-dimension one subspaces of
$\cM_V$. This special limit point turns out to be
a particularly nasty singularity
as it lies on an intersection of many such divisors. We will assume
there are in fact $h^{2,1}(Y)=\dim \cM_V$ such divisors intersecting
transversely at this limit point. If this is not the case then one may
blowup using standard methods in algebraic geometry to reduce back to
this case.
We may now consider the monodromy matrices $M_k$ which act on
$H^3(Y,\Z)$ as we go around each of these divisors. 

Mapping back to $X$ this limit point should be the large radius limit
as every component of the K\"ahler form tends to infinity. The
monodromy about this limit is $B\to B+v$, where $v\in H^2(X,\Z)$.

This fixes the mirror map as follows. First we need to switch back to
the dual language of periods defined in (\ref{eq:periods}). We will
find one period which we denote $t_0$ which is completely invariant
under the monodromies $M_k$. We also find periods $t_k$ such that
\begin{equation}
\begin{split}
  M_k: t_k &\mapsto t_k+t_0\\
  M_j: t_k &\mapsto t_k, \qquad\text{for $k\neq j$}
\end{split}
\end{equation}
where $k=1,\ldots,h^{2,1}(Y)$. The fact we may do this is a special
property of the limit point we have chosen and defines the property
that it can represent the mirror of a large radius limit point. This
is explained in more detail in \cite{Mor:gid}.

We now give the mirror map:
\begin{equation}
  (B+iJ)_k = \frac{t_k}{t_0},	\label{eq:mm}
\end{equation}
where $B+iJ = \sum (B+iJ)_ke_k$ is the complexified K\"ahler form on
$X$ expanded over a basis $e_k\in H^2(X,\Z)$.

This is the only map possible which gives the correct monodromy and
reflects the projective symmetry of the homogeneous coordinates $t_a$.

The canonical example is that of the quintic threefold as computed in
\cite{CDGP:}. In this case $X$ is the quintic hypersurface in $\P^4$
and thus $Y$ is $X/(\Z_5)^3$ according to proposition \ref{prop:GP}.
This case is fairly straight-forward as $\cM_V$ is only one dimensional
since $h^{1,1}(X)=h^{2,1}(Y)=1$.
In this case one can compute the periods and use (\ref{eq:pre}) to
compute
\begin{equation}
  \cF = (t_0)^2\Bigl(5t^3 + \ff{33}2t^2 - \ff{25}{72}t
  +\frac{25i}{12\pi^3}\zeta(3) + \frac{2875i}{72\pi^3}e^{2\pi it}
  +O(e^{4\pi it})\Bigr),   \label{eq:CDGP:}
\end{equation}
where $t=t_1/t_0$ can be viewed as the single component of the
complexified K\"ahler form on $X$.

Note we indeed get the correct leading term $5t^3$ from the
intersection theory but there are an infinite number of quantum
corrections. The quadratic and linear terms in $t$ are physically
meaningless whereas the constant term proportional to $\zeta(3)$ is a
loop term correcting the metric. 

The power series in $q=e^{2\pi it}$ corresponds to the worldsheet
instanton corrections. An instanton in the worldsheet quantum field
theory (\ref{eq:nlsm}) corresponds to a holomorphic map from
$\Sigma$ into the target space $X$ \cite{DSWW:}. At tree-level in string theory
such objects are therefore ``rational curves'' (i.e. holomorphic
complex curves of genus zero). 

This is an important subject and any respectable review of $N=2$
theories in string theory should go into some detail about these
rational curves. We will not do this however as there are already a
number of reviews of this subject.
As is explained in numerous places elsewhere,
the quantum corrections may be used to {\em count\/} the numbers of
rational curves 
in $X$. Indeed the 2875 appearing in (\ref{eq:CDGP:}) corresponds to the
number of {\em lines\/} on a quintic surface. The interested reader
should consult \cite{CK:mbook}, for example, for much more information
about this vast subject.

One rough and ready way to appreciate why rational curves should make
an appearance in mirror symmetry is as follows. We have already argued
that the truly stringy geometry of $X$ must somehow mix up the notion
of 0-cycles, 2-cycles, 4-cycles, etc as we move away from the large
radius limit. These worldsheet instanton corrections near the large
radius limit can be thought of as the way that 2-cycles (i.e.,
rational curves) start to mix into our notion of 0-cycles (i.e.,
points). This stringy geometry which can mix the notion of points and
rational curves has yet to be understood properly.


\subsection{$\cM_V$ in the heterotic string}  \label{ss:hetV}

Now we will consider the moduli space $\cM_V$ in terms of the
heterotic string compactified on $S_H\times E_H$. For a field theorist
with a bias towards gauge theories this is actually the most
useful way of viewing the resulting $N=2$ theories in four dimensions
as we now explain. 


\subsubsection{Supersymmetric abelian gauge theories} \label{sss:U1}

An abelian gauge theory of $\GU(1)^{n+2}$ in flat space is based on
the action 
\begin{equation}
  \int  d^4x\,\Bigl(\frac1{\lambda^2}\sum_{a=1}^{n+2}\|F^a\|^2+\ldots\Bigr),
	\label{eq:YM}
\end{equation}
where $\lambda$ is the gauge coupling constant. 
If this is an $N=2$ supersymmetric theory then $n+1$ of the $\GU(1)$'s
are associated to vector multiplets and the extra $\GU(1)$ is the
``graviphoton'' coming from the supergravity multiplet.
If the gauge particles
are actually fundamental strings then the coupling constant should be
given by the string dilaton as in equation (\ref{eq:qft}) and the
discussion following this equation. In the
heterotic string, the dilaton lives in one of the $n+1$ vector
multiplets and pairs up with the axion to form the complex field
\begin{equation}
  s = a + \frac{i}{\lambda^2},  \label{eq:S}
\end{equation}
where $a$ is the axion. We should note the difficulty of trying to
find such a theory by compactifying a type II string. Here the
dilaton lives in a hypermultiplet which cannot couple to the vector
bosons in the desired way.
Thus the gauge coupling cannot be interpreted as a type II
string coupling --- gauge bosons cannot be fundamental strings.

The fact that such a term is expected in the action immediately tells
us the form of the prepotential governing $\cM_V$. This is perhaps
easiest to see if we compactify the heterotic string first on $S_H$
times a circle to get an $N=1$ theory in five dimensions, and then
compactify on a circle to get our desired four-dimensional theory. The
theory in five dimensions will have generic gauge symmetry
$\GU(1)^{n+1}$ as compactifying on a circle gives a $\GU(1)$ via the
Kaluza--Klein mechanism.

The interesting term in the five-dimensional theory is the
Chern-Simons-like term (\ref{eq:CS5}). What will this reduce to when
we compactify on the circle down to four dimensions? The answer is
that we will replace the vector field $A$'s by four-dimensional real
scalar $a$'s to form a term 
\begin{equation}
  \int d^4x\, \kappa_{ebc}a^eF^b\wedge F^c.
\end{equation}
But this is the famous CP-violating term of a gauge theory and the
scalar field is playing the r\^ole of an axion. If we want the kinetic
term in the standard form (\ref{eq:YM}) then the {\em only\/}
scalar which is allowed to play the r\^ole of an axion 
in a theory with $N=2$ supersymmetry is the axion
partner of the dilaton, namely the $a$ in (\ref{eq:S}). In addition
this axion is not allowed to appear as a coefficient in front of field
strengths associated with the $\GU(1)$ gauge boson in the same
multiplet as the axion and dilaton. This would lead to rather
unorthodox terms proportional to $\lambda^{-4}$ in the action under
supersymmetrization.  

This implies immediately that the superpotential in five dimensions is
of the form $\cF_5=st^it^j\gamma_{ij}$, for some matrix $\gamma_{ij}$,
for $i,j=1,\ldots,n$ and where the $t^i$'s are the
five-dimensional moduli fields associated to the vector supermultiplets
other than the dilaton.

Of course we expect this cubic superpotential to be corrected when we
are in four dimensions just as the cubic potential was corrected for
the type IIA compactifications in section \ref{ss:mir}. This time
however the corrections will not be $\alpha'$-corrections due to
worldsheet instantons but they will be $\lambda$-corrections due to
gauge instantons in spacetime.

The cubic superpotential $\cF=st^it^j\gamma_{ij}$ which is exact in
five dimensions and correct to leading order in four dimensions is
rather constraining. We may also note that in order for the kinetic
term for the photons to 
be positive-definite it is required that $\gamma_{ij}$ have signature
$(+,-,-,-,\ldots)$ \cite{GST:J}. 
Running through a lengthy process using the definitions of special
K\"ahler geometry this leads to a moduli space for our
four-dimensional theory locally of the form \cite{FvP:Ka}
\begin{equation}
  \cM_{V,0}^{\textrm{Het}} = \frac{\Sl(2,\R)}{\GU(1)} \times
                                 \frac{\SO_0(2,n)}{\SO(2)\times\SO(n)},
	\label{eq:Het0}
\end{equation}
{\em before quantum corrections.}

The first thing to note about this space is that it is a product of
two symmetric spaces --- just the kind of thing we would expect if we
had more supersymmetry. The second thing to note is that the second term looks
a lot like Narain's moduli spaces as we discussed in section
\ref{sss:EH}. This term represents just what we would expect if we
look at the stringy moduli space of vector bundles of rank $n-2$ over
a 2-torus, together with deformations of the torus itself. This is
excellent news. It means that if we identify the first term with the
dilaton and axion then we have a very natural interpretation of this
moduli space in terms of the data discussed in section \ref{sss:EH}.

To get the moduli space perfectly correct we need to worry about the
global form. If the second term really is the Narain moduli space of
the bundle $V_E\to E_H$ discussed in section \ref{sss:EH} then we
should really expect it to be of the form
\begin{equation}
  \GO(\Upsilon)\backslash\GO(2,n)/(\GO(2)\times\GO(n)),
	\label{eq:Narf}
\end{equation}
where $\Upsilon$ is the lattice of signature $(2,n)$ given by
$\Gamma_{2,2}\oplus L$ and where $L$ is the Cartan lattice of the
structure group of $V_E$ with {\em negative\/} definite signature. 

One might also be tempted to replace the first term of (\ref{eq:Het0})
by the expression
\begin{equation}
  \Sl(2,\Z)\backslash \Sl(2,\R)/\GU(1).
\end{equation}
This $\Sl(2,\Z)$ would certainly respect the axion shift symmetry
$a\to a+2\pi$ which we expect to be correct but it would also imply
some strong-weak 
coupling duality for $N=2$ theories in four dimensions. This does not
exist in general. The problem is that moduli space (\ref{eq:Het0})
ignores quantum corrections and it therefore only correct as the
dilaton tends to $-\infty$. The only part of $\Sl(2,\Z)$ which
preserves this limit is the axion shift symmetry.

If we could find a type IIB string compactification dual to this
heterotic model then we could compute the prepotential exactly, just as
we did by mirror symmetry for the type IIA string. This would allow us
to compute the nonperturbative corrections to the moduli space arising
from quantum corrections due to $\lambda$. 

Oddly enough it is much more natural to ask first for a type IIA
string dual to the heterotic model we desire.


\subsubsection{Heterotic/Type IIA duality}  \label{sss:HIIA}

If the type IIA string compactified on $X$ is dual to the heterotic
string model giving the gauge theory of section \ref{sss:U1} then we
know a surprising amount of the geometry of $X$ with very little
effort.

The fact that the prepotential is
\begin{equation}
 \cF_0=st^it^j\gamma_{ij}  
\end{equation}
to leading order tells us about the cup product structure of
$H^2(X,\Z)$ or equivalently, the intersection form on $H_4(X,\Z)$. In
particular from (\ref{eq:kcap}) it tells us that the 4-cycle $S$
representing the complexified dilaton $s$ satisfies
\begin{equation}
  S\cap S\cap D =0,
\end{equation}
for any $D$ (whether it be associated to $s$ or a $t^i$). 
This implies that $S\cap S$ is empty.

One may now proceed \cite{KLM:K3f,AL:ubiq,me:lK3} to show that 
\begin{enumerate}
    \item $S$ can be represented by an algebraic surface embedded in $X$.
    \item $S$ is a K3 surface.
    \item Moving $S$ parallel to itself (as suggested by $S\cap
S=0$) sweeps out all of $X$. That is, {\em $X$ is a K3-fibration}. 
\end{enumerate}
As this is reviewed at length in \cite{me:lK3} we will not repeat the proof
here.

It is not hard to show that in order for $X$ to be a \CY\ manifold
with $\SU(3)$ holonomy it must have finite (or trivial) fundamental
group $\pi_1(X)$. For a K3 fibration $X\to W$, this implies that the
base $W$ also has finite $\pi_1$. Thus if $W$ is a smooth space of
complex dimension one, it must be isomorphic to $\P^1$.

Anyway, not only do we now know that $X$ is a K3-fibration, we
also know exactly which modulus of the complexified K\"ahler form
corresponds to the dilaton-axion. We know that the element of $H_4(X)$
corresponding to $S$ is the homology class of a generic K3 fibre. We
need the component of the K\"ahler form which controls the size of a
2-cycle which is dual (via intersection theory) to this K3 fibre. For
simplicity we could assume that $X$ as a K3-fibration has a global
section.\footnote{This section need not be unique and in the example
in section \ref{sss:ex1} it will not be. Its homology class and hence
its area is unique however.} That is, we have an embedding $W\to X$
which is an ``inverse'' of the fibration projection. This section acts
as a 2-cycle dual to the K3 fibre. We have thus shown

\begin{prop}
If a type IIA string on $X$ is dual to a heterotic string on a K3
surface times a torus, then $X$ must be a K3 fibration. Assuming this
fibration has a section then the area of this section (and the
corresponding component of the $B$-field) maps to the dilaton (and
axion) on the heterotic side.   \label{prop:Hbase}
\end{prop}
We refer to \cite{me:lK3} for a careful statement of the assumptions
which go into this proposition.

People often loosely refer to the area of the section as the ``area of
the base''.
If $X$ does not have a section then this duality can still work --- we
just have to work a little harder to determine the dilaton. We will
always assume there is a section.

At this point it is worthwhile to consider a sketchy picture of
instanton corrections in this dual pair. On the heterotic side we have
spacetime instanton effects\footnote{The observant reader will note
that we had assumed that we had an abelian gauge theory. Therefore we
don't really have any instantons in the gauge theory. We will see
in section \ref{sss:YM} that, if we want, there really is a 
nonabelian gauge theory lurking here.} which produce effects of the order
$\exp(-ns)$ in correlation functions. In the type IIA picture one gets
exactly the same effects thanks to the above mapping by wrapping
worldsheet instantons around the section of the fibration. Thus {\em
spacetime instanton effects in the heterotic string are exchanged with
worldsheet instanton effects in the type IIA string.} 

One can consider this statement to be rather profound. It shows that
neither the worldsheet picture nor the spacetime picture of the
quantum field which ``models'' string theory can be more fundamental
than the other. At least in the sense of instanton corrections, the
two pictures may be interchanged.

\iffigs
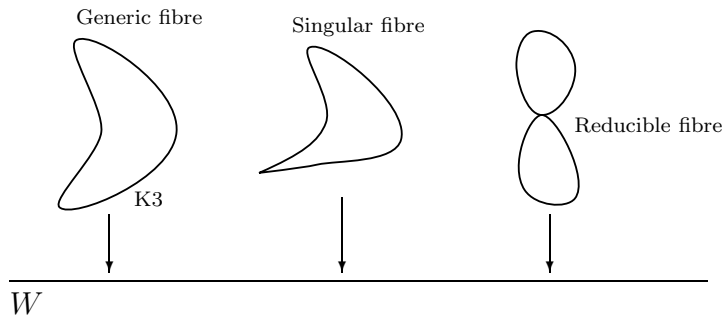
\begin{figure}
\begin{center}
\setlength{\unitlength}{0.00058300in}%
\begin{picture}(6324,2760)(1489,-4288)
\thinlines
\put(1501,-3961){\line( 1, 0){6300}}
\put(2401,-3361){\vector( 0,-1){525}}
\put(4501,-3211){\vector( 0,-1){675}}
\put(6376,-3361){\vector( 0,-1){525}}
\put(2101,-1786){
}
\put(3751,-2986){
}
\put(6301,-2461){
}
\put(6301,-2461){
}
\put(2101,-1636){\makebox(0,0)[lb]{\smash{\scriptsize Generic fibre}}}
\put(1501,-4261){\makebox(0,0)[lb]{\smash{$W$}}}
\put(4051,-1711){\makebox(0,0)[lb]{\smash{\scriptsize Singular fibre}}}
\put(6601,-2611){\makebox(0,0)[lb]{\smash{\scriptsize Reducible fibre}}}
\put(2626,-3286){\makebox(0,0)[lb]{\smash{\scriptsize K3}}}
\end{picture}
\end{center}
  \caption{A K3 fibration.}
  \label{fig:Kfib}
\end{figure}
\fi

Now we have discussed one of the moduli in $\cM_V$, let us find the
others. Note first that although $X$ is a K3 fibration, not all of its
fibres need be K3 surfaces. We only demand that the {\em generic\/}
fibre is a K3 surface. We refer to fig.~\ref{fig:Kfib}
for a picture of the K3 fibration.
There are three ways to obtain contributions towards $H^2(X)$
in terms of $X$ as a K3 fibration. Let us list the different
generators of $H^2(X)$  in the language of deformations of the K\"ahler form:
\begin{enumerate}[I:]
\item Deform the area of the section $W\to X$.
\item Deform the areas of curves within the generic K3 fibres.
\item Deform the independent volumes of the irreducible components of
a reducible bad fibre.
\end{enumerate}

In terms of elements of $H_4(X)$, the contributions of type II are
obtained by taking a 2-cycle $C_i$ in a generic fibre and then sweeping it
out by moving over $W$ to produce a 4-cycle which we denote $D_i$. In
this way, the 
intersection pairing $C_i\cap C_j$ between 2-cycles in the K3 fibre is
copied into the intersection numbers $S\cap D_i\cap D_j$ for $X$. The
$C_i$'s in the K3 fibre are not just any old 2-cycle. They have to be
{\em algebraic curves\/}, i.e., holomorphically embedded. It can be
shown \cite{me:lK3} that the intersection form with a K3 surface of
algebraic curves is an indefinite quadratic form of signature
$(+,-,-,-,\ldots)$. 

This shows that the moduli coming from contributions of type I and II
to $H^2(X)$ from the K3 fibrations form the special
K\"ahler geometry with prepotential $\cF=st^it^j\gamma_{ij}$ to
leading order as required in section \ref{sss:U1}. Indeed, one may
prove the following (as is done in section 3.4 of \cite{me:lK3} for example):
\begin{prop}
The moduli space of the K\"ahler form and $B$-field for a type IIA 
string on an algebraic K3 surface $S$ is given by
\begin{equation}
  \GO(\Upsilon)\backslash\GO(2,n)/(\GO(2)\times\GO(n)),
	\label{eq:Narf2}
\end{equation}
where $\Upsilon=\Pic(S)\oplus\Gamma_{1,1}$ and $\Pic(X)$ is the
``Picard lattice'' given by the algebraic curves in
$S$ together with their intersection form. The integer $n$ is given by
the dimension of the Picard lattice.
	\label{prop:p}
\end{prop}
This gives a precise isomorphism
between the moduli of type II above and the Narain factor of the
moduli space of the heterotic string.

There is one more result we may state here which will be useful later
on. Given the Narain moduli space of the heterotic string on $T^2$ as
given in (\ref{eq:Narf}) we can see that $\Upsilon$ must contain
$\Gamma_{2,2}$ and so $\Pic(S)$ will contain $\Gamma_{1,1}$. This is
actually a necessary and sufficient condition for $S$ to be {\em an
elliptic K3 surface with a least one section\/} \cite{me:lK3}. The
fibration structure of each K3 fibre extends to the following
statement about the whole of $X$:

\begin{prop}
If a type IIA string on $X$ is dual to a heterotic string on a K3
surface times a torus, and we see the full moduli space of the
heterotic torus, then $X$ is an elliptic fibration over some complex surface
with at least one section.
	\label{prop:F1}
\end{prop}

So finally what about contributions of type III?  These 4-cycles can be
associated with components of reducible fibers which do not intersect
the section. This lack of intersection with $S$ violates the expected
special K\"ahler geometry from section \ref{sss:U1}. It turns out that
these moduli will be something to do with the full nonperturbative
physics of the heterotic string --- more than is described by the
effective action discussed in section \ref{sss:U1}. We will have more
to say about these type III divisors later.


\subsubsection{Enhanced gauge symmetry}   \label{sss:YM}

We now want to deal with the important subject of enhanced nonabelian
gauge symmetries in the effective four-dimensional uncompactified
dimensions. To simplify our discussion we will tackle only the subject
of {\em simply-laced\/} Lie algebras, i.e., the ``ADE'' series of Lie
algebras whose roots are all the same length. We will also ignore the
subject of the global topology of the gauge group. That is we will not
concern ourselves too much with whether or not a gauge group is really
$\SU(2)$ or $\SO(3)$ for example. We refer the reader to \cite{me:lK3,
AM:frac,AKM:lcy} for details about these subtleties which we are ignoring.

To leading order we have a factor looking like (\ref{eq:Narf}) in the
moduli space which we recognize as the Narain moduli space for some
vector bundle, $V_E$, on a 2-torus $E_H$. The moduli here may be regarded as
deformations of the flat metric on the torus itself together with
deformations of the flat bundle. As discussed in section \ref{sss:EH},
the parameters controlling the bundle
are known as ``Wilson lines''. They measure the holonomy of the
bundle as we go around non-contractable loops within $E_H$.

As observed in section \ref{ss:N2}, the observed gauge group in the
uncompactified dimensions which remains unbroken by the compactification
process can be regarded as the centralizer of the holonomy acting on
the ten-dimensional primordial gauge group. For generic values of the
Wilson lines the holonomy of $V_E$ is $\GU(1)^{n-2}$, where $n-2$ is
the rank of the structure group $\cG_E$ (which we assume to be
simply-laced) of $V_E$. This holonomy is
simply the Cartan subgroup of $\cG_E$ and so the unbroken part of the
gauge symmetry is $\GU(1)^{n-2}$. (Note that the compactification
process on $E_H$ {\em adds\/} four more $\GU(1)$'s to bring the total
to $n+2$ as in section \ref{sss:U1}.)

The interesting question arises as to what happens when the holonomy
of the bundle $V_E$ is {\em not\/} generic. If we switch off some of
the Wilson lines, we might expect the structure group of $V_E$ to
decrease allowing for a larger centralizer. That is, the observed
gauge symmetry in four dimensions should become larger.

The idea is that the moduli space 
$\GO(2,n)/(\GO(2)\times\GO(n))$ is viewed as the Grassmannian of
space-like (positive) 2-planes $\mho\subset\R^{2,n}$. One may also embed
the lattice $\Upsilon$ into this same $\R^{2,n}$. The desired moduli
space (\ref{eq:Narf}) is then this Grassmannian divided out by the
automorphisms of the lattice $\Upsilon$. The rule is then as follows:
\begin{prop}
The observed gauge group in uncompactified space has rank $n+2$. The
roots of the semi-simple part of this gauge group correspond to
elements of $\Upsilon$ which have length squared $-2$ and which are
orthogonal to $\mho$.
\end{prop}

A few points are worth noting:
\begin{enumerate}
\item At a generic point in the moduli space $\mho$ is orthogonal to
no such elements of $\Upsilon$ and so the gauge group is $\GU(1)^{2+n}$ as
expected.
\item This rule is completely derivable from classical geometry for
the case that the roots are in $L\subset\Upsilon$, where $L$ is the
root lattice of $\cG_E$. Picking up roots in the rest of $\Upsilon$ is
a stringy effect --- the analogue of the $\SU(2)$ gauge symmetry one
sees on a circle of self-dual radius (see \cite{Gins:lect} for
example).
\item The maximal rank of the semi-simple part of the observed gauge
group is $n$. There are always at least two $\GU(1)$ factors which are
not enhanced to nonabelian groups. This is because the GSO projection
of the supersymmetric half of the heterotic string projects out the
would-be vector bosons which would like to enhance these gauge group
factors.
\item This Grassmannian picture for the moduli space is only true to
leading order. We can expect quantum corrections to break anything
--- including the nonabelian enhanced gauge symmetry.
\end{enumerate}

Now we would like to map this picture of gauge symmetry enhancement
back into the language of the type IIA string compactified on
$X$. What do we need to do to $X$ to get an enhanced gauge symmetry?

This is explained in great detail in \cite{me:lK3}. First of all note
that the factor (\ref{eq:Narf}) of the moduli space corresponds
exactly to the K\"ahler form parameters of ``type II'' above. We know this
because of the special K\"ahler geometry discussed in section
\ref{sss:U1} and the intersection numbers discussed in section
\ref{sss:HIIA}. This means that moving around in this Narain component
of $\cM_V$ corresponds to changing the size (and $B$-field) of the
algebraic curves in the generic K3 fibres of $X$.

The result is \cite{W:dyn,me:enhg,me:en3g,KMP:enhg,me:lK3}
\begin{prop}
Let a set of algebraic genus zero curves collapse to zero area in
every K3 fiber in $X$. Thus $X$ acquires a curve of singularities. In
addition set the corresponding components of the $B$-field to
zero. Then one obtains a nonabelian enhanced gauge symmetry. The ADE
classification of curves one may collapse in a K3 surface corresponds
to the ADE classification of the resulting Lie gauge groups.
\end{prop}
Again we need to note a few points:
\begin{enumerate}
\item We are assuming that there is no monodromy in these curves in
the K3 fibres as we move around the base $W$. If there is monodromy
one can obtain non-simply-laced gauge symmetries which we do not wish
to discuss here.
\item We also assume that the overall volume of each K3 fiber is
generic. By tuning the volume to the right values one may enhance the
gauge symmetry further.
\end{enumerate}

One usual way of picturing the appearance of a nonabelian gauge
symmetry is as follows. The type IIA string theory contains 2-branes
in its spectrum (as discussed in many other lectures at this
school). These 2-branes may be ``wrapped'' around the 2-spheres living
in the K3 fibres. The mass of the resulting solitons in the
four-dimensional theory is given by the area (and $B$-field) of these
2-spheres. In the limit that these spheres shrink to zero size we
obtain new massless states in the theory. These massless states may
lie in either hypermultiplets or vector multiplets. Which type is
determined by the moduli space of the 2-cycle that shrank down to zero
size. Witten showed \cite{W:MF} that isolated curves give rise to
hypermultiplets and curves that live in families parametrized by other
curves give vectors. Thus, in our case where we are shrinking down
whole families of curves in order to obtain a \CY\ threefold with a
singular curve we expect extra vectors. These vectors are the
``W-bosons'' which enhance the gauge group to a nonabelian group.

The case we have considered here is actually a special case of
acquiring a singular curve in $X$ and so must be considered to be a
special case of acquiring nonabelian gauge symmetry. Consider the
projection given by the K3-fibration $\pi:X_1\to W$ when $X_1$ is a
singular space made by shrinking down a particular curve (or set of
curves) within every K3 fibre. Let $C_{\text{sing}}\subset X_1$ be the
resulting singular curve within $X_1$. The restriction of the fibration
\begin{equation}
  \pi|_{C_{\text{sing}}}:C_{\text{sing}}\to W
\end{equation}
is an isomorphism. 

Suppose that we can find another family of curves
within $X$ which can be shrunk down to form another singular space
$X_2$ with a singular set $C_{\text{sing}}'\subset X_2$. The
projection under $\pi$ of a general singular set may or may not be
surjective onto $W$. {\em In particular we may have that the image under
$\pi$ is a point (or a set of points) in $W$.}
It is not hard to see that the fibre over such a point in $W$ is
peculiar and could not possibly be a smooth K3 fibre. Indeed we are
talking about contributions of ``type III'' to the moduli space of
vector multiplets when we shrink such 2-cycles down. We therefore
claim that a singular curve lying over a {\em point\/} in $W$ must
correspond to a {\em nonperturbative\/} enhanced gauge
group. We will see examples of nonperturbative gauge groups in
section~\ref{ss:Hclas}.


\subsubsection{An example} \label{sss:ex1}

Now that we have spoken rather abstractly about duality let us give an
example which illustrates most of what we have discussed above. This
example first appeared in \cite{KV:N=2}.

We begin by describing the \CY\ threefold $X$ on which we will
compactify the type IIA string. Let $X$ be the hypersurface
\begin{equation}
   x_0^2+x_1^3+x_2^{12}+x_3^{24}+x_4^{24}=0
\end{equation}
in the weighted projective space $\P^4_{\{12,8,2,1,1\}}$ with
homogeneous coordinates
\begin{equation}
  [x_0,x_1,x_2,x_3,x_4] \cong [\lambda^{12}x_0,\lambda^8x_1,
     \lambda^2x_2,\lambda x_3,\lambda x_4].
\end{equation}
Note that this satisfies the \CY\ condition (\ref{eq:CYc}). We also
need to note that this \CY\ threefold is not smooth. In particular,
putting $\lambda=i$ we obtain
\begin{equation}
  [x_0,x_1,x_2,x_3,x_4] \cong [x_0,x_1,-x_2, ix_3, ix_4].
\end{equation}
which produces a $\Z_4$ singularity at $[x_0,x_1,0,0,0]$ which is a
single point in $X$. Similarly
putting $\lambda=-1$ puts a $\Z_2$ singularity along
$[x_0,x_1,x_2,0,0]$, which is a curve in $X$ (containing the previous
$\Z_4$ fixed point).

These quotient singularities need to be blown up if we want a nice
smooth \CY\ threefold for $X$. For the singular curve in $X$
fixed by $\Z_2$ we may replace each point in this curve by a
$\P^1$. The homogeneous coordinates of this $\P^1$ may be considered
to be $[x_3,x_4]$ (which are not now allowed to vanish simultaneously ---
we have removed the singularity after all!).

Actually we may view $[x_3,x_4]$ as the coordinates of $W\cong\P^1$
and project in the obvious way
\begin{equation}
  \pi:X\to W.
\end{equation}
Let us denote a given point on $W$ by $\mu$. That is, let $x_4=\mu
x_3$. Then the inverse image of a point in $W$ under $\pi$ is 
\begin{equation}
   x_0^2+x_1^3+x_2^{12}+x_3^{24}(1+\mu^{24}) =0  \label{eq:K3fib}
\end{equation}
in the weighted projective space $\P^3_{\{12,8,2,1,\}}$. This is a K3
surface as required. This is most easily seen by putting $x_3'=x_3^2$
giving us an equation in $\P^3_{\{6,4,1,1\}}$. Thus we have written
$X$ as a K3-fibration. 

We may now play the same trick again on each K3 fibre. Each K3 fibre
has a $\Z_2$ singularity in it (as a side effect of the $\Z_4$
singularity in the original threefold). This may be resolved by
replacing it with a $\P^1$ which we denote $C$. Thus the fibre itself may be
written as a bundle over $C\cong\P^1$ with fibre given by a cubic equation in
$\P^2_{\{3,2,1\}}$ --- namely an elliptic curve.

Thus our final smooth $X$ consists of a K3-fibration over $W\cong\P^1$
where each K3 fibre is itself an elliptic fibration over another $C\cong\P^1$.
All these fibrations have sections allowing us to identify $W$
and $C$ as the bases of fibrations with subspaces of $X$.  

Now we may describe $H^2(X)$, or equivalently $H_4(X)$, in terms of
this K3 fibration in the language of section \ref{sss:HIIA}.
\begin{enumerate}[I:]
\item We have the size of the section $W$. This gives one
vector multiplet.
\item We may vary the sizes of the section $C$ of each K3 fibre and we
may vary the size of each elliptic fibre of these K3's. This gives two
more vector multiplets.
\item The only bad K3 fibres occur where $\mu^{24}=-1$ in
(\ref{eq:K3fib}). The resulting polynomial does not factorize and so
this bad fibre is still irreducible. Therefore we obtain no more
vector multiplets associated with bad fibres.
\end{enumerate}
So we have a theory with three vector multiplets (indeed,
$h^{1,1}(X)=3$). 

We may now write down the form of the moduli space to leading order
using proposition~\ref{prop:p}. First we need the Picard lattice of
the generic K3 fibre. There are two generators: the elliptic fibre, $e$ and
the $\P^1$ section $f$. It is not hard to show that $e\cap e=0$,
$e\cap f=1$, and $f\cap f=-2$. This intersection matrix is isomorphic
to $\Gamma_{1,1}$. Thus $\Upsilon\cong\Gamma_{2,2}$ and $n=2$ in 
(\ref{eq:Narf2}).

Let us try to find a heterotic string interpretation of this moduli
space. Going back to the discussion around equation (\ref{eq:Narf}) we
see that we have the simplest case where $L$, the Cartan lattice of
$\cG_E$, is empty and indeed the rank of $\cG_E$ is $n-2=0$. The
vector moduli space is purely described in terms of deformations of
the dilaton-axion and the Narain moduli space of the 2-torus $E_H$
with {\em no\/} bundle degrees of freedom. This accounts for all three
vector moduli.

In other words, all the the primordial gauge group in ten dimensions
must have been sucked up with the bundle on the K3 surface $S_H$ leaving
nothing left for $E_H$ to play with. To describe exactly what this
bundle on $S_H$ is requires a knowledge of the hypermultiplet moduli
space and so we won't be able to discover this until section~\ref{sss:E8l}.

We get enhanced gauge symmetries in the following ways. We may shrink
down the section $f$ in every K3 fibre. The undoes the second blow-up
we did when resolving at the start of this section. It produces a
single curve of ``$A_1$'' singularities within $X$. It corresponds to
putting the space-like 2-plane $\mho$ perpendicular to the single vector
$s\in\Upsilon$. Either way, we get an $\SU(2)$ gauge symmetry.

We may also squeeze out a rank 2 gauge symmetry --- either
$\SU(2)\times\SU(2)$ or $\SU(3)$ by tuning the vector moduli
further. This can be seen by noting that $A_1\oplus A_1$ and $A_2$ can
both be embedded in $\Gamma_{2,2}$ and we may arrange $\mho$ to be
orthogonal to either.  This corresponds to shrinking the elliptic
fiber, $e$, down to an area of order 1 as well as tuning the size of
$f$. The precise details are given in \cite{me:lK3}.

Now let us turn our attention to the type IIB picture. Using
proposition \ref{prop:GP} we see that $Y$ is given by
$X/(\Z_6\times\Z_{12})$ where the generators of the quotienting group
are given by
\begin{equation}
\begin{split}
g_1:[x_0,x_1,x_2,x_3,x_4]&\mapsto [x_0,x_1,x_2,e^{\frac{2\pi i}{24}}x_3,
e^{-\frac{2\pi i}{24}}x_4]\\
g_2:[x_0,x_1,x_2,x_3,x_4]&\mapsto [x_0,x_1,e^{\frac{2\pi i}{12}}x_2,x_3,
e^{-\frac{2\pi i}{12}}x_4].
\end{split}
\end{equation}
The general form of $Y$ may be written as a quotient of $X$ with
defining equation
\begin{equation}
  x_0^2+x_1^3+x_2^{12}+x_3^{24}+x_4^{24}+
  \alpha\, x_0x_1x_2x_3x_4 +\beta\, x_2^6x_3^6x_4^6+\gamma\,
  x_3^{12}x_4^{12}=0.  \label{eq:Y}
\end{equation}
The three parameters $\alpha$, $\beta$ and $\gamma$ then give the three
deformations of complex structure of $Y$ (as $h^{2,1}=3$).

Knowing the details of the mirror map allows us to map these
parameters to the complexified K\"ahler form of the type IIA
description. One may determine this using the ``monomial-divisor''
mirror map of \cite{AGM:II,AGM:mdmm} when one has a hypersurface in a
weighted projective space. 
This particular model was also studied in \cite{HKTY:}.
The upshot is that if $X$ is in the ``\CY\
phase'' where the areas of all possible algebraic curves are large then
essentially 
\begin{itemize}
\item Letting $x=\beta/\alpha^6\to0$ will take the size of the
elliptic fibre off to infinity.
\item Letting $y=4/\gamma^2\to0$ sends the size of the section
$W$ off to infinity.
\item Letting $z=4\gamma/\beta^2\to0$ sends the area of the
rational curve $f$ within each K3 fibre off to infinity.
\end{itemize}
The parameters $(x,y,z)$ are chosen so that the interior of the
K\"ahler cone of $X$
is described asymptotically by $x\ll1$, $y\ll1$ and $z\ll1$.
Away from this limit these parameters can get mixed up and everything
is less clear although well-understood.


\subsubsection{Quantum corrections to $N=2$ gauge theories} \label{sss:SW}

So far we have discussed purely the classical limit of the heterotic
string theory where we assume the dilaton is such that the coupling is
very weak and that the prepotential is purely cubic.

Thanks to the duality of the heterotic string to the type IIB string
we may try to continue our analysis of the heterotic string away from
this classical limit. This is an enormous subject but we will be very
brief here. Our intention is to give only a flavour of the subject.

Let us explain what happens in terms of the example of the previous
section. In particular let us study what happens to the would-be
$\SU(2)$ gauge theory which appears when every K3 fibre of $X$
contains an $A_1$ singularity.

First of all we mentioned that we could actually get the gauge group
to be $\SU(2)\times\SU(2)$ or $\SU(3)$ if we tuned the size of the K3
fibre suitably. Let's not concern ourselves with this fact here and
let us instead assume that the parameter $\alpha$ (or equivalently $x$) in
the last section is at any generic value. Now we can ask ourselves if
anything interesting happens to $Y$ as we vary $y$ and $z$. In
particular, the most obvious question to ask is whether $Y$ is ever
singular. 

$Y$ is singular whenever $f=\partial f/\partial x_0=\ldots \partial
f/\partial x_4=0$ has a solution for (\ref{eq:Y}). With a little
algebra we find that this has a fairly simple solution for $y=1$. In
this case we have 12 singular points in $Y$ lying in the subspace
$x_0=x_1=x_2=0$. We know that varying $y$ has something to do with
varying the dilaton in the heterotic string so this suggests that
something curious happens in our model when the heterotic string
coupling is of order 1. While this sounds interesting it is a bit too
exotic for our purposes here! We would rather discover something
interesting which happens near weak coupling.

The next simplest solution one finds is when we have singular points
in the larger subspace $x_0=x_1=0$. This demands that
\begin{equation}
  (1-z)^2-yz^2=0.   \label{eq:A1d}
\end{equation}
If our heterotic string has zero coupling we set $y=0$ and so this
has a solution when $z=1$. One may show that
$z=1$ is exactly the value required to make the little curves $f$
in each K3 fibre of $X$ acquire zero size \cite{AGM:sd}. So this
must be exactly where we expect to see enhanced $\SU(2)$ gauge
symmetry. To summarize we expect to see an $\SU(2)$ gauge symmetry
whenever $y=0$ and $z=1$.

Now we may probe into nonzero coupling by letting $y$ acquire a small
nonzero value. The odd thing to note is that (\ref{eq:A1d}) then has
{\em two\/} solutions for $z$ near 1. Somehow our single $\SU(2)$
theory has split into two interesting things for nonzero heterotic dilaton.

At this point we could easily go off and explore the wonders of these
quantum corrections. This subject is generally called ``Seiberg--Witten''
theory \cite{SW:I,SW:II}. These lecture notes would be dwarfed by a
full treatment of this subject so instead we will refer
to \cite{Pes:SW}, for example, for a review.

Here we will just review some basic properties. In its basic form
Seiberg--Witten 
theory is not a theory which includes gravity. It is a very
interesting question as to how one can remove gravity from the
four-dimensional theory we have constructed. One might regard the
removal of gravity as a rather regressive thing to do --- after all it
was precisely because string theory contains gravity that string
theory became so popular in the first place. Nevertheless going to a
limit where gravity can be ignored provides a very useful way of
making contact between what is known about string theory and quantum
field theory. Indeed this process has often dominated work in string
theory in recent years.

\iffigs
\begin{figure}
\begin{center}
\setlength{\unitlength}{0.00041700in}%
\begin{picture}(10824,4824)(889,-4873)
\thinlines
\put(8701,-1861){\circle*{80}}
\put(10501,-1861){\circle*{80}}
\put(1201,-61){\line( 0,-1){4800}}
\put(901,-3990){\line( 1, 0){4200}}
\multiput(2101,-3886)(100,0){20}{\makebox(3.3333,23.3333){.}}
\put(7501,-3661){\framebox(4200,3600){}}
\put(7426,-2011){\vector(-2,-1){3600}}
\put(1651,-211){
}
\put(4951,-4261){\makebox(0,0)[lb]{\smash{$z$}}}
\put(3076,-4386){\makebox(0,0)[lb]{\smash{$1$}}}
\put(901,-1261){\makebox(0,0)[lb]{\smash{$y$}}}
\put(10501,-1636){\makebox(0,0)[lb]{\smash{$\Lambda^2$}}}
\put(8551,-1636){\makebox(0,0)[lb]{\smash{$-\Lambda^2$}}}
\put(7726,-3436){\makebox(0,0)[lb]{\smash{$u$}}}
\end{picture}
\end{center}
  \caption{The rigid limit of an $\SU(2)$ theory.}
  \label{fig:u}
\end{figure}
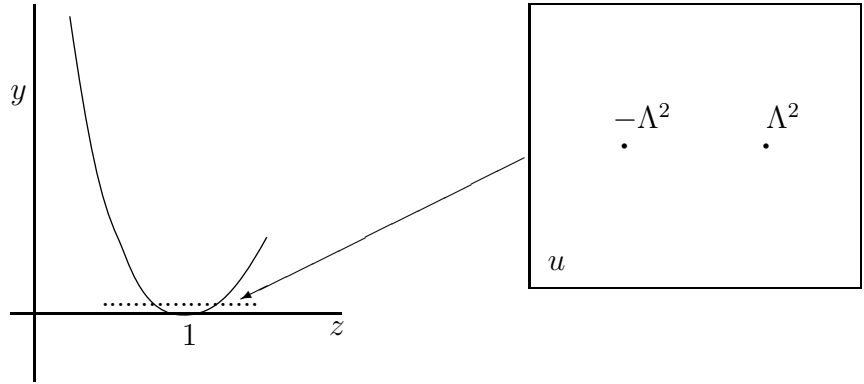
\fi

In order to switch off gravity we need to take the string coupling to
zero. As we discussed in the type IIA language this corresponds to
taking the area of the section $W$ to infinity. In type IIB language
we are taking $y\to0$. If we were to do this process alone then
everything would become rather trivial. Instead let us ``zoom in'' on
the splitting effect that we saw above. In particular let us rescale
$z-1$ as we take $y\to0$ so that we fix the location of the two
solutions of (\ref{eq:A1d}) at some fixed scale determined by a
constant traditionally called $\Lambda^2$. This leaves us with one
complex parameter $u$, where the $u=\pm\Lambda^2$ at the
discriminant. We show this in figure~\ref{fig:u}. This scale $\Lambda$ encodes
the effective coupling constant of the gravity-free Yang--Mills theory
which remains.  This process is explained in detail in
\cite{KKL:limit,KKV:geng}.

In a way this limit is one of the cleanest ways of viewing the
process of ``dynamical scale generation'' in quantum field theory. We
desire to zoom in on the part of the moduli space where gravity is
weakly coupled but the structure of the $\SU(2)$ gauge theory of
interest forces us to fix a scale. This is the same scale 
which appears when computing the running of a coupling constant in an
asymptotically free theory!

The two main statements of Seiberg--Witten theory for $\SU(2)$ are
\begin{enumerate}
\item The gauge symmetry $\SU(2)$ never appears. It is broken by
quantum effects (assuming $\Lambda$ is nonzero).
\item At $u=\pm\Lambda^2$ massless solitons appear. These are the
remnants of the ``W-bosons'' which appeared classically to enhance
the gauge symmetry.
\end{enumerate}

There is one aspect of this ``zooming in'' process which is of
great interest when discussing the geometry of $N=2$ theories. Namely,
the structure of special K\"ahler geometry changes. If one considers
the geometry of the moduli space with {\em no\/} gravity then
(\ref{eq:metric}) becomes \cite{ST:rigid,Gates:rigid}
\begin{equation}
  K = -\Img\left(\bar t^i\frac{\partial \cF}{\partial t^i}\right).
	\label{eq:Rmetric}
\end{equation}
where $t^i$ are {\em affine\/} coordinates.
This form of special K\"ahler geometry is often referred to as
``rigid'' special K\"ahler geometry while that of section \ref{ss:sK}
is called ``local'' special K\"ahler geometry.

The key point, as discussed in \cite{Craps:S,Frd:SK} for example, is
that while local special K\"ahler geometry is associated to the moduli
space of complex \CY\ threefolds, rigid special K\"ahler geometry is
associated to the moduli space of complex {\em curves}. Thus we should
expect the theory of $N=2$ supersymmetric field theories without
gravity to be associated to Riemann surfaces in much the same way that
these theories with gravity were associated to \CY\ threefolds.

This is pretty much exactly what Seiberg--Witten theory
\cite{SW:I,SW:II} does. An $\SU(2)$ gauge theory is associated with an
elliptic curve for example.

The exact way in which this curve appears in the limit of the \CY\
threefold as we decouple gravity is not at all clear. A fairly
systematic way of doing this construction was explained in
\cite{KKV:geng} in the case that $Y$ is constructed using toric
geometry. See also \cite{KLM:hSW} for an earlier analysis of this
problem and \cite{CKYZ:locM}, for example, for further discussion.

The geometry of the \CY\ threefold makes an explicit
appearance for $N=2$ theories with gravity --- it is the \CY\
threefold $Y$ on which the type IIB string is compactified. The manifest
geometry of the Riemann surface in the case of Seiberg--Witten theory
is a little more obscure. Possibly the best suggestion for a direct
picture in which this curve appears was given by Witten \cite{W:SWM}
in terms of M-theory and world-volume theories of D-branes. 


\subsubsection{Breaking T-Duality}  \label{sss:Tsnap}

Our discussion of the moduli space of the type IIA picture and the
heterotic picture for $\cM_V$ were in excellent agreement so long as
we ignored quantum effects. In both cases we had a ``Narain'' factor
in the form of the symmetric space given in (\ref{eq:Narf}). In the
language of the heterotic string this consisted of the moduli of the
2-torus $E_H$ together with the degrees of freedom of the Wilson lines
of the flat bundle $V_E$. The group $\GO(\Upsilon)$ gave the
T-dualities of the heterotic string on a torus.

\iffigs
\begin{figure}
  \centerline{\epsfxsize=9cm\epsfbox{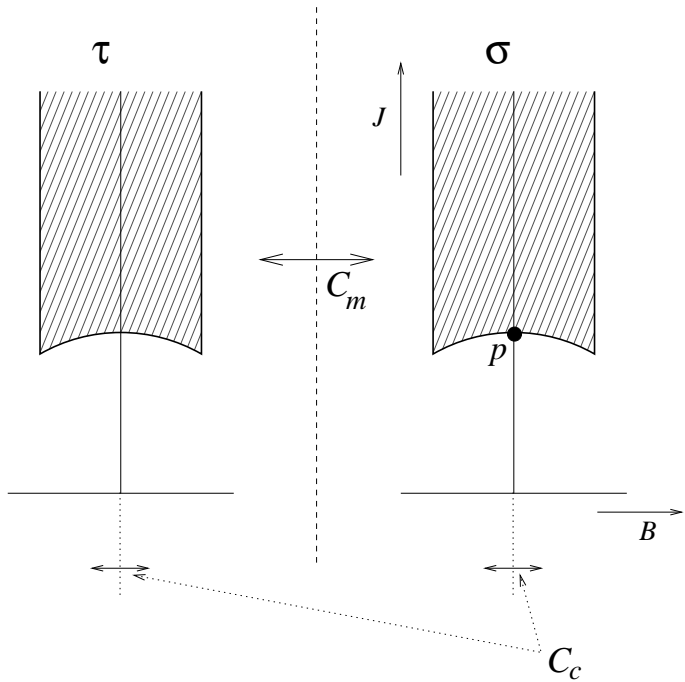}}
  \caption{Moduli Space of a Torus.}
  \label{fig:T}
\end{figure}
\fi

In particular if we consider the example of section \ref{sss:ex1} then
we have a Narain factor of the form
\begin{equation}
  \GO(\Gamma_{2,2})\backslash\GO(2,2)/(\GO(2)\times\GO(2))\cong
   (C_m\times C_c)\backslash\left(\frac{\mathsf{H}_\sigma}{\Sl(2,\Z)}
   \times \frac{\mathsf{H}_\tau}{\Sl(2,\Z)}\right).
\end{equation}
Here we have used the standard decomposition of the Grassmannian into a
form which makes it more recognizable for our purposes. We have two
copies of the upper half-plane $\mathsf{H}\cong\Sl(2,\R)/\GU(1)$ which
we parameterize by complex numbers $\sigma$ and $\tau$ respectively. The
groups $C_m$ and $C_c$ are both isomorphic to $\Z_2$ and are generated
by
\begin{equation}
\begin{split}
  g_m:(\tau,\sigma)&\mapsto(\sigma,\tau)\\
  g_c:(\tau,\sigma)&\mapsto(-\bar\tau,-\bar\sigma),
\end{split}
\end{equation}
respectively. We refer to \cite{Giv:rep}, for example, for details of
this isomorphism. We depict this moduli space in
figure~\ref{fig:T}.

The interpretation of this moduli space in terms of $E_H$ is
straight-forward. We let $\tau$ denote the complex structure in the
standard way and we let $\sigma$ denote the single component of
$B+iJ$.

Thus the $\Sl(2,\Z)$ action on $\tau$ is the standard modular
invariance of a 2-torus. The $\Sl(2,\Z)$ acting on $\sigma$ is
composed of the familiar $B\to B+1$ symmetry as well as a $J\to1/J$ T-duality.
Note that $C_m$ is ``mirror symmetry'' for a 2-torus as was first
seen in \cite{DVV:torus}. $C_c$ can be thought of as a complex
conjugation symmetry of the theory.

This is all very well but we have noticed in the previous section that
this picture of the moduli space is subject to quantum
corrections. That is, this Narain picture of the moduli space of $E_H$
is {\em not\/} exact. We will now argue that the effect of these
quantum corrections is to completely ruin the description of the
moduli space as a quotient and so any notion of T-duality for $E_H$ is lost.

\iffigs
\begin{figure}
  \centerline{\epsfxsize=8cm\epsfbox{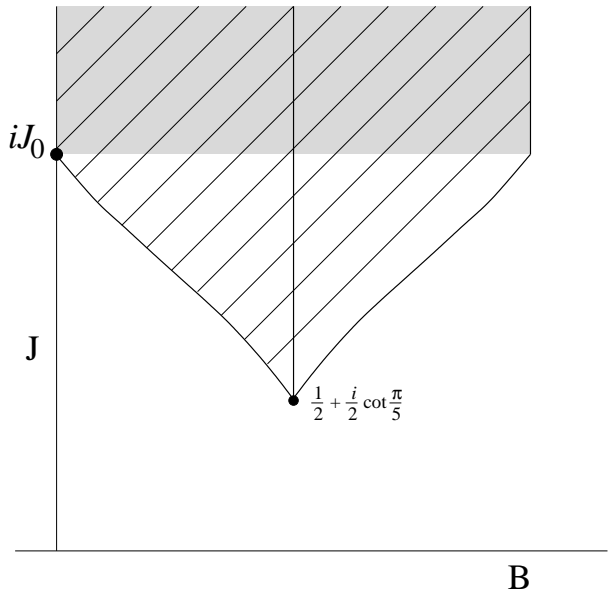}}
  \caption{Moduli Space of the Quintic.}
  \label{fig:Q}
\end{figure}
\fi

To argue this let us discuss what can go wrong with T-duality
arguments in another example. We will consider the classical quintic
hypersurface in $\P^4$ as was analyzed in \cite{CDGP:}. The \CY\
manifold has $h^{1,1}=1$. When we flatten out the single complex
coordinate describing $B+iJ$ we obtain the moduli space depicted in
figure~\ref{fig:Q}. Now although this moduli space looks similar to
the fundamental region of $\mathsf{H}/\Sl(2,\Z)$ there is a big
difference. There is no action of any discrete group on $\mathsf{H}$
for which the region in figure~\ref{fig:Q} is a fundamental
region. One may see this as follows. Note that there is an angle of
$2\pi/5$ formed at the lowest point in this region. One should
therefore need 5 fundamental regions touching at this point. Indeed
one may find such regions and they are pictured in figure 5.2 of
\cite{CDGP:}. One can also see that the $B\to B+1$ symmetry should
allow us to translate these fundamental regions one unit to the left
or right to form new fundamental regions. The problem is that doing
this translation gives a region which overlaps in an open set with one
the regions we built earlier by rotating by $2\pi/5$. Thus these
supposed fundamental regions do not {\em tessellate\/} in $\mathsf{H}$
and therefore cannot be derived in terms of a group action on
$\mathsf{H}$.

Indeed we argued in section \ref{s:gen} that $N=2$ theories in four
dimensions do not generically have locally symmetric moduli space. It was in the 
context of symmetric spaces that we saw the natural appearance of T-duality.
It should not therefore be a surprise that we do not find the
true analogue of a modular group for the quintic threefold.
 
We should therefore expect that the quintic threefold represents the
generic case of a \CY\ moduli space. In particular once we turn on
the heterotic string coupling, i.e., give finite size to the section
$W$ of the example in section \ref{sss:ex1} the Narain description of
the moduli space is lost. This is argued in \cite{AP:T}.

So if the heterotic string on a torus does {\em not\/} respect
T-duality how should we really describe the moduli space? The
principle should be the same as that for the quintic threefold. One
should begin with a weakly coupled heterotic string on a large circle
or torus. Here one unambiguously sees the geometry of the
compactification. Now move about the moduli space of
compactifications. In this we can label every point in the moduli
space by a set of moduli (such as radii) for the torus. One problem we
have to be careful about is that we may follow loops in the moduli
space which allow us to identify more than one torus with a given
point in the moduli space of theories. We {\em must\/} avoid this by
putting cuts in the moduli space. If we do not put in such cuts then
generically one would expect to be able to identify every possible
torus with each point in the moduli space! (Note that since the
classical $\Sl(d,\Z)$ symmetry of a $d$-torus is lost one must
describe the torus directly in terms of data which chooses a fundamental
region of the classical moduli space of flat metrics.)

Once we have completed this labelling process (the details of which
depend on a choice of cuts) we have defined every possible torus to
be considered. Tori excluded by the process, such a circle of radius
less that $\sqrt{\alpha'}$, do not exist and should not be
considered. It is only the accidental T-duality of the weakly-coupled
string that led us to believe that we could make real sense of small
tori.

\begin{difficult}
This example consisted of a moduli space $\cM_V$ which became locally a
symmetric space on its boundary at infinite distance corresponding to
some classical limit. It is interesting to note that there are other
known examples where a subspace of $\cM_V$ can be locally
symmetric. For example consider the so-called Z-orbifold $T^6/\Z_3$
with 27 fixed points. The rational curves in this space (after blowing
up) conspire to only give certain quantum corrections to $\cM_V$. The
effect of this is to make the prepotential $\cF$ exactly cubic if none
of the 27 blow-up modes are switched on \cite{Drk:Z}. The
result is that we get a slice of the moduli space (at finite distance)
of the form
\begin{equation}
  \cM_{V,\mathrm{orb}}
  \cong\GU(3,3;\Z)\backslash\GU(3,3)/(\GU(3)\times\GU(3)).
\end{equation}
Moving away from this subspace there are instanton corrections and the
symmetric space structure is lost.
\end{difficult}

Note that in general we lose the {\em classical\/} $\Sl(2,\Z)$
symmetry of the complex structure of the torus in addition to any
T-duality. How can this be?

The moduli space of a 2-torus of volume one is determined by 
considering the ways of making a 
a lattice of area one, dividing out by rotations, and then dividing out
by the modular group $\Sl(2,\Z)$. This gives us the familiar form
$\Sl(2,\Z)\backslash\Sl(2,\R)/\SO(2)$. If we declare that quantum
effects break this structure then quantum effects must be having a
drastic effect on this construction of the torus. As well as breaking the
$\Sl(2,\Z)$ invariance, we are also modifying the $\Sl(2,\R)/\SO(2)$
part. It is as if we are breaking the picture of the 2-torus as a
Riemannian manifold. Hopefully once stringy geometry is
better-understood it will be more clear what is happening here.

It is worth mentioning that there are two distinct types of
U-dualities discussed in the literature. One is an ``internal duality''
statement where one says that a string theory of type $\Scr{S}_1$ (e.g.,
type IIA, $E_8\times E_8$ heterotic etc.) compactified on $X_1$ with
coupling $\lambda_1$ is dual to a string theory of the same type
$\Scr{S}_1$ compactified on $X_2$ with coupling
$\lambda_2$. Alternatively one has an ``external duality'' where one
says that a string theory of type $\Scr{S}_1$ compactified on $X_1$ with
coupling $\lambda_1$ is dual to a string theory of a {\em different\/} type
$\Scr{S}_2$ compactified on $X_2$ with coupling
$\lambda_2$.

Our discussion of the breaking of T-dualities (and by implication
U-dualities) was in the context of internal dualities. In particular
we were fixing our string as an $E_8\times E_8$ heterotic string. What
happens when the external duality relating an $E_8\times E_8$ heterotic
string on a given torus and a given choice of Wilson lines to a
$\spnh$ heterotic string on another torus and set of Wilson lines?

Our discussion of mapping out the moduli space should apply again. Map
out the moduli space of tori and Wilson lines as above using the $E_8\times
E_8$ heterotic string. Now do the same thing with the $\spnh$
heterotic string. Note that the starting point for the large torus
will not be the same limit point in moduli space as the former
case. This means that every point in the moduli space will now
have two labels --- one for each heterotic string. One should not
obtain small radii for {\em either\/} heterotic string interpretation.

Thus strictly external U-dualities need not be broken by quantum
effects. The precise mapping between $(X_1,\lambda_1)$ and 
$(X_2,\lambda_2)$ can be expected to be modified however.


\section{The Moduli Space of Hypermultiplets} \label{s:hyp}

Now we come to the considerably more tricky subject of trying to map
out the moduli space of hypermultiplets for our $N=2$ theories in four
dimensions. In the case of the vector multiplet moduli space, the type
IIB string compactified on $Y$ gave an exact model. For the
hypermultiplets there is no exact model. This makes the subject much
more difficult and potentially much more interesting!


\subsection{Related Dimensions}   \label{ss:dim}

The purpose of these lectures is to discuss some special properties of
$N=2$ theories in four dimensions. It turns out to be very useful to
be aware of some other closely-related theories in both higher and
lower dimensions than four to help gain insight into the hypermultiplet
moduli space.

\subsubsection{$N=(1,0)$ in six dimensions}   \label{sss:d=6}

Imagine compactifying the heterotic string on a K3 surface $S_H$. This
would yield a theory with $N=(1,0)$ supersymmetry in six
dimensions. We refer the reader to \cite{SW:6d} for a good discussion
of many aspects of such theories.
This has an $R$-symmetry of $\Sp(1)$. We discussed the
supermultiplets of such theories in section \ref{ss:N2}. Such a theory
may then be compactified on a 2-torus to yield our familiar $N=2$
theory in four dimensions. Upon dimensional reduction, the $N=(1,0)$
supermultiplets in six dimensions become $N=2$ supermultiplets in four
dimensions as follows:
\begin{itemize}
\item A six-dimensional {\em hypermultiplet \/} becomes a
hypermultiplet in four dimensions. 
\item A six-dimensional {\em vector\/} multiplet becomes a vector
multiplet in four dimensions.
\item A six-dimensional {\em tensor\/} multiplet becomes a vector
multiplet in four dimensions.
\end{itemize}
In particular the hypermultiplet moduli space of a heterotic string
compactified on a K3 surface $S_H$ is exactly the same as the
hypermultiplet moduli space of a heterotic string compactified on
$S_H\times E_H$. This is consistent with our earlier comment that all
the hypermultiplet information comes from the K3 surface $S_H$.

It is therefore quite common to analyze the hypermultiplet moduli
space in terms of six-dimensional physics rather than four-dimensional
physics. Having said that, our duality statements might now sound a bit
peculiar. We want to say something to the effect that we can model the
hypermultiplet moduli space of a heterotic string on a K3 surface in
terms of a type IIA string on a \CY\ threefold $X$ but the former is 
six-dimensional while the latter is four-dimensional.

It is important to note that we cannot necessarily completely ignore
the 2-torus $E_H$ in the product $S_H\times E_H$. In effect we can
think of arriving at our six-dimensional theory by beginning in four
dimensions and decompactifying $E_H$. To do this we certainly need the
full moduli space of $E_H$ and from proposition~\ref{prop:F1} this in
turn implies that the \CY\ threefold $X$ is {\em an elliptic fibration
with a section}. Assuming this is the case, we may model the 
six-dimensional physics of the heterotic string on $S_H$ in terms of 
the type IIA string on $X$ by implicitly decompactifying $E_H$.

\iffigs
\begin{figure}[t]
  \centerline{\epsfxsize=12cm\epsfbox{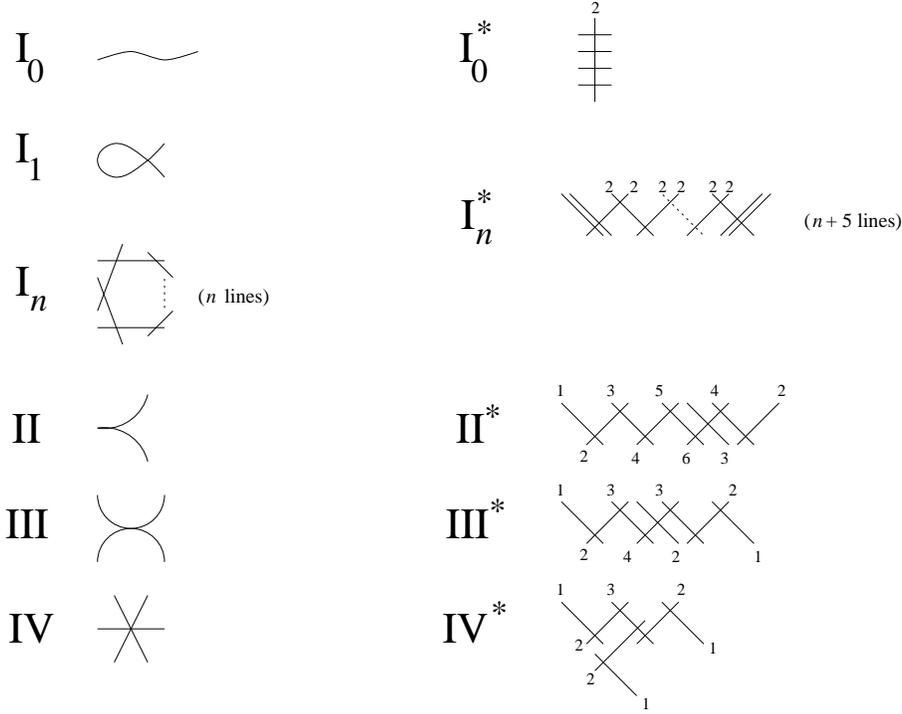}}
  \caption{Classification of elliptic fibres.}
  \label{fig:Kod}
\end{figure}
\fi

This mechanism of using type IIA strings on $X$ to model six-dimension
physics is known as ``F-theory''. The reader should be warned that
there are at least two other ways of defining F-theory common in the
literature. One is to treat F-theory as twelve dimensional (although
whether it lives in $\R^{2,10}$ or $\R^{1,11}$ is unclear). Another
way is to view it as a type IIB string compactification with a varying
dilaton. We refer to \cite{Vafa:F,MV:F} for more details. The type IIA
definition of F-theory is well-suited for our purposes of linking the
subject to four dimensions.

Let us denote the elliptic fibration as $p:X\to\Theta$, where $\Theta$
is a complex surface. We also know we have a K3-fibration $\pi:X\to
W$, where $W\cong\P^1$, and a fibration $\Theta\to W$ with generic
fibre given by $\P^1$. That is, $\Theta$ is a ``ruled surface''. If
$\Theta$ is a smooth $\P^1$-bundle over $W$, it is the ``Hirzebruch
surface'' $\HS n$. Here the section $W\hookrightarrow\Theta$ has
self-intersection $-n$ within $\Theta$.
Blowing up $\HS n$ at a few points replaces some of
the smooth $\P^1$-fibres by chains of $\P^1$'s. 

It is common to then draw $X$ (representing a complex dimension as a
real dimension) in the following form. We may use the
plane of the paper to represent $\Theta$ by letting the horizontal
direction represent the section and the vertical direction represent
the $\P^1$-fibre. That is, the ``ruling'' of the ruled surface
$\Theta$ is given by vertical lines. Now over a (complex) codimension one
subspace of $\Theta$ the elliptic fibration $p:X\to\Theta$ will
degenerate. We may draw this ``discriminant'' locus as a set of curves
and lines in the plane of the paper.

Kodaira has classified the possibilities for how an elliptic fibre may
degenerate in the case of one parameter family \cite{Kod:ell}. We show
the possibilities in figure~\ref{fig:Kod}. With the exception of
$\mathrm{I}_0$ which is the smooth elliptic case, and $\mathrm{II}$ 
which is an elliptic curve with a cusp, each line in the figure represents
a rational curve. This curve may appear with a multiplicity given by
the small numbers in the figure. This classification can
be used to label the generic points on the irreducible
components of the discriminant locus.

\iffigs
\begin{figure}[t]
  \centerline{\epsfxsize=11cm\epsfbox{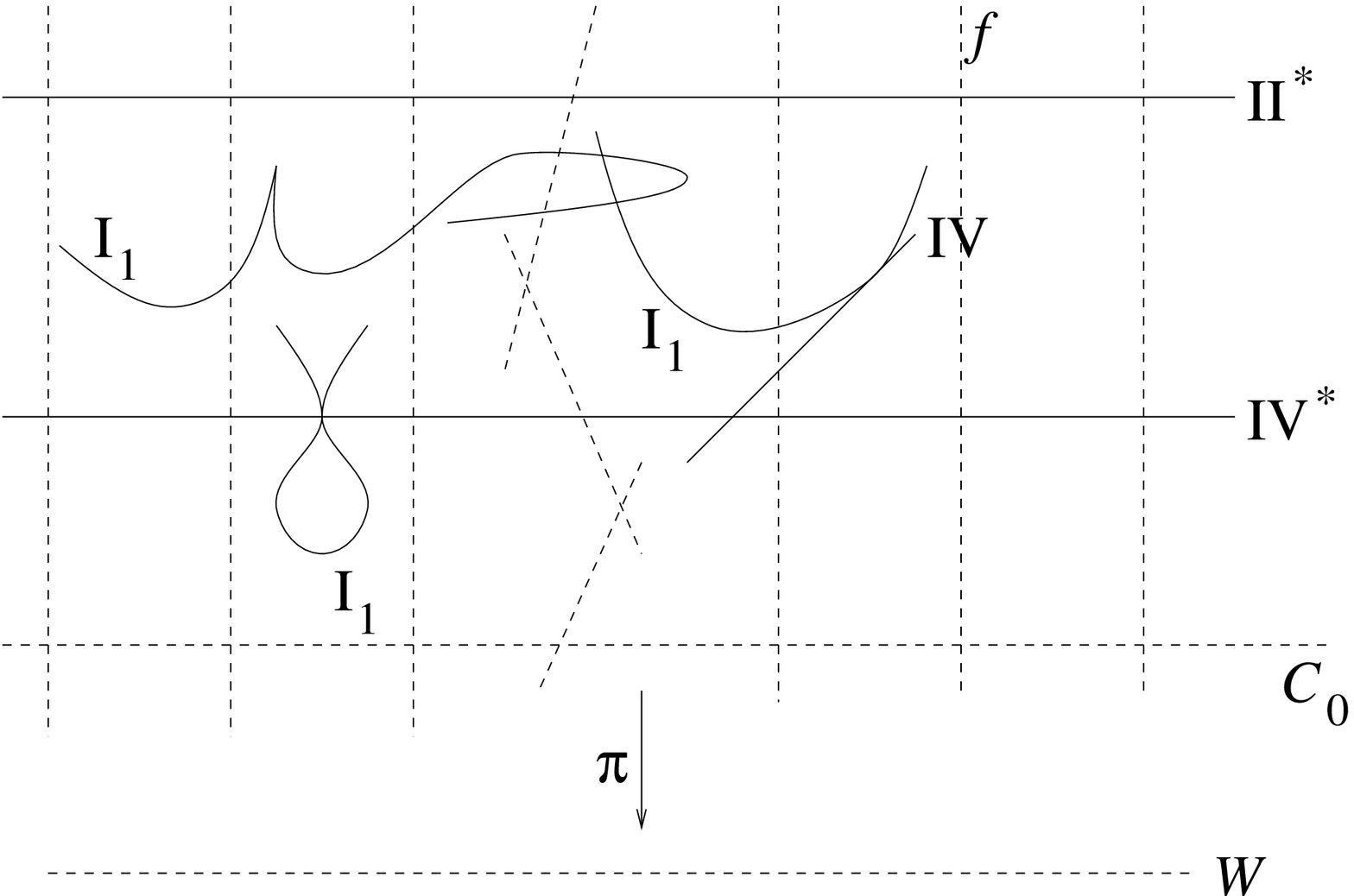}}
  \caption{A typical elliptic fibration.}
  \label{fig:eleg}
\end{figure}
\fi

The result is that one obtains a picture somewhat typically like
figure~\ref{fig:eleg} for $X$ in the form of an elliptic fibration.
In this figure the dotted lines represent lines ($\P^1$'s) within
$\Theta$. $C_0$ is a section and $f$ is a generic $\P^1$ fibre. Note 
that this notation is consistent with the $f$ which appeared in 
section \ref{sss:ex1}. At one
point over $W$ we have put a fibre as a chain of three $\P^1$'s. The
solid lines represent the discriminant locus. Each irreducible
component is labelled by its Kodaira type. When these components
collide, the elliptic fibration will degenerate further and the
resulting fibre need not lie in Kodaira's classification.

Since we wish to study the moduli space of hypermultiplets we are
particularly interested in the deformations of complex structure of
$X$. When we draw $X$ as an elliptic fibration, the complex structure
is encoded in the discriminant locus. Thus, deformations of $X$ are
given simply by the deformations of the discriminant locus.

It will also be worthwhile to note how the K\"ahler form data appears
in the elliptic fibration. Deforming the K\"ahler form may either
affect areas in the fibre direction (i.e., the area of the generic
fibre as well as areas within the chains of special Kodaira fibres) or
affect areas within $\Theta$. As we decompactify $E_H$ to go from 4
dimensions to 6 dimensions one can show that the areas in the fibre
direction become meaningless \cite{MV:F2,me:lK3}. Our discussion
of the types of supermultiplets in four and six dimensions given at the
start of this section leads one to conclude:
\begin{itemize}
\item Using the K\"ahler form to vary areas in the fibre direction
corresponds to moduli in a six-dimensional {\em vector\/}
supermultiplet.
\item Using the K\"ahler form to vary areas in $\Theta$
corresponds to moduli in a six-dimensional {\em tensor\/}
supermultiplet.
\end{itemize}


\subsubsection{$N=4$ in three dimensions}   \label{sss:d=3}

Imagine taking our $N=2$ theory in four dimension and compactifying
further on a circle. This leads to a theory in three dimensions with
$N=4$ supersymmetry. This has an $R$-symmetry of
$\SO(4)\cong\Sp(1)\times\Sp(1)$ (up to irrelevant discrete factors)
which implies that the moduli space should factorize into a product of
two quaternionic K\"ahler spaces. These three dimensional theories
have two different types of ``hypermultiplets'' whose moduli spaces
cannot mix. In the literature one often refers to one of these types of
hypermultiplets as ``vector multiplets'' to reflect their
four-dimensional origin. However, one should be aware that, 
within the context of the three-dimensional physics, such a 
distinction is arbitrary.

Note that the hypermultiplet moduli space $\cM_H$ from four
dimensions comes through unscathed into the three dimensional picture
whereas our vector multiplet moduli space becomes ``quaternionified''
in the compactification. It is remarkable how resilient $\cM_H$ is! It
is unchanged as we compactify on circles a theory in six dimensions
with $N=(1,0)$ supersymmetry down to three dimensions. Compare this
with the capricious vector multiplet moduli space which is
non-existent in 6 dimensions, real in 5 dimensions, complex in 4
dimensions and quaternionic in 3 dimensions!

Because the vector multiplet moduli space becomes a hypermultiplet
moduli space upon compactification to three dimensions, this picture
provides a potentially useful way of using our knowledge of special
K\"ahler manifolds to uncover some of the mysteries of quaternionic
K\"ahler manifolds.

Suppose we wish to study $\cM_H(X)$ for a type IIA string compactified on
the \CY\ threefold $X$. Consider instead the moduli space $\cM_V(Y)$
of the type IIA string compactified on $Y$, the mirror of
$X$. Compactifying further on $S^1_R$, a circle of radius $R$, the special
complex K\"ahler space $\cM_V(Y)$ becomes a quaternionic K\"ahler
space which we will denote $\cM_V(Y)_\H$. Since the type IIA string on
a circle of radius $R$ is supposedly T-dual to the type IIB string on
a circle of radius $1/R$, the type IIA string on $Y\times S^1_{R}$
should be dual to the type IIB string on $Y\times S^1_{1/R}$. Using
mirror symmetry this is then dual to the type IIA string on $X\times
S^1_{1/R}$. The space $\cM_V(Y)_\H$ must now represent the factor of
the moduli space containing deformations of complex structures of
$X$. That is, it descended from $\cM_H(X)$ upon compactification on
the circle. Having said that, $\cM_H(X)$ is unchanged by this circle
compactification and so
\begin{equation}
  \cM_H(X) \cong \cM_V(Y)_\H.
\end{equation}

\begin{difficult}
We already questioned the validity of T-duality for the heterotic
string in section \ref{sss:Tsnap}. It is natural to question whether
T-duality is valid for the type II strings when we have only modestly
extended supersymmetry. The crude statement that the type IIA string
compactified on $Y\times S^1_{R}$ is dual to the type IIB string
compactified on $Y\times S^1_{1/R}$ is almost certainly incorrect. It
is true however that one should expect this to be exact when the
strings are very weakly coupled. Most analyses of strings using this
statement of T-duality such as \cite{SS:3dU} do use only
weakly-coupled strings. We will not try to elucidate the exact meaning
of T-duality in type II strings in these lectures.
\end{difficult}

Determining $\cM_V(Y)_\H$ from the complex space $\cM_V(Y)$ is not easy.
An interesting attempt at this problem was made some time ago by
Cecotti et al.\ 
in \cite{CFG:II}.\footnote{see also \cite{FS:quat} for further
analysis along these lines.} This paper assumed that the moduli space
$\cM_V$ was 
determined by a prepotential that was exactly cubic. 
Particular attention was paid to the cases where $\cM_V$ is a
symmetric space. If one then ignored
quantum corrections upon compactification on a circle, this
symmetric space was mapped via the so-called ``$c$-map'' to
another symmetric space. For example one might have something like%
\footnote{The map $c$ is not intended to be viewed as a map of
topological spaces! We are replacing one space by another.} 
\begin{equation}
  c:\frac{\SU(3,3)}{\mathrm{S}(\GU(3)\times\GU(3))}\to
                \frac{E_{6(+2)}}{\SU(2)\times\SU(6)}.
\end{equation}

A notable case of the $c$-map is
\begin{equation}
  c:\frac{\Sl(2,\R)}{\GU(1)}\times\frac{\SO_0(2,n-2)}{\SO(2)\times\SO(n-2)}
 \to \frac{\SO_0(4,n)}{\SO(4)\times\SO(n)}.
\end{equation}
We will revisit this briefly in section \ref{sss:univ}.

Of course, unless we pick a very special model to examine\footnote{See
\cite{FHSV:N=2} for such an example.}, there {\em
will\/} be quantum corrections and the analysis of \cite{CFG:II} will
not be directly applicable. However, this method may provide a good starting
point for the analysis of the quaternionic K\"ahler moduli spaces
as it does give the asymptotic behaviour where quantum effects
can be neglected.

An exact version of the $c$-map was elucidated by Seiberg and Witten
\cite{SW:3d} in the case of rigid special K\"ahler geometry. As
discussed in section \ref{sss:SW}, $\cM_V(Y)$ is described in this
limit by the deformation of a complex {\em curve}
$C_{\mathrm{SW}}$. Seiberg and Witten's remarkably simple result is
then
\begin{prop}
In the case that $\cM_Y(V)$ is a rigid special K\"ahler space,
$\cM_V(Y)_\H$ is simply the hyperk\"ahler space given by an
abelian (i.e., complex algebraic torus) fibration over $\cM_V(Y)$
where the fibre is given by the Jacobian
$H^1(C_{\mathrm{SW}},\GU(1))$. In addition the volume of the fibre is
determined by $R$, the radius of the circle on which one compactifies.
	\label{prop:SW3}
\end{prop}


\subsection{Extremal Transitions}   \label{ss:ex}

Since direct analysis of the hypermultiplet moduli space is so
formidable the most prudent course of action is to try to squeeze as
much information out of our knowledge of the vector multiplet moduli
space as we possible can.

This is facilitated by the occurrence of phase transitions or
``extremal transitions''. We go to a funny point in moduli space where
vector moduli disappear and new hypermultiplet moduli appear. We may
then pretend that we actually did this process in reverse and claim
that we know something about what happens when we move around in the
moduli space of hypermultiplets!

\subsubsection{Conifolds}   \label{sss:coni}

Let us consider the simplest type of extremal transition first --- the
``conifold'' of \cite{CGH:con}. We may understand this both from the
point of view of geometry and from the point of view of field theory
as explained in \cite{Str:con,GMS:con}.

We begin with the geometrical picture. Consider the type IIB string
compactified on the \CY\ manifold $Y$. We move about the moduli space
of vector multiplets by deforming the complex structure of $Y$. Let us
consider a one-dimensional family of such $Y$'s and denote an element
of this family by $Y_t$ where $t$ parameterizes the family. At a
special point in this part of the moduli space, say $t=0$, $Y$ may
become singular. The simplest thing that can happen as $t\to0$ is that an
$S^3$ can contract to a point. Locally such a singularity would look
like the hypersurface
\begin{equation}
  w^2+x^2+y^2+z^2=0,
\end{equation}
in $\C^4$. This is called a ``conifold singularity''. 

Locally such a conifold point can be resolved by replacing the point
by a $\P^1$ (see, for example, \cite{CGH:con} for a nice explanation of
this). Since the K\"ahler form controls the areas of $\P^1$'s such a
resolution might be pictured as a deformation of K\"ahler form. In
other words we have turned a degree of freedom from a deformation of
complex structure into a deformation of K\"ahler form.

Globally this picture does not work quite this simply. We need to
consider the case of $P$ disjoint $S^3$'s, each shrinking to a point at
$t=0$. If $Y_t$ represents the smooth $Y$ for a generic value of $t$
then a simple application of the Mayer-Vietoris sequence gives a
relationship between the homology of $Y_t$ and the homology of $Y_0$.
See \cite{Clem:} for a full description of this process.

Now resolve the resulting $P$ conifold points by adding $\P^1$'s
and call the resulting smooth manifold $Y'$. Another application of
the Mayer-Vietoris sequence gives a relationship between the homology
of $Y_0$ and the homology of $Y'$. Combining these results we obtain
\begin{equation}
\begin{split}
0\to H_4(Y_t)\to H_4(Y')\labto{f_1} &\,\Z^P\to H_3(Y_t)\to H_3(Y_0)\to 0\\
0\to H_3(Y')\to H_3(Y_0)\to &\,\Z^P\labto{f_2} H_2(Y')\to H_2(Y_t)\to 0.
\end{split}
\end{equation}

Let us denote by $Q$ the rank of the map labelled $f_1$. By Poincar\'e
duality the rank of $f_2$ must also then be $Q$. Note that $Q$
represents the dimension of the kernel of the map $\Z^P\to H_3(Y_t)$,
i.e., the number of homology relations between the $P$ 3-spheres in
the smooth $Y$. The above exact sequences give
\begin{equation}
\begin{split}
  b_2(Y') &= b_2(Y_t) + Q\\
  b_3(Y') &= b_3(Y_t) - 2(P-Q).
\end{split}
\end{equation}
That is, as we go through the conifold transition, we lose $P-Q$
vector multiplets and gain $Q$ hypermultiplets. Note that $P>1$ is
required for this transition to make sense and so a single conifold point
is not sufficient.

From the point of view of field theory this process is a
supersymmetric variant of the Higgs mechanism. As we wander about the
moduli space of vector multiplets it is possible that some
hypermultiplets suddenly become massless. Indeed, Strominger
\cite{Str:con} noted that the singularities in the moduli space
metric associated to a conifold are exactly the same as seen by Seiberg
and Witten when a hypermultiplet becomes massless.

Suppose $P$ hypermultiplets become massless and that these
hypermultiplets are charged under $P-Q$ of the $\GU(1)$ gauge
symmetries in our original theory. We may try to give these new
hypermultiplets vacuum expectation values which would then
spontaneously break this $\GU(1)^{P-Q}$ gauge symmetry. Our $N=2$
gauge theory in four dimensions has the standard gauge theory
couplings and so these broken gauge symmetries must ``eat up'' some
Goldstone bosons in order to become massive. What's more they must do
this in a way consistent with $N=2$ supersymmetry. The only way this
can happen is for us to lose $P-Q$ of our $P$ new massless
hypermultiplets leaving us with $Q$ new hypermultiplets. This is the
field theory picture for losing $P-Q$ vector multiplets and gaining
$Q$ hypermultiplets.

Since we obtain $Y'$ via the Higgs mechanism, this is often referred to
as the ``Higgs phase''.
Since $Y_t$ has more $\GU(1)$'s (massless photons) it is referred
to as the ``Coulomb phase''. That is, the Higgs phase is the one with
more hypermultiplets and the Coulomb phase is the one with more vector
multiplets. 

The conifold transition is just the simplest example of all kinds of
extremal transitions which may occur.

\subsubsection{Enhanced gauge symmetry}  \label{sss:YMh}

The Higgs phase transition of the preceding section was a little
boring because there was no nonabelian gauge symmetry at the phase
transition point. We know how to get enhanced gauge symmetry (at least
in some limit) from section \ref{sss:YM}. We need to consider the type
IIA string compactified on $X$, where $X$ has a curve of ADE
singularities. 

This is easy to arrange using the elliptic fibration language of
section \ref{sss:d=6}. We can describe the situation using the
``Weierstrass form'' of the elliptic fibration which is standard when
discussing F-theory. Let $s$ and $t$ be affine complex coordinates on
some patch of the base $\Theta$. We may then write the elliptic
fibration as
\begin{equation}
  y^2 = x^3 + a(s,t)x + b(s,t).
\end{equation}
The discriminant is then given by $\Delta=4a^3+27b^2$.
The geometry of such fibrations was discussed in detail in
\cite{me:lK3} and so we will be brief here.

\def\pII{\vphantom{\rm II}}
\begin{table}
$$\begin{array}{|c|c|c|c|c|}
\hline
L&K&N&\mbox{Fibre}&\mbox{Sing.}\\
\hline
\geq0&\geq0&0&{\rm I}_0&\\
0&0&>0&{\rm I}_N&A_{N-1}\\
\geq 1&1&2&{\rm II}&\\
1&\geq2&3&{\rm III}&A_1\\
\geq 2&2&4&{\rm IV}&A_2\\
\geq 2&\geq 3&6&{\rm I\pII}^*_0&D_4\\
2&3&\geq 7&{\rm I\pII}^*_{N-6}&D_{N-2}\\
\geq 3&4&8&{\rm IV}^*&E_6\\
3&\geq 5&9&{\rm III}^*&E_7\\
\geq 4&5&10&{\rm II}^*&E_8\\
\hline
\end{array}$$
\caption{Weierstrass classification of fibres.}
\label{tab:Mir}
\end{table}

Let us assume $a$ and $b$ are independent of $t$ for the time
being. We wish to put a line of interesting fibres along
$s=0$. Table~\ref{tab:Mir} lists the resulting fibres where $a\cong
s^L$, $b\cong s^K$ and $\Delta\cong s^N$ near $s=0$. The final column
denotes the resulting singularity {\em if all the components of the
fibre not intersecting the section are shrunk down to zero area}. Note
that the fibres $\mathrm{I}_0$, 
$\mathrm{I}_1$ and II only have one component and thus
cannot produce a singularity.

This results in an explicit description of an extremal transition
involving nonabelian gauge symmetry. Begin with a type IIA string on a
smooth \CY\ threefold $X$ where all the components of all the fibres have
nonzero area. Now shrink down all the components of the fibres which
do not hit the section. This will result in curves of ADE
singularities producing some gauge group $\cG$. We may then be free to
deform the discriminant by a deformation of complex structure to
smooth the threefold.

Let us recast this transition in terms of the language of a heterotic
string compactified on $S_H\times E_H$. The process begins by a
deformation of the K\"ahler form of $X$ which is thus a deformation of the
bundle over $E_H$ (or $E_H$ itself). That is, we vary Wilson
lines over $E_H$. We then obtain the gauge group $\cG$ by switching
these lines ``off''. The deformation of complex structure of $X$ then
corresponds to deforming the bundle over the K3 surface $S_H$ to reabsorb the
enhanced gauge symmetry $\cG$ into a bundle.

This extremal transition therefore appears as reducing the structure
group of the bundle $V_E\to E_H$ and increasing the structure group of
$V_S\to S_H$.

We begin in the ``Coulomb'' branch where $\cG$ is broken to its Cartan
subgroup $\GU(1)^{\rank(\cG)}$. We end up in the ``Higgs'' branch
where $\cG$ may be completely broken. This process therefore decreases
the number of vector multiplets as one would expect.

An interesting point to bear in mind is that the gauge group $\cG$ can
be broken by quantum effects, i.e., effects due to
$\lambda$-corrections in the heterotic string and
$\alpha'$-corrections (specifically worldsheet instantons wrapped
around the base $W$) in the type IIA string. Even though $\cG$ is
broken however it does not mean that the phase transition cannot
happen. Quantum effects cannot obstruct motion in the moduli space and
these extremal transitions most certainly exist in terms of \CY\
threefolds.

What tends to happen, as explained in \cite{SW:II}, is that the phase
transition point does not happen at a point of enhanced gauge symmetry
(which need not exist) but rather at a point where some
solitons become massless. Only if quantum effects are ignored would
these solitons actually produce the enhanced gauge symmetry.

In a particularly interesting class of examples the extremal transition
can become more complicated. One may have more than one Higgs phase
joining on to the Coulomb branch. This is actually understood both in
terms of field theory and in terms of the geometry of \CY\
threefolds. An example of a field theory with two Higgs branches was
discussed in \cite{SW:II}. The geometry was explained in
\cite{MS:five} based on an earlier observation by Gross
\cite{Gross:obs}.

As mentioned above, when we go to the six-dimensional picture of this
field theory, the degrees of freedom associated to the areas of the
elliptic fibration $p:X\to\Theta$ become frozen. That is, the vector
supermultiplets associated to the above gauge groups lose their
moduli. Because of this we lose the Coulomb branch of the theory. In
other words there are special points in $\cM_H$ where we may acquire
enhanced gauge symmetry but there is never any phase transition
associated with such events.

\subsubsection{Massless Tensors}   \label{sss:tens}

Having said that we lose the standard Higgs-Coulomb phase transitions
associated to enhanced gauge symmetry when we look at six dimensional
$N=(1,0)$ theories, one may ask if we have any transitions at all. There are
indeed still interesting phase transitions in six dimensions as was
explained in \cite{SW:6d}.

Going to the six dimensional decompactification limit of the four
dimensional theories may freeze out the K\"ahler form degrees of
freedom associated to the fibres of $p:X\to\Theta$, but there are
still K\"ahler degrees of freedom remaining within $\Theta$ itself.

Since these degrees of freedom are present as moduli in six dimensions
and descend to vector multiplet moduli in four dimensions, they must
be associated to scalars living in the six-dimensional {\em tensor\/}
multiplets \cite{MV:F2}.

Note that the scalar fields in tensor multiplets have only one real
degree of freedom. There is no modulus associated to varying the
$B$-field on $\Theta$. Effectively the periodicity of $B$ tends to
zero as we decompactify the four dimensional theory to six
dimensions. The geometry of the tensor moduli space is given by the
real special K\"ahler geometry of section \ref{sss:BC}.

The six-dimensional phase transitions are then between a phase
spanned by hypermultiplets, which we still call the Higgs phase, and a
phase spanned by tensor multiplets, which is called the Coulomb phase
for consistency with the four-dimensional picture.

In terms of F-theory on a \CY\ threefold $X$ this phase transition is
really nothing more than the conifold transition we discussed in
section \ref{sss:coni}. We will give an example here to explicitly
give the geometry of the elliptic fibration. For more details on the
geometry we refer to \cite{me:lK3}.

Consider an elliptic fibration whose local Weierstrass form is
\begin{equation}
  y^2 = x^3 + s^4x + s^5t.   \label{eq:Xi1}
\end{equation}
This has a type $\mathrm{II}^*$ fibre running along $s=0$ and so one
would associate this to an $E_8$ gauge group. At $t=0$ something
special happens. The elliptic fibration degenerates so badly that the
only fibre that would smooth the space out would actually be complex
dimension two rather than some algebraic curve. To avoid this one may
blow up the point $s=t=0$ in the base to introduce a new rational
curve into $\Theta$. One is certainly not always free to do this! Blowing
up any old point in $\Theta$ would usually result in breaking the \CY\
condition. It is only because (\ref{eq:Xi1}) is so singular that one
can do this.

The form (\ref{eq:Xi1}) is therefore precisely at the phase transition
point. We may go off into the Higgs phase by deforming the equation,
and thus the complex structure of $X$. We may go off into the Coulomb
phase by blowing up the base $\Theta$ at $s=t=0$.


\subsection{The classical limit}   \label{ss:Hclas}

The preceding section on extremal transitions gives us invaluable
information about specific points in $\cM_H$ --- those which allow
phase transitions into new dimensions in $\cM_V$. We now explore the
other part of $\cM_H$ which is accessible. We will look at the
boundary where all quantum effects may be ignored.

We have asserted that the heterotic string
compactified on $(V_S\to S_H) \times (V_E\to E_H)$ is dual to the
type IIA string compactified on a \CY\ threefold $X$. If we could go
to a limit in the moduli space where the $\alpha'$-corrections to the
heterotic string and the $\lambda$-corrections to the type IIA string
were simultaneously switched off then we should be able to map
the two respective moduli spaces of hypermultiplets {\em exactly\/} onto 
each other.  

In order to completely ignore $E_H$ and its bundle we will assert that
we are in the F-theory situation where $X$ is a K3 fibration and an
elliptic fibration with a section. We will also demand that $S_H$ is
itself an elliptic surface with a section. This latter demand kills
many moduli and one might ask whether one really needed to impose such
a drastic constraint. As we will see, it appears to be necessary to
get a simple description of the classical moduli spaces.

In proposition \ref{prop:Hbase} of section \ref{sss:HIIA} we showed
that the dilaton of the 
heterotic string is mapped to the area of the $\P^1$ base of $X$ as a
K3-fibration. While we tried to be quite rigorous in showing
proposition \ref{prop:Hbase}, there is a quicker (but dirtier) way
showing the same thing. Suppose that $X$ were not a K3-fibration over
$\P^1$ but simply a product of a K3 surface times an elliptic
curve. This would yield an $N=4$ theory in four dimensions. It is also
dual to a heterotic string on $T^6$. One may then use a simple
dimensional reduction argument \cite{Duff:S,AM:Ud,Duff:tri} to show
that the coupling of the heterotic string is given by the area of the
elliptic curve on which the type IIA string was compactified. The same
argument shows that the coupling of the type IIA string is given by
the area of one of the $T^2$'s in the heterotic 6-torus.

If we assume that $T^2\times Q$ (for any space $Q$) is equivalent to a
$Q$-fibration over $\P^1$ as far as areas are concerned then this
simple dimensional reduction argument reproduces proposition
\ref{prop:Hbase}. It also implies that {\em the coupling of the type
IIA string is determined by the area of the section of the K3 surface
$S_H$, as an elliptic fibration, on which the heterotic string is 
compactified}.   

We will assume this statement is true even though this argument
considerably lacks rigour. See \cite{AP:hetcor} for a more thorough
treatment of this question.

In order to make the type IIA string very weakly coupled we are
therefore required to make the section of $S_H$ very large on the
heterotic side. This will eliminate $\lambda$-corrections on the type
IIA side. Now in order to remove the $\alpha'$-corrections on the
heterotic side we are required to make the K3 surface $S_H$ very
large. Since we have made the section of $S_H$ large we have already 
fulfilled this requirement partially.

If we assume that $S_H$ is a completely generic elliptic K3 surface
with a section, then the only other area we need care about is that of
the generic elliptic fibre. If both the section and the fibre have
large area then every minimal 2-cycle in $S_H$ will be large, unless we
have chosen to be close to a special point in the moduli space of Ricci-flat
metrics where a 2-cycle shrinks down to zero size.

How exactly we take the area of the generic fibre of $S_H$ to be
infinite was first explained in \cite{FMW:F} following an observation
in \cite{MV:F}. It was then explored more fully in \cite{AM:po,me:hyp}. We
refer the reader to \cite{AM:po,me:hyp} for details of the following
argument. We will approach this problem as an algebraic geometer
would. For a discussion of the link of this approach with a more
metric-minded picture see \cite{me:MvF}.

\iffigs
\begin{figure}[t]
  \centerline{\epsfxsize=14cm\epsfbox{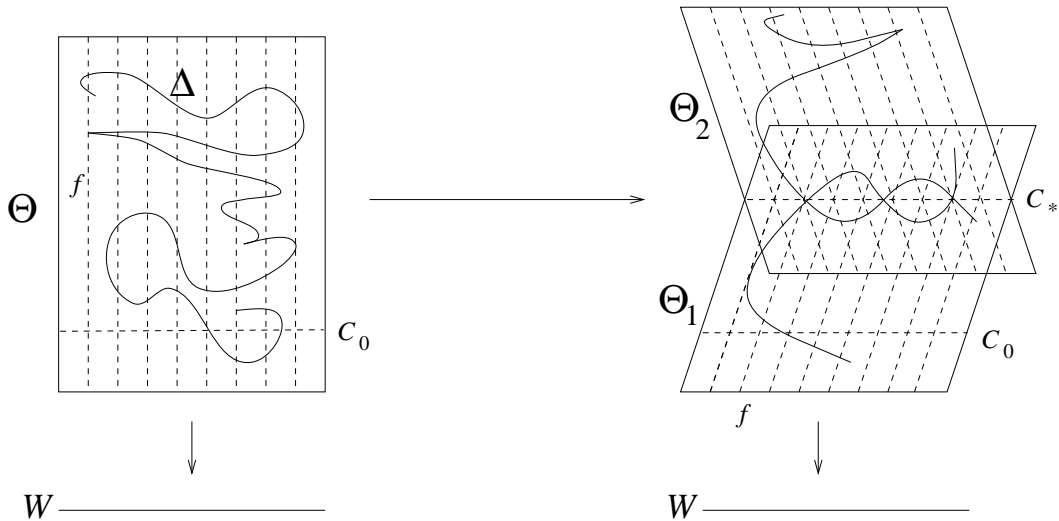}}
  \caption{The $E_8\times E_8$ stable degeneration.}
  \label{fig:sdeg}
\end{figure}
\fi

The basic idea is that taking the areas of the fibres of $S_H$ to be
large corresponds to a deformation of complex structure of $X$. There
is therefore some limiting complex structure of $X$ which represents
$S_H$ at infinite size. We may construct this by considering a
one-dimensional family of $X$'s. Let $w:\cX\to D$ be a fibration of some
4-dimensional complex manifold $\cX$ over some complex disc $D$. Let
$u$ be a complex parameter for $D$. If $u\neq0$ then the the fibre
$w^{-1}(u)$ will be a smooth \CY\ threefold in the class $X$. When
$u=0$, our fibre $X_0$ will be singular. It is $X_0$ which will
correspond to $S_H$ with a generic elliptic fibre of infinite area.

\subsubsection{The $E_8\times E_8$ heterotic string} \label{sss:E8l}

In order to proceed further we need to specify whether we are talking
about the $E_8\times E_8$ heterotic string or the $\spnh$ heterotic
string. We will deal with the $E_8\times E_8$ case first.

A picture of what happens as $X$ turns into $X_0$ is depicted in figure
\ref{fig:sdeg} for the case of the $E_8\times E_8$ heterotic string.
What happens is that the \CY\ threefold $X$ ``breaks in two'' to give
a reducible space $X_1\cup X_2$ intersecting along a complex surface
$S_*$. This surface is an elliptic fibration over a $\P^1$ which we
denote $C_*$ in figure~\ref{fig:sdeg}. The surface $S_*$ is in fact a
K3 surface and is isomorphic to $S_H$!

The way one shows this is via an adiabatic argument where one thinks
of $S_H$ as a slowly-varying elliptic fibration. One then focuses
attention on one elliptic fibre and pretends that the heterotic string
compactified on this single fibre is dual to F-theory on a K3
surface. Such an adiabatic argument might be considered a little
dangerous when trying to obtain exact results. The fact that we indeed
recover a K3 surface $S_*$ in the stable degeneration shows that the
result is in fact exact. The only way of mapping the moduli space of
$S_H$ onto the moduli space of $S_*$ is to identify them with each
other!

Having determined the heterotic K3 surface $S_H$ from the degeneration
$X\to X_1\cup X_2$, we should now like to determine the bundle data
$V_S$. This may be done by a very direct but rather technical
process. Whereas $X\to W$ was a K3-fibration, each of $X_1\to W$ and
$X_2\to W$ is a fibration with fibre given by a ``rational elliptic
surface'' (sometimes called an ``$E_9$ Del Pezzo Surface''). Each
rational elliptic surface is itself an elliptic fibration over a
$\P^1$ (the vertical dotted lines in figure~\ref{fig:sdeg}).

We now need to introduce the notion of the ``Mordell--Weil'' group
$\Phi$ of an elliptic fibration. If we have an elliptic fibration with
a given section $\sigma_0$ we may associate $\sigma_0$ with the
identity element of $\Phi$. Any further sections give further elements
of $\Phi$. $\Phi$ has a group structure given by the obvious
$S^1\times S^1$ structure of the elliptic curve.

Nontrivial elements of the Mordell--Weil group of the rational elliptic
surfaces intersect $S_*$ at points. The locus of all these points
generates curves $C_1$ and $C_2$ within $S_*$ associated to $X_1$ and
$X_2$ respectively. These curves $C_i$ each specify an $E_8$ bundle
over $S_*$. The way that these ``spectral curves'' (or ``cameral
curves'' to be more precise) determine the bundles is beyond the scope
of these lectures. We refer to \cite{Don:F,FMW:F,FMW:ell,me:hyp} for
details. See also \cite{BM:Fbun} for a discussion of this problem from
a toric point of view.

Note also that we have the R-R degrees of freedom in the type IIA
string from 3-cycles which are invariant under monodromy in $D$ around
$u=0$. Some of these R-R degrees of freedom are essential in determining the
$E_8$ bundle structure. They translate into specifying a line bundle
over the spectral curve. The remaining R-R degrees of freedom describe
much of the $B$-field degree of freedom of the heterotic string on
$S_H$ \cite{me:hyp}. The ``lost'' R-R degrees of freedom which are not
invariant around the stable degeneration $u=0$ can be matched up with the
deformations of $S_H$ which kill the elliptic fibration and/or the
section \cite{AP:hetcor}.

While we will not explain here how to determine the $E_8$ bundles
exactly we will list some of the interesting results we discover in
this classical limit. There are a plethora of possibilities! As is
common we will refer to the characteristic class of the bundle in
$H^4(S_{H},\Z)$ as ``$c_2$'' even when this bundle is not a $\GU(n)$-bundle.

\begin{enumerate}
\item We may deform a smooth vector bundle so that all of its
curvature is concentrated at points. The fundamental such point has
$c_2=1$ and is known as a ``point-like instanton''
\cite{W:small-i}. It was shown in \cite{AD:tang} that such objects can
naturally be thought of as an ideal sheaf of a point.
These point-like instantons
produce a phase transition as described in section
\ref{sss:tens} \cite{SW:6d,MV:F2}. That is, once we deform a bundle to
obtain such an instanton, we obtain a new massless tensor which we may use to
move down into the Coulomb phase.
\item We may acquire ADE singularities in $S_H$. If the bundle is
suitably generic in this case nothing interesting happens.
\item We may acquire ADE singularities in $S_H$ and let point-like
instantons collide with these singularities. All possible cases were
determined in \cite{AM:po}. For example, a collection of $k$ point-like
$E_8$ instantons 
on a $\C^2/\Z_m$ (that is, type ${A}_{m-1}$) quotient singularity,
where $k\geq2m$,
yields $k$ new tensor directions in the Coulomb branch and a local
contribution to the gauge symmetry of\footnote{It is possible that
the actual group is a discrete quotient of this. This comment also applies
to later examples of this nature.}
\begin{equation}
  \cG\cong\SU(2)\times\SU(3)\times\ldots\times\SU(m-1)\times
     \SU(m)^{k-2m+1}\times\SU(m-1)\times\ldots\times\SU(2).
		\label{eq:Gan}
\end{equation}
One may show that the case $k<2m$ reduces to the case
obtained by replacing $m$ with the integer part of $k/2$.
\item One may put fractional point-like instantons on orbifold
points. That is, one may concentrate all the curvature of a vector
bundle at an orbifold point such that the remaining holonomy is a 
discrete group which embeds into group associated to the orbifold 
singularity. Note that for such a bundle we need not have a local integral 
contribution to $c_2$.
Many possibilities were discussed in \cite{AM:frac}. The
interesting feature here is that the finite part of the Mordell--Weil
group of the fibration $p:X\to\Theta$ plays an important r\^ole. Also in this
case, the specific embedding of the holonomy in $E_8\times E_8$ must be
specified.

For example, suppose we take $S_H$ to have a singularity of the form 
$\C^2/\Z_2$ (or $A_1$) and take the $B$-field associated to this to be
zero. Then we build the simplest point-like instanton
on this which has monodromy $\Z_2$ and breaks $E_8$ to
$(E_7\times\SU(2))/\Z_2$. Such an instanton then has $c_2=\ff12$ and
produces no Coulomb branch or new gauge group enhancement.

\item If however we take the same $\C^2/\Z_2$ singularity but now
break $E_8$ to 
$\Spin(16)/\Z_2$ then the resulting instantons have $c_2=1$ and each
produces a nonperturbative contribution of $\SU(2)$ to the gauge
group.

\item One may ``embed the spin connection in the gauge group'' to
break $E_8\times E_8$ to $E_8\times E_7$ and then take the limit where
one again acquires a $\C^2/\Z_2$ singularity with zero $B$-field. This
was analyzed in \cite{AD:tang}. In this case one obtains a point-like
instanton with $c_2=\ff32$ and a nonperturbative contribution of
$\SU(2)$ to the gauge group. No new massless tensors appear.
\end{enumerate}

By counting point-like instantons one may also arrive at the following
\cite{MV:F2} (see also \cite{me:lK3} for more details)
\begin{prop}
A type IIA string compactified on an elliptic fibration (with section)
over the Hirzebruch surface $\HS n$ is dual to an $E_8\times E_8$
heterotic string compactified on $(V_S\to S_H)\times(V_E\to E_H)$
where $V_S=V_S^{(1)}\oplus V_S^{(2)}$. The bundles $V_S^{(1)}$ and 
$V_S^{(2)}$ are
then $E_8$ bundles where $c_2(V_S^{(1)})=12-n$ and $c_2(V_S^{(2)})=12+n$ .
\end{prop}
The example of section~\ref{sss:ex1} corresponds to an elliptic
fibration over $\HS 2$. Thus it corresponds to an
$E_8\times E_8$ bundle on $S_H$ whose $c_2$ is split $(10,14)$.

Some of the above results may also be approached using toric methods. We
refer to \cite{CPR:E8-17} for some examples. See also \cite{PR:hetmir}
for an interesting conjecture concerning mirror symmetry and these results.

Note that in addition to nonperturbatively enhanced gauge symmetry and
new massless tensor multiplets, one may also acquire new massless
hypermultiplets nonperturbatively. Although these hypermultiplets are
massless, they need not provide new directions in the hypermultiplet
moduli space. In order to do so they must give massless fields which
remain massless when we try to use the fields to move in the moduli
space. The usual Higgs mechanism as described above dictates which
hypermultiplets remain massless even when one tries to move off into a
Higgs branch.

Although we have only specified the F-theory rules for analyzing
enhanced gauge symmetry and extra massless tensors, there is an
assortment of rules for determining the hypermultiplet spectrum and
its transformation rules under the gauge symmetry. This is a
fascinating subject which links the theory of Lie algebras to the
geometry of elliptically fibred \CY\ threefolds. We will not discuss
this subject here as it is still a little incomplete. We refer the
reader to \cite{BSV:D-man,IMS:5deg,CPR:mtor,AKM:lcy} for more details.

A quantum field theory with $N=(1,0)$ supersymmetry in six dimensions
coupled to gravity may have chiral anomalies coming from both gravity
and Yang--Mills. One of the remarkable facts about the F-theory
description of these six-dimensional theories is that a massless
spectrum is always generated such that all these anomalies cancel. See
\cite{me:lK3} for an example of this. Why the geometry of \CY\ 
threefolds should know about these anomalies is currently a mystery.


\iffigs
\begin{figure}[t]
  \centerline{\epsfxsize=14cm\epsfbox{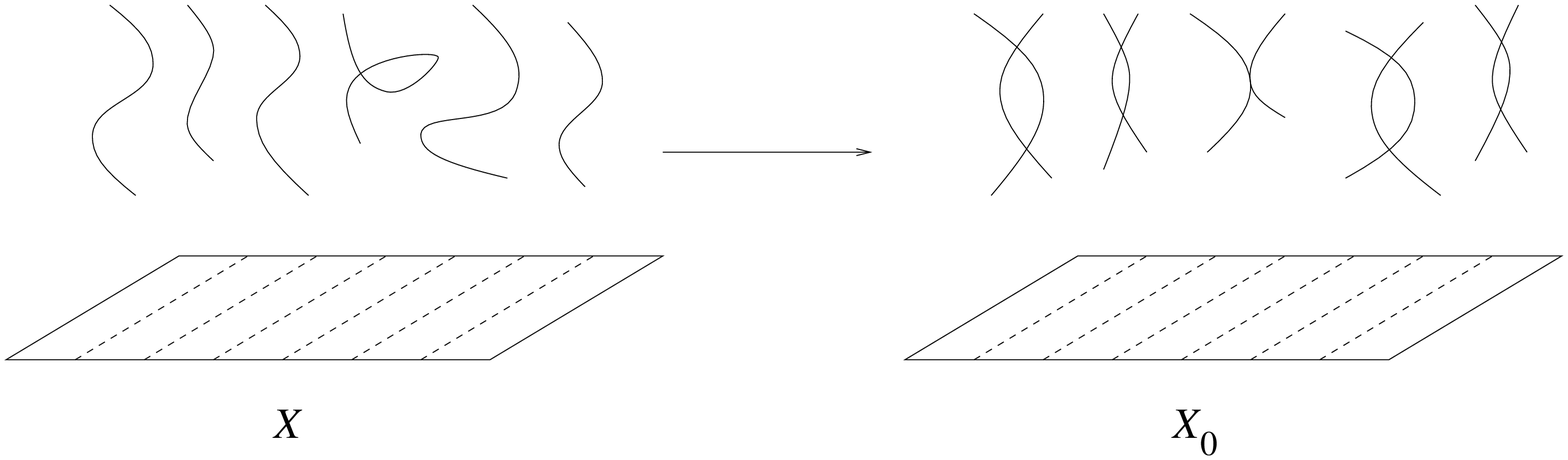}}
  \caption{The $\spnh$ stable degeneration.}
  \label{fig:sdegs}
\end{figure}
\fi

\subsubsection{The $\spnh$ heterotic string}  \label{sss:s32l}

Since we have discussed many of the peculiar properties of the
classical limit of an $E_8\times
E_8$ heterotic string on various bundles on a K3 surface, we should now
be able to have just as much fun with the $\spnh$ heterotic
string. Unfortunately at the present point in time there has been less
attention paid to this string, at least in the context of F-theory.

Having said that the $\spnh$ string is more amenable to analysis in
terms of open strings --- the $\spnh$ is believed to be dual to the
type I open string. This allows $D$-brane technology to be used as was
done in \cite{W:small-i,BLPSSW:so32,DM:qiv,BI:6d1,BI:6d2}.

There is a stable degeneration in the $\spnh$ case but it is quite
different to the $E_8\times E_8$ case \cite{me:sppt,AM:po}. This time, the
elliptic fibres break in half as opposed to the base. A generic
elliptic fibre becomes two rational curves intersecting at two points
(i.e., an $\mathrm{I}_2$ fibre in Kodaira's notation) as depicted in
figure~\ref{fig:sdegs}. At some of the fibres these two rational
curves only intersect at a single point. Thus $X$ becomes a reducible
space $X_0=X_a\cup X_b$ where $X_a\cap X_b$ is a double cover of the
base branched over some subspace. This intersection is again a K3
surface which we take to be equivalent to $S_H$.

Another difference between the $E_8 \times E_8$ heterotic string and the
$\spnh$ heterotic string is that in the latter case the bundle data
has yet to be elucidated. Determining the exact way the $\spnh$ vector
bundle data is encoded in $X_a$ and $X_b$ may not be particularly
difficult and is a problem which should be investigated. Here is a
collection of some known results:

\begin{enumerate}
\item We may deform a smooth vector bundle so that all of its
curvature is concentrated at points. The fundamental such point has
$c_2=1$ and is known as a ``point-like instanton''
\cite{W:small-i}. $k$ such instantons coincident at a smooth point in
$S_H$ will yield an enhanced gauge symmetry of $\Sp(k)$.
\item We may acquire ADE singularities in $S_H$. If the bundle is
suitably generic in this case nothing interesting happens.
\item We may acquire ADE singularities in $S_H$ and let point-like
instantons collide with these singularities. All possible cases were
determined in \cite{AM:po,BI:6d2}. For example, consider a collection
of $k$ point-like $\spnh$ instantons on a $\C^2/\Z_m$ (that is, type
${A}_{m-1}$) quotient singularity. If $m$ is even and $k\geq2m$,
then we have $\ff12m$ new tensor directions in the Coulomb branch and 
a local contribution to the gauge symmetry of
\begin{equation}
\Sp(k)\times\SU(2k-8)\times\SU(2k-16)\times\ldots
\quad\ldots\times\SU(2k-4m+8)\times\Sp(k-2m).
\end{equation}
If $m$ is odd and $k\geq2m-2$,
then we have $\ff12(m-1)$ new tensor directions in the Coulomb branch and 
a local contribution to the gauge symmetry of
\begin{equation}
\Sp(k)\times\SU(2k-8)\times\SU(2k-16)\times\ldots
\quad\ldots\times\SU(2k-4m+4).
\end{equation}
For smaller values of $k$ we refer to \cite{AM:po}.
\item Suppose we put a point-like instanton with $\Z_2$ monodromy on
an $A_1$ singularity such that $\spnh$ in broken to
$\GU(16)/\Z_2$. A minimal such instanton has $c_2=1$ and gives no new
gauge symmetry or tensors \cite{BLPSSW:so32}.
\item One may produce a peculiar point-like instanton called a
``hidden obstructer'' which may live anywhere in $S_H$, has $c_2=4$,
and produces a massless tensor leading to a Coulomb phase \cite{me:sppt}.
\end{enumerate}

Again by counting point-like instantons one may arrive at the
following \cite{MV:F2,me:sppt}
\begin{prop}
A type IIA string compactified on an elliptic fibration (with section)
over the Hirzebruch surface $\HS n$ is dual to a $\spnh$
heterotic string compactified on $(V_S\to S_H)\times(V_E\to E_H)$ with 
$4-n$ hidden obstructers and where $c_2(V_S)=8+4n$.
\end{prop}


\subsection{Into the interior}   \label{ss:spooky}

So far we have danced around the edges of the moduli space $\cM_H$
where we may ignore both the $\alpha'$-corrections to the heterotic
moduli space and the $\lambda$-corrections of the type II moduli
space. Surprisingly little is known about what happens if one ventures
into the interior of the moduli space. We collect here briefly the few known
results.


\subsubsection{The hyperk\"ahler limit}

We already mentioned this in section \ref{sss:d=3}. In effect we may
look at the ``first order'' behaviour as we move away from the classical
limit. 

In any of the examples where we had a perturbative gauge symmetry we
may ask what happens if we allow this theory to interact (i.e., allow
some coupling or some effective scale $\Lambda$ to be nonzero)
while keeping the effective gravitational coupling zero. This would
lead to a field theory limit which is described by a hyperk\"ahler
moduli space. This is the ``rigid limit'' of the quaternionic k\"ahler
manifold in the same sense as we had a rigid special K\"ahler limit of
a special K\"ahler manifold. Proposition \ref{prop:SW3} by Seiberg
and Witten gives a powerful tool in this respect.

In terms of the heterotic compactification picture we go to the
hyperk\"ahler limit by rescaling the overall size of $S_H$ to
infinity. In order to get something interesting we simultaneously
scale down some minimal 2-spheres to keep their areas finite. The
result is that we end up describing a heterotic string on an ALE
space.

The analysis of such systems is perhaps best done by using various
dualities involving D-branes along the lines of
\cite{Sei:Bdyn}. Because of this we will regard this subject as
beyond the scope of these lectures.

We will give one interesting result however. Suppose one were to
consider perhaps the simplest case of $k$ point-like instantons moving
around an ALE space of type $A_{m-1}$. One can then show
\cite{Sen:kkm,W:hADE,Roz:hyp3d,AP:hetcor,Mayr:3d} that
the resulting hyperk\"ahler moduli space with $k+m-1$ quaternionic
dimensions is the same as you would get from the $c$-map of section
\ref{sss:d=3} applied to the rigid limit of $\cM_V$ for a theory with
gauge symmetry $\SU(m)\times\GU(1)^k$. In other words, suppose our desired
moduli space is the hyperk\"ahler limit of $\cM_H$ which is given by
the type IIA string on $X$. Then the type IIA string compactified on
$Y$, the mirror of $X$, would yield a gauge symmetry of
$\SU(m)\times\GU(1)^k$. 

We know from section \ref{sss:E8l} that when we go to the classical
limit of this theory we will get a gauge group of the form
(\ref{eq:Gan}). That is, we are in the Higgs branch of a field theory
associated to the gauge group (\ref{eq:Gan}).

From section \ref{sss:d=3} this implies that in the three-dimensional
picture, mirror symmetry exchanges a field theory with gauge group
$\SU(m)\times\GU(1)^k$ with a field theory with gauge group given by
(\ref{eq:Gan}). This is a statement of ``Intriligator--Seiberg mirror
symmetry''. See \cite{IS:3dmir} for many examples of such mirror pairs
and \cite{HOV:mir,AP:hetcor} for further discussion of this example.

Clearly analysis of this hyperk\"ahler limit is much easier than a
discussion of the quaternionic K\"ahler $\cM_H$ in its full
glory. This is essentially because one ends up studying field theory
(without gravity) rather than full string theory.


\subsubsection{Mixed instantons}    \label{sss:mixed}

Both the type IIA and type IIB strings suffer from
$\lambda$-corrections when studying $\cM_H$. In \cite{BBS:5b} it was
argued that one could study the associated instantons by considering
maps of certain cycles into the \CY\ space. 
These cycles represent the world-volume of D-brane solitons.
In a way therefore these
$\lambda$-corrections could be modeled by something that looks like a
generalization of worldsheet instantons.

In the case of a the type IIA string on a \CY\ space $X$, one needs to
consider ``supersymmetric'' or ``special Lagrangian'' minimal 3-cycles
embedded in $X$. (On a related note, such 3-cycles have also achieved
prominence from the mirror conjecture of \cite{SYZ:mir}.) Because
counting these 3-cycles is very difficult, this approach to computing
the quantum corrections has not to date been very useful. Indeed, it
will probably be easier to compute the quantum corrections in some
other way and then use this to predict the number of 3-cycles --- just
as was done for rational curves.

For the type IIB string, the instanton $\lambda$-corrections come from
{\em even\/}-dimensional cycles in $Y$, including rational
curves. Remember that we also have worldsheet instanton corrections
coming from rational curves in $Y$. Thus it would appear at first that
in order to compute the quantum corrections to $\cM_H$ we should count
the rational curves in terms of worldsheet instantons and then add to
this the contribution of rational curves from D-1-brane worldsheets.

It was shown in \cite{AP:hetcor} that this is not the full story. The
subtleties of our discussion of quantum corrections in section
\ref{ss:who} turn out to have real significance. We only really understand
worldsheet instantons when $\lambda=0$ and we only understand the
D-brane instantons when $\alpha'=0$. We have no right to trust either
of these pictures when we set both $\lambda$ and $\alpha'$ to be nonzero.

By analyzing a heterotic string on $S_H\times E_H$ which is dual to
the type IIB string on $Y$, one may show that there are many quantum
corrections which correspond to instantons which depend on many
different combinations of $\alpha'$ and $\lambda$ \cite{AP:hetcor}. It
is as if we had instantons which are both worldsheet and spacetime
simultaneously. 

One very rough way of saying what happens is that the type IIB string
in ten dimensions has an $\Sl(2,\Z)$ symmetry which permutes the
fundamental string with ``$(p,q)$-strings'' for any relatively prime
$(p,q)$. One then needs to add up the contribution from instantons
from all of these $(p,q)$-strings. On closer inspection this
description as it stands is flawed. Firstly, S-duality, like any
U-duality, is broken when we have only modestly extended
supersymmetry. This was shown explicitly for the type IIB string on
$Y$ in \cite{BGHL:IIBS}. Secondly we do not really have a formulation
of $(p,q)$-strings which allows one to make much sense of a
computation of instanton corrections.

Understanding these mixed instanton corrections may be one of the most
challenging problems for our current definitions of string theory. It
may be that we need to replace our basic formulation of string theory
to be able to make sense of this problem.


\subsubsection{Hunting the universal hypermultiplet}   \label{sss:univ}

We will close our discussion of the hypermultiplet moduli space by
further demonstrating how troublesome analysis of $\cM_H$ can be. We
want to analyze the question of whether the dilaton belongs to some
special hypermultiplet which may have some universal properties for
any $\cM_H$. We will begin by a quick review of some general facts about
quaternionic geometry.

It is well-known that we may put patches of complex coordinates on a
complex manifold $M_{\C}$. That is, we may take some open
neighbourhood in $M_{\C}$ with a homeomorphism to some open subset of
$\C^n$. Then 
do this for a collection of patches covering $M_{\C}$ such that the
coordinates are related by elements of $\Gl(n,\C)$ between patches. We
may also consider complex {\em submanifolds\/} of $M_{\C}$. The patches on
such submanifolds map holomorphically to the patches of $M_{\C}$.

Unfortunately this does not work at all as nicely for quaternionic
K\"ahler manifolds $M_{\H}$. We refer to section 14.F of
\cite{Besse:E} for more details and references. One might suppose that
one could consider patches homeomorphic to an open subset of $\H^n$
such that these coordinates were related by elements of
$\Sp(1).\Gl(n,\H)\subset\Gl(4n,\R)$.  We multiply by $\Sp(1)$ on the
left and by $\Gl(n,\H)$ on the right to try to match the holonomy
structure discussed in section \ref{ss:hol}.  These would be patches
of ``quaternionic coordinates''. Unfortunately the only spaces which
can admit such a structure are necessarily locally projectively
equivalent to quaternionic projection space $\H\P^n$
\cite{Kul:Hman}. The hypermultiplet moduli spaces one encounters in
string theory are not expected to be of this specific form. In other
words we would not expect the quaternionic structure of $\cM_H$ to be
``integrable''. 

For a typical $\cM_H$ one cannot think in terms of quaternionic
coordinates. While it is true that the scalars in a hypermultiplet
give a quaternion, these scalars only give {\em tangent directions\/}
in the moduli space. There is no way to integrate such a quaternionic
structure a nonzero distance along such directions. In other words if
one tries to start at a generic point in space and then integrate
along the tangent directions given by the 4 massless scalars of a
chosen hypermultiplet then one will lose the hypermultiplet
structure. The four scalars one ends up with will not be mapped purely
into each other by the $\Sp(1)$ $R$-symmetry.

There is also generically a lack of existence of quaternionic
submanifolds in a generic quaternionic K\"ahler manifold, by which we
mean the following. If one considers the tangent bundle at a given
point $\cM_H$ one can certainly see a quaternionic structure. One may
pick a quaternionic subspace of this and try to integrate along these
quaternionic directions to map out a submanifold. After integrating a
nonzero distance one will generically discover that one has rotated
out of the desired quaternionic structure. In other words, the
$\Sp(1)$ part of the holonomy will no longer have a closed action
within the new tangent directions. 

Having said this, if one chooses the starting point and tangent
directions carefully one can sometimes integrate to find closed
manifolds which {\em are\/} compatible with the quaternionic
structure. We may call such rare objects quaternionic submanifolds. We
emphasize that finding quaternionic submanifolds of a quaternionic
manifold is a {\em much\/} harder problem than finding complex submanifolds of
a complex manifold.

In \cite{CFG:II} the notion of a ``universal hypermultiplet'' was
introduced. If one ignores $\lambda$-corrections to a type II
compactification one might argue from the conformal field theory that
the hypermultiplet in which the dilaton lives somehow decouples from
the rest of the theory. If this were the case then one could find this
universal hypermultiplet by studying any particularly simple
example. Consider compactifying the type II string on a 6-torus to
obtain a theory in four dimensions with $N=8$ supersymmetry. Now imagine
what would happen to the moduli space if one embedded the $\GU(2)$
$R$-symmetry of $N=2$ into the $\GU(8)$ $R$-symmetry of $N=8$. 
It was argued in \cite{CFG:II} that this leads to a natural
embedding
\begin{equation}
\frac{E_{7(+7)}}{\SU(8)} \supset \frac{\Sl(2,\R)}{\GU(1)} \times
			\frac{\SU(2,1)}{\mathrm{S}(\GU(2)\times\GU(1))}.
	\label{eq:8to2}
\end{equation}
The right-hand-side is therefore a possible moduli space for an $N=2$
system (embedded in an $N=8$ system). Clearly the first factor would
be $\cM_V$ and the second factor would be $\cM_H$. This would suggest
that if a universal hypermultiplet exists it must be of the form
$\SU(2,1)/\mathrm{S}(\GU(2)\times\GU(1))$. 

Even this simplest of examples shows that one cannot expect the
universal hypermultiplet to appear as a {\em factor\/} in
the moduli space. Equation (\ref{eq:8to2}) represents an embedding of the
universal hypermultiplet into the moduli space which does not
factorize. One should therefore immediately question the validity of
saying that the dilaton can be decoupled in a special way from the
other fields (even when quantum effects are ignored).

One might argue that the failure of the universal
hypermultiplet to appear as a factor might be due to an excess of
supersymmetry in the above example. This is not so as we see
shortly. The best we might 
hope for then is that the dilaton lives in a hypermultiplet which can
be integrated at least at some special points in $\cM_H$ to give a
quaternionic submanifold of dimension one.

Let us consider a class of genuine $N=2$ examples. We know from the heterotic
string that there are many cases where $\cM_H$ can be described
asymptotically (as the K3 surface gets large) by the moduli space of
K3 surfaces with bundles. In many of these cases we may freeze the
bundle moduli as well as some of the deformations of the K3 itself by
pushing point-like instantons into singularities and moving off in the
corresponding Coulomb branch. An example of this was studied in
\cite{AP:hetcor}. This implies that many examples of $\cM_H$ look
asymptotically like 
\begin{equation}
  \cM_H \sim \GO(\Lambda_{4,n})\backslash\GO(4,n)/(\GO(4)\times\GO(n)),
	\label{eq:my-u}
\end{equation}
for some $n$ and some lattice $\Lambda_{4,n}$. Indeed in a few special
examples such as \cite{FHSV:N=2} there are no quantum corrections and
this moduli space is exact (see \cite{OS:CYsim} for the classification
of this type of example).

Now it is known \cite{Gray:Hsub} that any quaternionic submanifold of
$\cM_H$ must be totally geodesic. From an old result of E.~Cartan, the
totally geodesic submanifolds of 
a symmetric space are always determined
by Lie triples which have been classified (see \cite{Faulk:Ltrip}
for example). This will actually allow for an embedding of the
universal hypermultiplet (assuming $n>1$):
\begin{equation}
\frac{\SO_0(4,n)}{\SO(4)\times\SO(n)}\subset
  \frac{\SO_0(4,2)}{\SO(4)\times\SO(2)}\cong
  \frac{\SU(2,2)}{\mathrm{S}(\GU(2)\times\GU(2))}\subset
  \frac{\SU(2,1)}{\mathrm{S}(\GU(2)\times\GU(1))}.
\end{equation}
Note however that (\ref{eq:my-u}) does not factorize in any way.

This embedding relies very much on the special properties of symmetric spaces.
The question we should address however is whether this delicate
embedding can be expected to remain when $\lambda$-corrections are
taken into account. If the deformation of $\cM_H$ produced by these
quantum corrections is sufficiently generic then this embedding will be
destroyed even if we were to allow for deformations of the universal
hypermultiplet itself. 

Until we know more about $\lambda$-corrections this is impossible to
address but for now it would seem to be most prudent to assume that
any notion of a universal hypermultiplet, even if only as a
quaternionic submanifold of $\cM_H$ rather than a factor, should be doubted.

Since it was the quaternionic structure that caused problems above one
might consider 
an alternative approach to finding the dilaton without trying to keep
it cooped up in a special hypermultiplet.

It is tempting to conjecture that (\ref{eq:my-u}) is
the universal behaviour of $\cM_H$ in the weakly-coupled limit. We can
then try something like a decomposition of this symmetric
space along the lines of \cite{W:dyn,AM:Ud} into a warped product such as
\begin{equation}
  \frac{\SO_0(4,n)}{\SO(4)\times\SO(n)}\cong
    \frac{\Sl(2,\R)}{\GU(1)}\times\frac{\SO_0(2,n-2)}{\SO(n-2)\times\SO(2)}
    \times(\R_+\times\R) \times\R^{2n},
\end{equation}
where we have pulled the dilaton out as the $\R_+$ factor. Actually
this decomposition is well-suited to understanding the stable
degenerations of section \ref{ss:Hclas}. We leave it as an interesting
exercise for the reader to interpret each factor (although see
\cite{AP:hetcor} for hints!).

Of course, this symmetric space is only the asymptotic form of the moduli
space $\cM_H$. The quantum corrections will make everything much
more difficult to analyze. Clearly we have much about $\cM_H$ to learn!

\section*{Acknowledgements}

It is a pleasure to thank R.~Bryant, D.~Morrison, R.~Plesser and
E.~Sharpe for numerous conversations and collaborations on topics
covered in these lectures.
I would also like to thank S. Kachru, J.~Harvey, K.~T.~Mahanthappa and
E.~Silverstein for organizing TASI99.
The author is supported in part by a research fellowship from the Alfred
P.~Sloan Foundation. 


\end{document}